\documentstyle[11pt]{article}
\begin{document}
\parskip=5pt plus 1pt minus 1pt

\begin{flushright}
\framebox{\large\bf hep-ph/9912358} \\
{\bf LMU-99-16} 
\end{flushright}

\vspace{0.2cm}

\begin{center}
{\Large\bf Mass and Flavor Mixing Schemes of Quarks and Leptons}
\end{center}

\vspace{0.4cm}
\begin{center}
{\bf Harald Fritzsch} 
~ {and} ~
{\bf Zhi-zhong Xing}
\footnote{E-mail: xing@hep.physik.uni-muenchen.de} \\
{\it Sektion Physik, Universit\"at M\"unchen,
Theresienstrasse 37, 80333 M\"unchen, Germany} 
\end{center}

\vspace{2.5cm}
\begin{abstract}
We give an overview of recent progress in the 
study of fermion mass and flavor mixing phenomena.
The hints exhibited by the quark and lepton mass spectra 
towards possible underlying flavor symmetries, from which realistic
models of mass generation could be built, are emphasized.
A variety of schemes of quark mass matrices 
at low and superhigh energy scales are described,
and their consequences on flavor mixing and $CP$ violation
are discussed. Instructive patterns of 
lepton mass matrices, which can naturally lead to large flavor 
mixing angles, are explored to interpret current data on
atmospheric and solar neutrino oscillations.
We expect that $B$-meson factories and long-baseline
neutrino experiments will soon shed more light on the dynamics
of fermion masses, flavor mixing and $CP$ violation.
\end{abstract}

\vspace{2cm}
\begin{center}
Invited Article to appear in {\sf Vol. 45} of 
{\sf Prog. Part. Nucl. Phys.} \\ 
Elsevier Science, The Netherlands (2000)
\end{center}

\newpage

\tableofcontents
\setcounter{footnote}{0}

\newpage

\section{Introduction}
\setcounter{equation}{0}
\setcounter{figure}{0}

Since its foundation in the 1960's the standard electroweak model,
which unifies the weak and electromagnetic interactions, has
passed all experimental tests. Neither significant evidence for the
departures from the standard model nor convincing hints for the
presence of new physics has been found thus far at HERA, LEP, SLC, Tevatron
and other high-energy facilities \cite{PDG}. In spite of the impressive
success of the standard model, many physicists believe that it
does not represent the final theory, but serves merely as an effective
theory originating from a more fundamental, yet unknown theoretical
framework.
For instance there is little understanding, within the standard model, 
about the intrinsic physics of the electroweak symmetry breaking, 
the hierarchy of charged fermion mass spectra, the vanishing or smallness 
of neutrino masses, and the origin of flavor mixing and $CP$ violation.
Any attempt towards gaining an insight into such problems 
inevitably requires significant steps to go beyond the standard model.

The investigations of fermion mass generation and flavor mixing
problems, which constitute an important part of today's particle
physics, can be traced back to the early 1970's, soon after the
establishment of the standard electroweak model. Since then many 
approaches have been developed, in the contexts of different
theoretical and phenomenological models. Regardless of the 
energy scales at which those models are built, the mechanisms
for fermion mass generation and flavor mixing can roughly be classified 
into four different types: 
(a) Radiative mechanisms \cite{Weinberg72}, 
(b) Texture zeros \cite{Fr77,Fr78}, 
(c) Family symmetries \cite{Harari78,Froggatt79}, and 
(d) Seesaw mechanisms \cite{Seesaw}. These mechanisms cannot be
regarded as disjoint from one another; rather they are related.
The mechanism (d) is especially related to a natural interpretation 
of the smallness of neutrino masses. 
Phenomenologically some striking progress has been made,
in particular with the help of the mechanisms (b) and (c), in
specifying the quantitative relationship between flavor mixing angles
and quark mass ratios \cite{FX99}. From the theoretical point of
view, however, our understanding of the fermion mass spectrum
remains quite unsatisfactory. In a model with effective 
higher-dimension fermion mass operators and non-abelian family
symmetries, for instance, the {\it observed} quark mass hierarchy
is interpreted by the {\it assumed} texture of U(1) flavor
charges or Higgs-field vacuum expectation values. Another
example is the determination of left-handed Majorana neutrino
masses from the {\it assumed} textures of Dirac and (or) right-handed 
Majorana neutrino mass matrices through the seesaw mechanism.
In many cases, the problem seems to be transferred from one
place to another, leaving the model itself with few 
testable predictions. Before a significant breakthrough 
takes place on the theoretical side, the phenomenological 
approaches will remain to play a crucial role in interpreting new
experimental data on quark mixing, $CP$ violation, and neutrino
oscillations. They are expected to provide useful hints towards
discovering the full dynamics of fermion mass generation and
$CP$ violation.

This article aims at giving an overview of recent progress in the 
phenomenological
study of fermion masses, flavor mixing and $CP$ violation, in particular with
respect to the searches for underlying discrete or continuous 
flavor symmetries which can lead to a realistic texture of
fermion mass matrices. We are motivated not only by the theoretical
significance of these topics, as briefly outlined above, but also by their 
experimental prospects at present and in the near future. The new
$e^+e^-$ and hadronic $B$-meson factories are about to open a new era 
to determine the flavor mixing parameters and the $CP$-violating
phases in the quark sector to an unprecedented degree of accuracy.
A variety of neutrino experiments, being done or to be done, are
expected to pin down the true mechanism of neutrino oscillations
and to provide a wealth of precise information about neutrino masses,
lepton flavor mixing angles, and even leptonic $CP$ violation. 
Such experimental developments, together
with those in searching for new particles at much higher energy scales,
will allow us to gain a deeper understanding of the standard 
model, especially the sector of Yukawa interactions.

The remaining parts of this article are organized as follows.
An overview of the fermion mass spectra is given in section 2.
We list the values of quark masses at the weak-interaction scale
$\mu = M_Z$ and highlight the hierarchy features of both
the charged lepton mass spectrum and the quark mass spectrum. For neutrinos
the upper mass bounds from the direct-mass-search experiments
and the mass-squared differences from the neutrino oscillation
experiments are summarized. The distinctions between Dirac and
Majorana neutrino masses are briefly described.

In section 3 we illustrate the main features of fermion flavor
mixing and $CP$ violation. The numbers
of flavor mixing and $CP$-violating parameters are counted, for
both quarks and leptons. The necessary and sufficient conditions
of $CP$ violation in the standard electroweak model are clarified.
A geometric description of $CP$ violation, in terms of the unitarity 
triangles in the complex plane, is introduced for both quark 
and lepton sectors. 
After classifying a variety of different parametrizations
of the $3\times 3$ flavor mixing matrix, we highlight a unique
one which is particularly favored for the study of quark mass
matrices and $B$-meson physics.

Section 4 is devoted to the realistic schemes of quark mass 
matrices. First of all we derive the flavor mixing and
$CP$-violating parameters from a generic Hermitian texture of
$3\times 3$ quark mass matrices. Two useful symmetry limits
of quark masses are taken into account, and the concept of
the light-quark triangle is introduced.
We then discuss an interesting Hermitian pattern with four
texture zeros, and explore its consequences on flavor mixing and
$CP$ violation in detail. In particular we show that the 
light-quark triangle is congruent with the rescaled unitarity
triangle to a good degree of accuracy. The Hermitian schemes of quark 
mass matrices with five texture zeros are briefly summarized.
The idea of flavor democracy is stressed, and the possible
breaking patterns of this symmetry are discussed in order to
generate the light quark masses and the flavor mixing
angles. We also look at the non-Hermitian textures of 
quark mass matrices by taking three typical examples, i.e.,
the nearest-neighbor mixing pattern, the triangular pattern,
and the pure phase pattern. Finally the running effects of
quark masses and flavor mixing parameters from superhigh energy
scales to the weak scale are illustrated in the framework
of the minimal supersymmetric standard model. A specific 
ansatz of quark and charged lepton mass matrices is proposed
at the scale of supersymmetric grand unified theories, and
its low-energy consequences on flavor mixing and $CP$ violation
is analytically calculated.

Section 5 is devoted to lepton mass
matrices and neutrino oscillations. 
In view of current experimental data on atmospheric
and solar neutrino oscillations, we make a generic
classification of possible textures of $3\times 3$
neutrino mass matrices. To be specific we study
a simple model of lepton flavor mixing and $CP$ violation
based on the breaking of the charged lepton flavor
democracy and the neutrino mass degeneracy. The
numerous consequences of this model, including 
nearly bi-maximal, bi-maximal and small-versus-large
mixing patterns, are explored in detail for neutrino oscillations.
The texture zeros of lepton mass matrices,
similar to those of quark mass matrices, are taken into account
to interpret the neutrino mass hierarchy and large lepton
flavor mixing. In particular we emphasize that a seesaw-invariant
texture of lepton mass matrices could naturally originate
from a grand unified theory. We present an illustrative
scheme of four-neutrino mixing, which can well accommodate the present
solar, atmospheric and accelerator neutrino oscillation data.
The scale dependence of the neutrino mass matrix is 
qualitatively described by using the renormalization-group
equations. Finally we point out the essential features of
leptonic $CP$ violation in both three- and four-neutrino
mixing scenarios, and discuss the possibility to measure
$CP$ and $T$ asymmetries in the long-baseline neutrino experiments.

The conclusion and outlook are given in section 6. 

\section{Overview of fermion mass spectra}
\setcounter{equation}{0}
\setcounter{figure}{0}

\subsection{Charged lepton and quark masses}

In the standard $\rm SU(3) \times SU(2) \times U(1)$ model of
strong, weak and electromagnetic interactions, it is the 
Higgs mechanism that provides a theoretically consistent framework
to generate masses for gauge bosons and fermions -- the latter 
acquire masses, after spontaneous breaking of the 
$\rm SU(2)$ gauge symmetry, through the Yukawa 
couplings and the vacuum expectation value of
the neutral Higgs field. This framework, however, can neither
predict the values of fermion masses nor interpret the observed
hierarchy of their spectra. Hence the three charged lepton masses
and six quark masses are free parameters of the 
standard model. The vanishing of three neutrino masses 
follows as a straightforward consequence of the symmetry structure of
the standard model. 

The physical mass of a charged lepton is just the pole of
its propagator and can directly be measured. 
We have \cite{PDG}
\begin{eqnarray}
m_e & = & 0.51099907 \pm 0.00000015 ~ {\rm MeV} \; , \nonumber \\
m_\mu & = & 105.658389 \pm 0.000034 ~ {\rm MeV} \; , \nonumber \\
m_\tau & = & 1777.05^{+0.29}_{-0.26} ~ {\rm MeV} \; .
\end{eqnarray}
The fact that the mass spectrum of charged leptons 
is almost entirely dominated by the tau-lepton mass strongly
suggests the existence of a ``rank-one'' limit of
the corresponding mass matrix:
\begin{equation}
M_{0l} \; =\; C^{~}_l \left ( \matrix{
0       & 0     & 0 \cr
0       & 0     & 0 \cr
0       & 0     & 1 \cr} \right ) \; \; ,
\end{equation}
in which $m_e = m_\mu =0$ and $m_\tau = C^{~}_l$. The
realistic mass matrix $M_l$ can be obtained if
proper perturbative corrections to $M_{0l}$, from 
which the light charged leptons $e$ and $\mu$ become massive
and non-degenerate, are taken into account \cite{Fr87}. 

Since quarks are confined inside hadrons, 
their masses cannot directly be measured. The only way to
determine the quark masses is through the study of their impact on 
hadron properties. The quark mass parameters in the QCD 
and electroweak Lagrangians depend both on the renormalization scheme
adopted to define the theory and on the scale parameter
$\mu$ -- this dependence reflects the fact that a bare quark is surrounded
by a cloud of gluons and quark-antiquark pairs. 
In the limit where all quark masses vanish, the QCD Lagrangian
has a $\rm SU(3)_L \times SU(3)_R$ chiral symmetry, under which left- 
and right-handed quarks transform
independently. The scale of dynamical chiral symmetry breaking,
$\Lambda_{\chi} \approx 1$ GeV, can be used to distinguish between light quarks
($m < \Lambda_{\chi}$) and heavy quarks 
($m > \Lambda_{\chi}$) \cite{Leutwyler82}.
To determine the quark mass values one may make use of the QCD 
perturbation theory at high energy scales, i.e., $\mu \gg \Lambda_{\chi}$,
where nonperturbative effects such as chiral symmetry breaking are
negligible.

Useful information on the mass ratios of light quarks can be obtained
from analyzing properties of the light pseudoscalar mesons with the
help of the chiral perturbation theory. For example, it has been
argued that $m_u/m_d$ and $m_s/m_d$ fulfil the following 
relation \cite{Leutwyler96}:
\begin{equation}
\frac{m_s/m_d}{\sqrt{1 - (m_u/m_d)^2}} \; =\; 22.7 \pm 0.08 \; .
\end{equation}
The absolute values of $m_u$, $m_d$ and $m_s$, usually normalized to
the scale $\mu = 1$ GeV, can be extracted from QCD sum rules. 
The lattice gauge theory is expected to be a powerful and accurate 
tool for computing meson masses directly from the QCD Lagrangian,
thus it provides another way to determine the
light quark masses. Conservatively we list the ranges of light quark
masses, allowed by current data \cite{PDG,Narison} 
and rescaled to $\mu = 1$ GeV in the modified minimal subtraction 
($\overline{\rm MS}$) scheme, as follows:
\begin{eqnarray}
m_u  ({\rm 1 ~ GeV}) & = & 2 - 6.8 ~ {\rm MeV} \; , \nonumber \\
m_d  ({\rm 1 ~ GeV}) & = & 4 - 12 ~ {\rm MeV} \; , \nonumber \\
m_s  ({\rm 1 ~ GeV}) & = & 81 - 230 ~ {\rm MeV} \; .
\end{eqnarray}
Note that the light quark masses under discussion are 
the {\it current} masses, which have nothing to do with the {\it constituent}
masses defined in nonrelativistic quark models.

A study of the spectrum and decays of hadrons containing heavy quarks allows
one to extract useful information about the heavy quark masses. 
The calculations may be done with the help of the heavy quark effective
theory, the QCD sum rules, the lattice gauge theory, etc.. 
Within the QCD perturbation theory one may define the position of the pole
in the quark propagator as the quark mass $m^{~}_{\rm pol}$, 
the so-called {\it pole} quark
mass, which is independent of the adopted renormalization scheme. 
One may also define the $\overline{\rm MS}$ running quark mass 
$m (\mu)$ by regulating the QCD
theory using dimensional regularization and subtracting the divergences
using the $\overline{\rm MS}$ scheme. The relation between the
pole quark mass and the running quark mass at the one-loop level of 
perturbative QCD corrections reads \cite{Gray90}
\begin{equation}
m^{~}_{\rm pol} \; =\; m (m^{~}_{\rm pol}) \left [ 1 ~ + ~ 
\frac{4}{3} \cdot \frac{\alpha_s (m^{~}_{\rm pol})}{\pi} \right ] \; ,
\end{equation}
where $\alpha_s (\mu)$ is the strong-interaction coupling constant.
The generously allowed ranges of the charm and bottom running masses in
the $\overline{\rm MS}$ scheme are gived by \cite{PDG}
\begin{eqnarray}
m_c (m_c) & = & 1.1 - 1.4 ~ {\rm GeV} \; , \nonumber \\
m_b (m_b) & = & 4.1 - 4.4 ~ {\rm GeV} \; ;
\end{eqnarray}
and the experimentally measured value of the top mass is \cite{PDG}
\begin{equation}
m_t \; =\; 173.8 ~ \pm ~ 5.2 ~ {\rm GeV} \; .
\end{equation}
Given the techniques used to extract the top mass at CDF 
and $\rm D\emptyset$ \cite{CDFD0},
the mass value in (2.7) should be interpreted as the top {\it pole} mass.

The quark mass values given in (2.4), (2.6) and (2.7) indicate the 
existence of a strong mass hierarchy in both $(u, c, t)$ and $(d, s, b)$ 
quark sectors. 
To get the relative magnitudes of different quark masses in a physically
meaningful way, one has to describe all quark masses in the same scheme and
at the same scale. It is instructive to consider the light and heavy quark masses
at the scale $\mu = M_Z$, the mass of the $Z$ boson, by adopting the 
$\overline{\rm MS}$ scheme. The advantage of choosing $M_Z$ as the reference
scale is two-fold: above $M_Z$ extensions of the standard model may naturally
appear; below $M_Z$ the strong-interaction coupling constant $\alpha_s$
is sizable and special attention has to be paid to the running and the
matching in passing a heavy quark threshold. For illustration
the ranges of six quark masses given above are listed at the 
scale $M_Z$ as follows:
\begin{eqnarray}
m_u (M_Z) & = & 0.9 - 2.9 ~ {\rm MeV} \; , \nonumber \\
m_c (M_Z) & = & 0.53 - 0.68 ~ {\rm GeV} \; , \nonumber \\
m_t (M_Z) & = & 168 - 180 ~ {\rm GeV} \; ; 
\end{eqnarray}
and
\begin{eqnarray}
m_d (M_Z) & = & 1.8 - 5.3 ~ {\rm MeV} \; , \nonumber \\
m_s (M_Z) & = & 35 - 100 ~ {\rm MeV} \; , \nonumber \\
m_b (M_Z) & = & 2.8 - 3.0 ~ {\rm GeV} \; .
\end{eqnarray}
The hierarchical pattern of the up-type quark masses ($m_u, m_c, m_t$) and
the down-type quark masses ($m_d, m_s, m_b$) is remarkable.
For each quark sector the mass spectrum is dominated by the mass
of the third-family quark. This property is similar to that of the
charged lepton mass spectrum. It implies that the quark mass matrix 
$M_{\rm u}$ or $M_{\rm d}$, like the charged lepton mass matrix $M_l$, 
is close to an interesting
``rank-one'' limit, in which the masses of the first two quark families
vanish ($C_{\rm u} = m_t$ and $C_{\rm d} = m_b$):
\begin{equation}
M_{0\rm q} \; =\; C_{\rm q} \left (\matrix{
0       & 0     & 0 \cr
0       & 0     & 0 \cr
0       & 0     & 1 \cr} \right ) \; \; .
\end{equation}
Such a symmetry limit may be very suggestive for studying the
realistic textures of quark mass matrices in the flavor 
basis \cite{Fr87}.
It is also worth mentioning two approximate quantitative relations of
quark masses: 
\begin{eqnarray}
\frac{m_u}{m_c} & \sim & \frac{m_c}{m_t} \; \sim \; \lambda^4 \; , 
\nonumber \\
\frac{m_d}{m_s} & \sim & \frac{m_s}{m_b} \; \sim \; \lambda^2 \; ,
\end{eqnarray}
where $\lambda \approx 0.22$. The larger hierarchy of the up-type
quark masses implies that they may have less significant contributions
to the quark flavor mixing angles. This feature, which will be
discussed in some detail in section 4, is expected to be
true for most realistic models of quark mass matrices. 

\subsection{Massive Dirac and Majorana neutrinos}

In the standard model neutrinos are assumed to be the exactly 
massless Weyl particles. This assumption agrees with all 
direct-mass-search experiments, which have set the upper
bounds on masses of the primary mass eigenstates 
$(\nu_1, \nu_2, \nu_3)$ of electron, muon and tau
neutrinos \cite{PDG}
\footnote{The limits are kinematically obtained from the tritium
$\beta$-decay $^3_1 {\rm H} \rightarrow ~ ^3_2 {\rm He} + e^- +
\bar{\nu}_e$, the $\pi^+ \rightarrow \mu^+ + \nu_\mu$ decay and the
$\tau \rightarrow 5\pi + \nu_\tau$ (or $\tau \rightarrow 3\pi
+ \nu_\tau$) decay, respectively.}:
\begin{eqnarray}
m^{~}_{\nu_1} & < & 15 ~ {\rm eV} \; , \nonumber \\
m^{~}_{\nu_2} & < & 0.17 ~ {\rm MeV} \; , \nonumber \\
m^{~}_{\nu_3} & < & 18.2 ~ {\rm MeV} \; .
\end{eqnarray}
However, the masslessness of neutrinos is not assured by 
any basic symmetry principle of particle physics.
Indeed most extensions of
the standard model (such as the grand unified theories)
allow the existence of massive neutrinos, although the
masses of three active neutrinos 
may be extremely smaller than those of their
corresponding charged leptons. 

If neutrinos have masses, they may be either Dirac or Majorana
particles. A massive Dirac neutrino field describes four
independent states -- left-handed and right-handed particle
states ($\nu^{~}_{\rm L}$ and $\nu^{~}_{\rm R}$) as well as
left-handed and right-handed antiparticle states 
($\bar{\nu}^{~}_{\rm L}$ and $\bar{\nu}^{~}_{\rm R}$). Among
them $\nu^{~}_{\rm L}$ and $\bar{\nu}^{~}_{\rm R}$ already
exist in the standard model and can take part in weak interactions.
The $\nu^{~}_{\rm R}$ and $\bar{\nu}^{~}_{\rm L}$ states 
need to be introduced into the standard model as necessary
ingredients to give the Dirac neutrino a mass, but they should
be ``sterile'' in the sense that they would not take part in
the normal weak interactions. A Dirac mass term, which conserves
the total lepton number but violates the law of individual
lepton flavor conservation, can be written as
\begin{equation}
- {\cal L}_{\rm Dirac} \; =\; \overline{\psi}_{\rm L} M^{\rm D}_\nu
\psi_{\rm R} ~ + ~ \overline{\psi}_{\rm R} M^{\rm D\dagger}_\nu
\psi_{\rm L} \; ,
\end{equation}
where $\psi \equiv \psi_{\rm L} + \psi_{\rm R}$ denotes a
column vector in family space of the neutrino interaction
eigenstates ($\nu_e$, $\nu_\mu$ and $\nu_\tau$), and $M^{\rm D}_\nu$
is the corresponding $3\times 3$ Dirac mass matrix. 

On the other hand, the neutrino $\nu$ might be a Majorana
particle, which has only two independent particle states 
of the same mass ($\nu^{~}_{\rm L}$ and $\bar{\nu}^{~}_{\rm R}$,
or $\nu^{~}_{\rm R}$ and $\bar{\nu}^{~}_{\rm L}$). By definition,
a Majorana neutrino is its own antiparticle:
$\nu^{\rm c} \equiv C \bar{\nu}^{\rm T} = e^{{\rm i}\Theta}\nu$,
where $C$ denotes the charge-conjugation operator and $\Theta$
is an arbitrary real phase. A Majorana mass term, which violates
both the law of total lepton number conservation and that of individual
lepton flavor conservation, can be written either as
\begin{equation}
-{\cal L}_{\rm Majorana} \; =\; \frac{1}{2} \left [
\overline{\psi}_{\rm L} M^{\rm M}_\nu \left (\psi^{\rm c}\right )_{\rm R}
~ + ~ \overline{\left (\psi^{\rm c}\right )}_{\rm R} M^{\rm M\dagger}_\nu
\psi_{\rm L} \right ] \; ,
\end{equation}
or as
\begin{equation}
-\tilde{\cal L}_{\rm Majorana} \; =\; \frac{1}{2} \left [
\overline{\left (\psi^{\rm c}\right )}_{\rm L} \tilde{M}^{\rm M}_\nu
\psi_{\rm R} 
~ + ~ \overline{\psi}_{\rm R} \tilde{M}^{\rm M \dagger}_\nu
\left (\psi^{\rm c} \right )_{\rm L} \right ] \; ,
\end{equation}
where $M^{\rm M}_\nu$ and $\tilde{M}^{\rm M}_\nu$ stand for
the symmetric $3\times 3$ mass matrices of active and sterile Majorana neutrinos,
respectively. The most general neutrino mass Lagrangian is the sum of
${\cal L}_{\rm Dirac}$, ${\cal L}_{\rm Majorana}$ and 
$\tilde{\cal L}_{\rm Majorana}$, in which $M^{\rm D}_\nu$,
$M^{\rm M}_\nu$ and $\tilde{M}^{\rm M}_\nu$ are $n\times n$ complex
matrices.

Although Dirac and Majorana neutrinos have different 
properties, it remains extremely difficult to distinguish between
them in practical high-energy experiments, if there are not
right-handed currents \cite{Kayser89}. 
This is easy to understand \cite{Fr76}: since the electroweak interactions
conserve helicity (i.e., a L- or R-state remains as the L- or R-state),
they are not sensitive to distinguish between Dirac and 
Majorana neutrino states.
Presumably the only possibility to identify the
type of neutrinos would be to measure the neutrinoless double beta
decay, which occurs through the exchange of
a Majorana neutrino between two decaying neutrons inside a nucleus
and thus violates the
lepton number by two units (e.g., $^A_Z X \rightarrow ~ ^A_{Z+2}X
+ 2e^-$). The latest upper bound on the effective
neutrino mass of the $(\beta\beta)_{0\nu}$ process is \cite{KK99}
\begin{equation}
\langle m_\nu \rangle \; =\; \sum_i \left ( m_i V^2_{ei} \right ) 
\; \leq \; 0.2 ~ {\rm eV} \; 
\end{equation}
at the $90\%$ confidence level, where $m_i$ denotes the Majorana
neutrino mass of the $i$-th family, and $V_{ei}$ is the element
of the lepton flavor mixing matrix $V$. In the framework of three
active neutrinos, $V$ links the neutrino mass eigenstates 
$|\nu_i \rangle$ (for $i=1,2,3$) to the neutrino flavor
eigenstates $|\nu_\alpha\rangle$ (for $\alpha=e,\mu,\tau$).

The recent observation of the
atmospheric and solar neutrino anomalies, particularly
that in the Super-Kamiokande experiment \cite{SK}, has provided 
indirect but strong evidence that neutrinos are
massive and lepton flavors are mixed. 
Analyses of the atmospheric neutrino deficit in the framework
of two-flavor neutrino oscillations yield the mass-squared
difference
\begin{equation}
\Delta m^2_{\rm atm} \; \sim \; 10^{-3} ~ {\rm eV^2} \; 
\end{equation}
with the mixing factor $\sin^2 2\theta_{\rm atm} > 0.8$. In
contrast, there exist two different oscillation mechanisms
yielding three possible solutions to the solar neutrino
problem: the long wave-length vacuum oscillation (``Just-so''
mechanism) with
\begin{equation}
\Delta m^2_{\rm sun} \; \sim \; 10^{-10} ~ {\rm eV^2} \; 
\end{equation}
and $\sin^2 2\theta_{\rm sun} \approx 1$; and the matter-enhanced
oscillation (Mikheyev-Smirnov-Wolfenstein or 
MSW mechanism \cite{MSW}) with
\begin{equation}
\Delta m^2_{\rm sun} \; \sim \; 10^{-6} ~ {\rm eV^2} \;
\end{equation}
and $\sin^2 2\theta_{\rm sun} \sim 10^{-3} - 10^{-2}$
(small-angle solution) or with
\begin{equation}
\Delta m^2_{\rm sun} \; \sim \; 10^{-5} ~ {\rm eV^2} \;
\end{equation}
and $\sin^2 2\theta_{\rm sun} \sim 0.65 - 1$ (large-angle
solution) \cite{Bahcall98}. In the framework of three-flavor neutrino
oscillations, the big hierarchy between $\Delta m^2_{\rm atm}$
and $\Delta m^2_{\rm sun}$ together with the no observation
of $\bar{\nu}_e \rightarrow \bar{\nu}_e$ oscillation in the
CHOOZ experiment \cite{CHOOZ} 
implies that the $\nu^{~}_3$-component in $\nu_e$
is rather small (even negligible) and the atmospheric neutrino 
oscillation decouples approximately from the
solar neutrino oscillation. If this
simplified picture is essentially true, then the solar and
atmospheric neutrino deficits are dominated respectively by 
the $\nu_\mu \rightarrow \nu_e$ and $\nu_\mu \rightarrow
\nu_\tau$ transitions with the mass-squared
differences
\footnote{Without loss of any generality we have assumed
$m_1 < m_2 < m_3$.}
\begin{eqnarray}
\Delta m^2_{\rm sun} & = & \Delta m^2_{21} \; \equiv \;
m^2_2 - m^2_1  \; , \nonumber \\
\Delta m^2_{\rm atm} & = & \Delta m^2_{32} \; \equiv \;
m^2_3 - m^2_2  \; .
\end{eqnarray}
Nevertheless, the
hierarchy of $\Delta m^2_{21}$ and $\Delta m^2_{32}$ (
$\approx \Delta m^2_{31}$) can shed little light on the absolute values
or the relative magnitudes of three neutrino masses. For example,
either the strongly hierarchical neutrino mass spectrum 
($m_1 \ll m_2 \ll m_3$) or the nearly degenerate one 
($m_1 \approx m_2 \approx m_3$) is practically allowed to reproduce the 
``observed'' gap between $\Delta m^2_{21}$ and $\Delta m^2_{32}$.

The LSND evidence \cite{LSND} 
for the conversion of neutrino flavors
is so far the only indication of neutrino oscillations which
is a {\it signal} instead of a {\it deficit}. Such evidence has been
seen for both $\bar{\nu}_\mu \rightarrow \bar{\nu}_e$ and
$\nu_\mu \rightarrow \nu_e$ oscillations with the mass-squared
difference
\begin{equation}
\Delta m^2_{\rm LSND} \; \sim \; 1 ~ {\rm eV^2} \; 
\end{equation}
and the mixing factor $\sin^2 2\theta_{\rm LSND} \sim 10^{-3} -
10^{-2}$. However, the LSND observation was not confirmed by the
recent KARMEN experiment \cite{KARMEN}, which is sensitive to most of the 
LSND parameter space. It has to be seen how the
disagreement between these two measurements is resolved.
Before a further check of the LSND result
which will be available in the coming years, the conservative
approach is to set it aside tentatively and to concentrate on
solar and atmospheric neutrino oscillations. The latter can 
therefore be interpreted by use of 
two mass-squared differences defined in (2.21). 
 
Indeed it is extremely difficult, if not impossible \cite{333}, 
to accommodate the solar, atmospheric and LSND data simultaneously
within the scheme of three-flavor neutrino oscillations. 
A more natural approach is to assume the existence of
a light sterile neutrino ($\nu_{\rm s}$ with mass $m_0$),
which may produce an additional mass-squared difference with
one of the three active neutrinos to fit $\Delta m^2_{\rm LSND}$.
Recent analyses of the four-neutrino mixing patterns suggest that
there exist two possible options \cite{4N}:
the solar, atmospheric and
LSND neutrino oscillations may be attributed respectively to 
(a) $\nu_e \rightarrow \nu_{\rm s}$, $\nu_\mu \rightarrow
\nu_\tau$ and $\nu_\mu\rightarrow \nu_e$ transitions; or to (b)
$\nu_e \rightarrow \nu_\tau$, $\nu_\mu \rightarrow \nu_{\rm s}$
and $\nu_\mu \rightarrow \nu_e$ transitions (see Fig. 2.1 for
illustration). Note that pattern (a) seems to be more favored
by the Super-Kamiokande data, but pattern (b) has not been ruled out.
Either pattern
involves two neutrinos in the eV mass range, which might be
a suitable candidate for the hot dark matter of the universe.
\begin{figure}[t]
\hspace*{-1.8cm}\begin{picture}(400,160)(-10,210)
\put(90,285){\line(1,0){130}}
\put(90,285.5){\line(1,0){130}}
\put(70,285){\makebox(0,0){$m_2$}}
\put(90,270){\line(1,0){130}}
\put(90,270.5){\line(1,0){130}}
\put(70,270){\makebox(0,0){$m_3$}}
\put(90,360){\line(1,0){130}}
\put(90,360.5){\line(1,0){130}}
\put(70,360){\makebox(0,0){$m_0$}}
\put(90,345){\line(1,0){130}}
\put(90,345.5){\line(1,0){130}}
\put(70,345){\makebox(0,0){$m_1$}}
\put(150,315){\makebox(0,0){$\Delta m^2_{\rm LSND}$}}
\put(150,325){\vector(0,1){20}}
\put(150,305){\vector(0,-1){20}}
\put(253,352.5){\makebox(0,0){$\Delta m^2_{\rm sun}$}}
\put(253,277.5){\makebox(0,0){$\Delta m^2_{\rm atm}$}}
\put(150,245){\makebox(0,0){(a)}}
\hspace*{-0.5cm}
\put(300,285){\line(1,0){130}}
\put(300,285.5){\line(1,0){130}}
\put(450,285){\makebox(0,0){$m_2$}}
\put(300,270){\line(1,0){130}}
\put(300,270.5){\line(1,0){130}}
\put(450,270){\makebox(0,0){$m_0$}}
\put(300,360){\line(1,0){130}}
\put(300,360.5){\line(1,0){130}}
\put(450,360){\makebox(0,0){$m_3$}}
\put(300,345){\line(1,0){130}}
\put(300,345.5){\line(1,0){130}}
\put(450,345){\makebox(0,0){$m_1$}}
\put(370,315){\makebox(0,0){$\Delta m^2_{\rm LSND}$}}
\put(370,325){\vector(0,1){20}}
\put(370,305){\vector(0,-1){20}}
\put(370,245){\makebox(0,0){(b)}}
\end{picture}
\vspace{-1.1cm}
\caption{Possible neutrino mass spectra to accommodate current
data on solar, atmospheric and LSND neutrino oscillations.}
\end{figure}
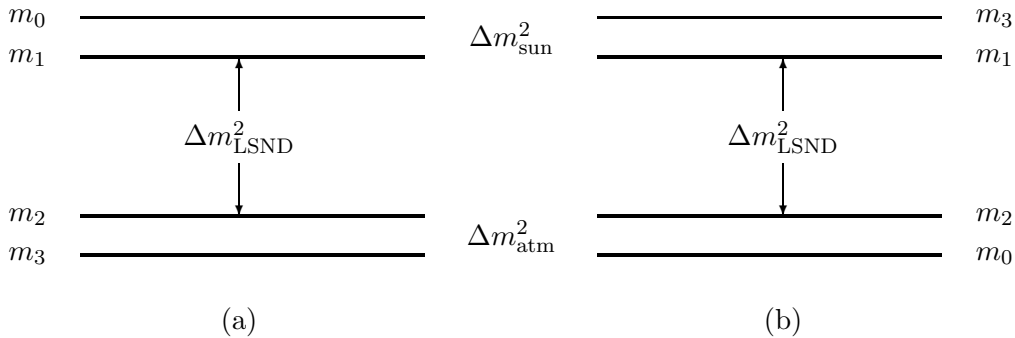

The indication that the mass density of the universe is smaller
than its critical value (i.e., $\Omega_{\rm m} < 1$) comes
in particular from direct observation of the existence of galaxies at very 
high redshift, the high baryon content of galaxies, 
the evolution of galactic clusters, and the 
high-redshift Type 1A supernovae. The present evidence points 
to $0.3 \leq \Omega_{\rm m} \leq 0.6$, whose main 
part is the dark matter \cite{HDM}. The dark matter is expected to
consist of both cold and hot components.
Neutrinos with a sum of their masses in
the range of a few eV could constitute the hot dark matter,
which is relativistic or at a temperature lower than 1 keV.
If the experimental results of solar, atmospheric and LSND
neutrino oscillations would be correct, then we should be left
with the hot dark matter composed of two types of massive neutrinos, 
as mentioned above. 

If the LSND result were invalid, it is possible that the
hot dark matter consists of three massive neutrinos
with a mass spectrum
\begin{equation}
m_1 \; \approx \; m_2 \; \approx \; m_3 \; \sim \; 2 ~ {\rm eV} \; .
\end{equation}
The near degeneracy of three neutrino 
masses is of particular interest, as it could arise from 
the breaking of a specific flavor symmetry in which the neutrinos have
an exact mass degeneracy \cite{FX96,FX98}:
\begin{equation}
M^{(0)}_\nu \; =\; m_i \left ( \matrix{
\eta^{~}_1 & 0 & 0 \cr
0 & \eta^{~}_2 & 0 \cr
0 & 0 & \eta^{~}_3 \cr} \right ) \; 
\end{equation}
with $m_1 = m_2 = m_3$ and $|\eta^{~}_i| =1$ for $i=1,2,3$. For Majorana 
neutrinos $\eta^{~}_i$ denote their $CP$ parities. Thus
$\eta^{~}_1 =\eta^{~}_2 =\eta^{~}_3$ means that $M^{(0)}_\nu$ has
an exact S(3) symmetry. The implications of $M^{(0)}_\nu$
and its symmetry breaking on lepton flavor mixing
and neutrino oscillations will be explored in some detail
in section 5. 
It should be noted, however, that it is impossible to accommodate 
current atmospheric and solar neutrino oscillation data and to satisfy 
the requirement of hot dark matter simultaneously, if three 
neutrinos have a hierarchical mass spectrum \cite{Smirnov96}.

We shall not mention some other astrophysical and cosmological 
hints that neutrinos are probably massive \cite{Others}. 
Instead let us ask the 
question why the masses of neutrinos, if not vanishing, are 
so small in comparison with those of the charged leptons
and quarks. This feature could be linked to the fact that neutrinos 
are the only known {\it neutral} fermions. A theoretically natural
interpretation of the smallness of neutrino masses, e.g., in
grand unified theories and most of other extensions of the
standard model, is to
assume that the active neutrinos are left-handed Majorana particles 
accompanied by very heavy right-handed (sterile) Majorana partners. The
latter may serve to reduce the masses of the former through the
well-known seesaw mechanism \cite{Seesaw}. In the spirit of the seesaw
mechanism, the left-handed Majorana neutrino mass matrix
$M^{\rm M}_\nu$ is given as
\begin{equation}
M^{\rm M}_\nu \; =\; ( M^{\rm D}_\nu )^{\rm T}
(\tilde{M}^{\rm M}_\nu )^{-1} (M^{\rm D}_\nu ) \; ,
\end{equation}
where $M^{\rm D}_\nu$ and $\tilde{M}^{\rm M}_\nu$ are the
Dirac neutrino mass matrix and the right-handed Majorana
neutrino mass matrix, respectively. If the mass eigenvalues
of $M^{\rm D}_\nu$ are of the order of charged lepton masses
and those of $\tilde{M}^{\rm M}_\nu$ are at the
level of a superhigh energy scale, then the active Majorana neutrinos
will acquire small (even tiny) masses. In some unified
theories of quarks and leptons $M^{\rm D}_\nu$ is usually
taken to be the same as the up-type quark mass matrix $M_{\rm u}$, and
the mass eigenvalues of $\tilde{M}^{\rm M}_\nu$ are more or
less of order $10^{13}$ GeV. For a review of various 
versions of the seesaw mechanism, we refer the reader to
Ref. \cite{Langacker98}.

\section{Flavor mixing and $CP$ violation}
\setcounter{equation}{0}
\setcounter{figure}{0}

\subsection{Number of fermion mixing parameters}

The quark mass matrices in the Lagrangian of Yukawa interactions,
$M_{\rm u}$ and $M_{\rm d}$, can be diagonalized by the
bi-unitary transformations:
\begin{eqnarray}
{\cal M}_{\rm u} & = & U^{\dagger}_{\rm u L} M_{\rm u} U_{\rm u R} \; =\;
{\rm Diag} \{ m_u, ~ m_c, ~ m_t \} \; , 
\nonumber \\
{\cal M}_{\rm d} & = & U^{\dagger}_{\rm d L} M_{\rm d} U_{\rm d R} \; =\;
{\rm Diag} \{ m_d, ~ m_s, ~ m_b \} \; .
\end{eqnarray}
Such transformations, equivalent to changing quark fields from the
basis of flavor eigenstates to that of mass eigenstates, introduce 
non-diagonal couplings into the Lagrangian of charged weak interactions,
in which only the left-handed quarks take part. This $3\times 3$
coupling matrix (the so-called Cabibbo-Kobayashi-Maskawa or CKM
matrix \cite{Cabibbo,KM}), given as
\begin{equation}
V \; =\; U^{\dagger}_{\rm u L} U_{\rm d L} \; ,
\end{equation}
describes the mixing of quark flavors. The unitarity is 
a constraint, imposed by the symmetry structure of the standard model,
on the flavor mixing matrix $V$. 

Although we have taken the number of quark families to be three,
a more general discussion can be made for $n$ families
of quarks. In this case the flavor mixing matrix
$V$ will be a $n\times n$ unitary matrix consisting of $n^2$ real
parameters: $n (n-1)/2$ of them may be taken as rotation angles
and the remaining are phase angles. Since the phases of quark
fields are arbitrary, one can redefine them so as to rearrange 
the phase parameters of $V$. This freedom allows 
$2n-1$ phase angles to be absorbed. Therefore 
$V$ can be described in terms of only 
$n^2 - (2n-1) = (n-1)^2$ parameters, among which $n (n-1)/2$
are the rotation angles and $(n-1)(n-2)/2$ are the phase angles. 
For the case $n=3$, we then arrive at
three mixing angles and one nontrivial phase, which is 
responsible for $CP$ violation.

The charged lepton and Dirac neutrino mass matrices in the flavor
basis, $M_l$ and $M^{\rm D}_\nu$, can be diagonalized in a similar way
by the bi-unitary transformations:
\begin{eqnarray}
{\cal M}_l & = & ~ U^{\dagger}_{l\rm L} M_l U_{l \rm R} \;\; =\;\;
{\rm Diag} \{ m_e, ~ m_\mu, ~ m_\tau \} \; , 
\nonumber \\
{\cal M}^{\rm D}_\nu & = & U^{\dagger}_{\nu \rm L} 
M^{\rm D}_\nu U_{\nu \rm R} \; =\;
{\rm Diag} \{ m_1, ~ m_2, ~ m_3 \} \; .
\end{eqnarray}
In the basis of mass eigenstates, one arrives at a $3\times 3$
flavor mixing matrix $V^{\rm D}_l$ in 
the Lagrangian of charged weak interactions,
in which only the left-handed leptons take part:
\begin{equation}
V^{\rm D}_l \; =\; U^{\dagger}_{l \rm L} U_{\nu \rm L} \; .
\end{equation}
Note that $V^{\rm D}_l$ links the neutrino mass eigenstates 
$(\nu_1, \nu_2, \nu_3)$ to the neutrino flavor eigenstates
$(\nu_e, \nu_\mu, \nu_\tau)$ in the basis that the charged lepton mass
matrix is diagonal. Similar to the quark mixing case,
the $3\times 3$ lepton mixing matrix $V^{\rm D}_l$ can be
generalized to the $n\times n$ one 
for $n$ families of leptons. The latter can then be parametrized
in terms of $n (n-1)/2$ rotation angles and $(n-1)(n-2)/2$
phase angles.

If neutrinos are Majorana particles, however, the situation
is different. In this case the neutrino mass matrix $M^{\rm M}_\nu$
has the property ${(M^{\rm M}_\nu)}^{\rm T} = M^{\rm M}_\nu$, 
i.e., $M^{\rm M}_\nu$ is in general a complex symmetric
$n\times n$ matrix. The diagonalization of $M^{\rm M}_\nu$
needs only a single unitary matrix $\hat{U}_\nu$ as follows
($n=3$, for example):
\begin{equation}
{\cal M}^{\rm M}_\nu \; =\; \hat{U}^{\rm T}_\nu
M^{\rm M}_\nu \hat{U}_\nu \; =\;
{\rm Diag} \{ m_1, ~ m_2, ~ m_3 \} \; .
\end{equation}
Accordingly the lepton flavor mixing matrix is given by
\begin{equation}
V^{\rm M}_l \; =\; U^{\dagger}_{l \rm L} \hat{U}_\nu \; ,
\end{equation}
where $U^{\dagger}_{l \rm L}$ has been given in (3.3) to diagonalize the
charged lepton mass matrix $M_l$. The $n\times n$ flavor
mixing matrix $V^{\rm M}_l$ consists totally of $n^2$ real
parameters, and $n (n-1)/2$ of them can always be chosen
as rotation angles. Unlike the quark or Dirac neutrino
mixing case, there is no freedom to redefine phases of
the Majorana neutrino fields, as Majorana particles are
their own antiparticles. Hence some phases of $V^{\rm M}_l$
can be absorbed only by redefining the charged lepton 
fields. The number of physical phase angles left in $V^{\rm M}_l$
is $n(n+1)/2 -n = n(n-1)/2$. Thus $V^{\rm M}_l$ can
be parametrized in terms of $n(n-1)/2$ rotation angles and
the same number of phase angles. For the case $n=3$, we
obtain the lepton mixing matrix with 3 rotation angles and
3 $CP$-violating phases. 

It should be noted, however, that the $n\times n$ Majorana-type
flavor mixing matrix $V^{\rm M}_l$ can always be represented
in the following form \cite{Langacker87}:
\begin{equation}
V^{\rm M}_l \; =\; V^{\rm D}_l P_\nu \; ,
\end{equation}
in which $V^{\rm D}_\nu$ is the $n\times n$ Dirac-type flavor mixing
matrix and $P_\nu$ is a $n\times n$ diagonal phase matrix with
$n-1$ nontrivial phase parameters
\footnote{The elements of $P_\nu$ can be written as
$(P_\nu)_{ij} = \delta_{ij} e^{{\rm i}\sigma_j}$ 
(for $i,j =1, \cdot\cdot\cdot, n$) with $\sigma_1 =0$.}.
The $CP$-violating 
phases in $P_\nu$, in addition to those in $V^{\rm D}_l$,
characterize the Majorana nature of $V^{\rm M}_l$.
It is straightforward to show
\begin{equation}
(V^{\rm M}_l)_{i\alpha} (V^{\rm M}_l)_{j\beta} 
(V^{\rm M}_l)^*_{i\beta} (V^{\rm M}_l)^*_{j\alpha}
\; =\; (V^{\rm D}_l)_{i\alpha} (V^{\rm D}_l)_{j\beta}
(V^{\rm D}_l)^*_{i\beta} (V^{\rm D}_l)^*_{j\alpha} 
\end{equation}
for arbitrary indices $i,j$ and $\alpha,\beta$. This
result implies that $V^{\rm M}_l$ and $V^{\rm D}_l$ have
the same physical effects in neutrino
oscillation probabilities, which depend only upon the
combinations in (3.8). We then arrive at the conclusion
that it is impossible to determine the
type of massive neutrinos by studying the phenomena of
neutrino oscillations. Only the experiments which probe transitions
between left-handed and right-handed neutrino states, like
the neutrinoless double beta decay,
could tell whether massive neutrinos are of the Majorana type or not.

In subsequent discussions about the lepton flavor mixing
and $CP$ violation, we shall make use of the notations
$V^{\rm M}_l$ and $V^{\rm D}_l$ only when it is necessary to
distinguish between the Dirac and Majorana neutrinos. Otherwise
$V_l$ will simply be adopted to represent either 
$V^{\rm D}_l$ or $V^{\rm M}_l$. If it is unnecessary
to distinguish the lepton mixing from the quark mixing, 
we shall just use $V$ to denote the flavor mixing matrix in general.

\subsection{Conditions for $CP$ violation}

The quark sector of the standard electroweak model with
three fermion families consists of
ten free parameters: six quark masses (associated directly
with $M_{\rm u}$ and $M_{\rm d}$) and four flavor mixing
parameters (related directly to $V$). Since $V$ is obtained by
the diagonalization of $M_{\rm u}$ and $M_{\rm d}$, the flavor
mixing parameters are expected to depend to a great extent on the
quark masses. These explicit relationships should be obvious in
an underlying theory of fermion mass generation,
which would be more fundamental than the standard model.
Within the standard model, the conditions for $CP$ violation 
can be expressed in terms of either the parameters of 
$M_{\rm u}$ and $M_{\rm d}$ or those of $V$. It is worth remarking
that a double-counting of the $CP$-violating conditions, in terms
of the parameters of both $M_{\rm u,d}$ and $V$, must be 
avoided.

We first discuss the necessary and sufficient conditions for
$CP$ violation at the level of quark mass matrices.
Although $M_{\rm u}$ and $M_{\rm d}$ are arbitrary $3\times 3$
matrices, the products $M_{\rm u} M_{\rm u}^{\dagger} \equiv H_{\rm u}$
and $M_{\rm d} M_{\rm d}^{\dagger} \equiv H_{\rm d}$ are Hermitian,
and each of them can be diagonalized by a single unitary transformation:
\begin{eqnarray}
U^{\dagger}_{\rm u L} H_{\rm u} U_{\rm u L} & = & 
{\cal M}^2_{\rm u} \; , \nonumber \\
U^{\dagger}_{\rm d L} H_{\rm d} U_{\rm d L} & = & 
{\cal M}^2_{\rm d} \; ,
\end{eqnarray}
where $U_{\rm u L}$ and $U_{\rm d L}$ have been given in
(3.1). It is obvious that $CP$ symmetry 
will be violated, if and only if there is
at least one nontrivial phase difference between $H_{\rm u}$ and 
$H_{\rm d}$. In other words, 
${\rm Im} \left (H_{{\rm u} ij} H^*_{{\rm d} ij} \right )
\neq 0$ (for $i,j =1,2,3$ and $i<j$) 
is the necessary and sufficient condition for $CP$ violation in
the standard model. If one defines a commutator for
$H_{\rm u}$ and $H_{\rm d}$, $[H_{\rm u},
H_{\rm d} ] \equiv {\rm i} ~ {\cal C}$, then it is easy to show
\begin{equation}
\frac{{\cal C}_{ii}}{2} \; =\; {\rm Im} \left (H_{{\rm u} ij} 
H^*_{{\rm d} ij} \right )  + 
{\rm Im} \left (H_{{\rm u} ik} H^*_{{\rm d} ik} \right ) \;
\end{equation}
for $i,j,k=1,2,3$ but $i\neq j\neq k$. 
Clearly ${\cal C}_{ii} \neq 0$ if $CP$ symmetry is violated \cite{FX99}.
Note that $CP$ symmetry would be conserved, if two quarks with
the same charge were degenerate in mass. The reason is simply
that the freedom induced by the mass degeneracy of two 
up- or down-type quarks allows one to rearrange the matrix 
elements of $H_{\rm u}$ or $H_{\rm d}$ and remove all possible phase
differences between $H_{\rm u}$ and $H_{\rm d}$ (or between
$M_{\rm u}$ and $M_{\rm d}$). 

Now we discuss the condition for $CP$ violation at the level of
the flavor mixing matrix. Of course $CP$ symmetry is violated, if
$V$ contains a nontrivial complex phase which cannot be removed
from the redefinition of quark-field phases. The most 
appropriate measure of $CP$ violation, due to the unitarity of
$V$, is the rephasing-invariant parameter 
${\cal J}$ \cite{Jarlskog85} defined through 
\begin{equation}
{\rm Im} \left (V_{i\alpha} V_{j\beta} V^*_{i\beta} V^*_{j\alpha}
\right ) \; =\; {\cal J} \sum_{k,\gamma} \left (\varepsilon_{ijk}
\varepsilon_{\alpha\beta\gamma} \right ) \; ,
\end{equation}
in which each Latin subscript ($i,j$ or $k$) runs
over the up-type quarks $(u,c,t)$ and each Greek subscript
($\alpha, \beta$ or $\gamma$) runs over the down-type quarks
$(d,s,b)$. In terms of the matrix elements of $V$, ${\cal J}^2$
can be given as 
\footnote{Note that there was a typing error 
in the original formula of ${\cal J}^2$ given in Ref. \cite{Sasaki}.}
\begin{eqnarray}
{\cal J}^2 & = & |V_{i\alpha}|^2 |V_{j\beta}|^2 
|V_{i\beta}|^2 |V_{j\alpha}|^2 - 
\frac{1}{4} \left (1 + |V_{i\alpha}|^2 |V_{j\beta}|^2 
+ |V_{i\beta}|^2 ~ |V_{j\alpha}|^2
\right . \nonumber \\
& & \left . 
- |V_{i\alpha}|^2 - |V_{j\beta}|^2 - |V_{i\beta}|^2
- |V_{j\alpha}|^2 \right )^2 \; . 
\end{eqnarray}
This result implies that the
information about $CP$ violation can in principle be extracted
from the moduli of
the flavor mixing matrix elements. 
Note again that any mass degeneracy between two quarks with the
same charge results in $CP$ conservation. 
If two up- or down-type quarks were degenerate in mass, 
then any linear combination
of their mass eigenstates via an arbitrary unitary matrix
would remain a mass eigenstate. This additional freedom 
allows us to redefine the phases of relevant quark fields and
eliminate the single nontrivial ($CP$-violating) phase in $V$.
The resultant flavor mixing matrix is real (with either three 
or two nonvanishing
mixing angles), then ${\cal J} =0$ holds, i.e., $CP$ symmetry is
conserved. In fact one can show, as done
in Ref. \cite{FX99}, that 
the magnitude of $\cal J$ is proportional to the
product of the mass-squared differences
$(m^2_i - m^2_j)$ and $(m^2_\alpha - m^2_\beta)$, 
where the subscripts $(i,j)$ and $(\alpha, \beta)$ run over
$(u,c,t)$ and $(d,s,b)$ respectively, for arbitrary
mass matrices $M_{\rm u}$ and $M_{\rm d}$. 
Therefore any
mass degeneracy between two quarks with the same charge
would lead to ${\cal J} =0$. 

In short, the conditions for $CP$ violation in the quark
sector can be set at either the level of quark
mass matrices or that of the flavor mixing matrix \cite{FX99}.
A double-counting, arising from a combination of the 
conditions obtained from two different levels, must be
avoided. It should in particular be noted that the determinant of $\cal C$ and 
the rephasing-invariant parameter $\cal J$, which 
are related to each other through \cite{Jarlskog85}
\begin{equation}
{\rm Det} ~ {\cal C} \; =\; -2{\cal J} 
\prod_{i<j} \left (m^2_i - m^2_j \right )
\prod_{\alpha < \beta} \left ( m^2_\alpha - m^2_\beta \right )
\; ,
\end{equation}
contain the same information about $CP$ violation.
In reality the quark masses of both up and down sectors
have been found to exhibit a clear hierarchy, and all
elements of the flavor mixing matrix are nonvanishing.
Therefore the realistic
condition for $CP$ violation is only associated with
the existence of one nontrivial phase parameter in $V$,
which in turn requires (at least) one nontrivial phase
difference
between $M_{\rm u}$ and $M_{\rm d}$ (or equivalently
between $H_{\rm u}$ and $H_{\rm d}$).

The conditions for $CP$ violation in weak interactions of
charged leptons and Dirac neutrinos can be discussed in the
same way as outlined above. Similar to $\cal J$, a 
rephasing-invariant measure of leptonic $CP$ asymmetry,
${\cal J}^{\rm D}_l$, can be defined in terms of elements of
the flavor mixing matrix $V^{\rm D}_l$. Therefore the
necessary and sufficient condition for leptonic $CP$ violation is
just ${\cal J}^{\rm D}_l \neq 0$.

If neutrinos are of the Majorana type, however, one has to
take care of two additional $CP$-violating phases from
$P_\nu$ in (3.7). In this case the necessary and sufficient 
condition for $CP$ violation is that there exist nontrivial
phase differences between the mass matrices $M_l$ and
$M^{\rm M}_\nu$, i.e., the flavor mixing matrix $V^{\rm M}_l$
contains at least one nontrivial complex phase. 
In analogy to the definition of
${\cal J}^{\rm D}_l$, 
a $CP$-violating parameter ${\cal J}^{\rm M}_l$ which is
independent of the phase rearrangement of charged lepton fields,
can be defined. It should be noted that 
${\cal J}^{\rm M}_l = 0$ is only a necessary 
condition for $CP$ invariance in weak interactions
of the charged leptons and Majorana neutrinos. Taking (3.8)
into account, we arrive at
\begin{equation}
{\cal J}^{\rm M}_l \; =\; {\cal J}^{\rm D}_l \; \equiv \;
{\cal J}_l \; .
\end{equation}
Hence only ${\cal J}_l$ appears in the transition
probabilities of neutrino oscillations.

Finally let us emphasize that the mass degeneracy of two
Majorana neutrinos may not lead to the vanishing of
$CP$ violation in a model with the family number $n\geq 3$, 
provided they have the opposite $CP$ parities.
It has been shown in Ref. \cite{Branco98}
that the $3\times 3$ flavor mixing matrix 
$V^{\rm M}_l$ consists of two rotation angles and one
$CP$-violating phase, if all three Majorana neutrinos
are degenerate in mass but one of them has a different
$CP$ parity from the other two. However, the
mass degeneracy of any two Dirac neutrinos would make
the $CP$-violating phase in $V^{\rm D}_l$ removable and
thus lead to $CP$ conservation.
It should be noted that there would be no 
neutrino oscillations, if neutrino masses were degenerate.

\subsection{Unitarity triangles of quark mixing}

A geometric description of the quark flavor mixing phenomenon 
turns out to be instructive. 
The unitarity of the $3\times 3$ flavor mixing matrix $V$ can be expressed
by two sets of orthogonality relations and two sets of 
normalization conditions for all nine matrix elements:
\begin{eqnarray}
\sum_\alpha \left (V^*_{i\alpha} V_{j\alpha} \right )
& = & \delta_{ij} \; , \nonumber \\
\sum_i \left (V^*_{i\alpha}V_{i\beta} \right ) 
& = & \delta_{\alpha\beta} \; ,
\end{eqnarray}
where the Latin indices run over the up-type quarks
$(u,c,t)$ and the Greek indices over the down-type quarks
$(d,s,b)$. In the complex plane the six orthogonality 
relations in (3.15) define six triangles, the so-called
unitarity triangles (see Fig. 3.1 for illustration). 
In general these six triangles have eighteen different 
sides and nine different inner (or outer) angles. 
Unitarity requires that all six triangles have the same
area amounting to ${\cal J}/2$ \cite{Jarlskog85},
where $\cal J$ is just the
rephasing-invariant measure of $CP$ violation defined in 
(3.11). If $CP$ were an exact symmetry, ${\cal J}=0$
would hold, and those unitarity triangles would collapse into
lines in the complex plane. 
\begin{figure}[t]
\begin{picture}(400,315)(20,0)
\put(70,300){\line(1,0){130}}
\put(120,310){\makebox(0,0){${\bf V}^*_{cb}{\bf V}_{tb}$}}
\put(70,300){\line(2,-5){12}}
\put(51,285){\makebox(0,0){${\bf V}^*_{cd}{\bf V}_{td}$}}
\put(200,300){\line(-4,-1){118}}
\put(137,268){\makebox(0,0){${\bf V}^*_{cs}{\bf V}_{ts}$}}
\put(80,294){\makebox(0,0){\scriptsize\bf 1}}
\put(150,294){\makebox(0,0){\scriptsize\bf 2}}
\put(87,278){\makebox(0,0){\scriptsize\bf 3}}
\put(130,245){\makebox(0,0){($\triangle_u$)}}
\put(275,300){\line(1,0){112}}
\put(345,310){\makebox(0,0){${\bf V}^*_{cs}{\bf V}_{cb}$}}
\put(275,300){\line(4,-1){117}}
\put(325,272){\makebox(0,0){${\bf V}^*_{ts}{\bf V}_{tb}$}}
\put(387,300){\line(1,-6){4.9}}
\put(415,282){\makebox(0,0){${\bf V}^*_{us}{\bf V}_{ub}$}}
\put(325,294){\makebox(0,0){\scriptsize\bf 2}}
\put(380,293){\makebox(0,0){\scriptsize\bf 9}}
\put(383,280){\makebox(0,0){\scriptsize\bf 6}}
\put(340,245){\makebox(0,0){($\triangle_d$)}}
\end{picture}

\begin{picture}(400,100)(20,0)
\put(85,300){\line(1,0){90}}
\put(127,310){\makebox(0,0){${\bf V}^*_{td}{\bf V}_{ud}$}}
\put(85,300){\line(1,-3){13}}
\put(68,278){\makebox(0,0){${\bf V}^*_{ts}{\bf V}_{us}$}}
\put(175,300){\line(-2,-1){77}}
\put(155,268){\makebox(0,0){${\bf V}^*_{tb}{\bf V}_{ub}$}}
\put(95,294){\makebox(0,0){\scriptsize\bf 4}}
\put(146,294){\makebox(0,0){\scriptsize\bf 5}}
\put(102.5,272.5){\makebox(0,0){\scriptsize\bf 6}}
\put(130,245){\makebox(0,0){($\triangle_c$)}}
\put(290,300){\line(1,0){85}}
\put(335,310){\makebox(0,0){${\bf V}^*_{ub}{\bf V}_{ud}$}}
\put(290,300){\line(2,-1){73}}
\put(313,270){\makebox(0,0){${\bf V}^*_{tb}{\bf V}_{td}$}}
\put(375,300){\line(-1,-3){12}}
\put(395,280){\makebox(0,0){${\bf V}^*_{cb}{\bf V}_{cd}$}}
\put(318,294){\makebox(0,0){\scriptsize\bf 5}}
\put(365,293.5){\makebox(0,0){\scriptsize\bf 7}}
\put(360,273){\makebox(0,0){\scriptsize\bf 1}}
\put(340,245){\makebox(0,0){($\triangle_s$)}}
\end{picture}

\begin{picture}(400,100)(20,0)
\put(70,300){\line(1,0){150}}
\put(130,310){\makebox(0,0){${\bf V}^*_{ud}{\bf V}_{cd}$}}
\put(70,300){\line(1,-4){6}}
\put(49,286){\makebox(0,0){${\bf V}^*_{ub}{\bf V}_{cb}$}}
\put(220,300){\line(-6,-1){143.7}}
\put(157,274){\makebox(0,0){${\bf V}^*_{us}{\bf V}_{cs}$}}
\put(79,294){\makebox(0,0){\scriptsize\bf 7}}
\put(150,294){\makebox(0,0){\scriptsize\bf 8}}
\put(81,283){\makebox(0,0){\scriptsize\bf 9}}
\put(130,253){\makebox(0,0){($\triangle_t$)}}
\put(255,300){\line(1,0){130}}
\put(330,310){\makebox(0,0){${\bf V}^*_{cd}{\bf V}_{cs}$}}
\put(255,300){\line(6,-1){134}}
\put(305,277){\makebox(0,0){${\bf V}^*_{ud}{\bf V}_{us}$}}
\put(385,300){\line(1,-5){4.5}}
\put(411,286){\makebox(0,0){${\bf V}^*_{td}{\bf V}_{ts}$}}
\put(325,294.5){\makebox(0,0){\scriptsize\bf 8}}
\put(379,294.5){\makebox(0,0){\scriptsize\bf 3}}
\put(382,284.5){\makebox(0,0){\scriptsize\bf 4}}
\put(340,253){\makebox(0,0){($\triangle_b$)}}
\end{picture}
\vspace{-8.6cm}
\caption{Unitarity triangles of the flavor mixing matrix 
in the complex plane. Each triangle is named in terms of 
the quark flavor that does not manifest in its three sides.}
\end{figure}
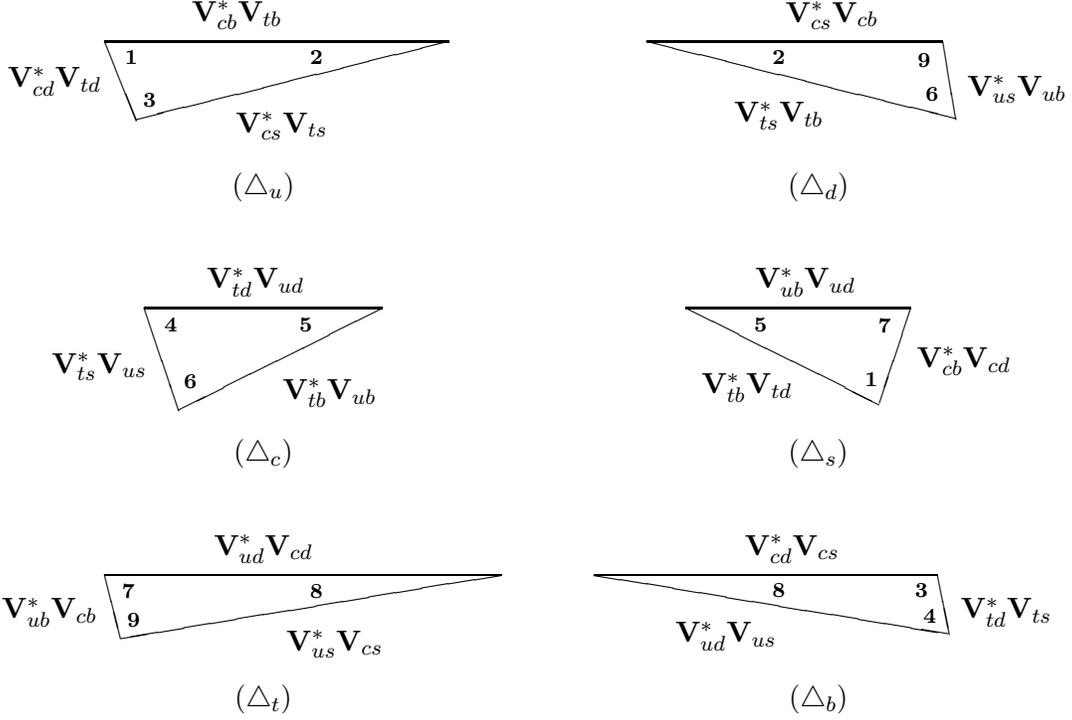

In additon to $\cal J$, there exist other two characteristic
quantities of $V$, resulting from its normalization conditions in
(3.15). They are the off-diagonal asymmetries ${\cal A}_1$ and
${\cal A}_2$, respectively about the axes $V_{ud}$-$V_{cs}$-$V_{tb}$ and
$V_{ub}$-$V_{cs}$-$V_{td}$ \cite{Xing96}:
\begin{eqnarray}
{\cal A}_1 & \equiv & |V_{us}|^2 - |V_{cd}|^2 \; =\;
|V_{cb}|^2 - |V_{ts}|^2 \; = \;
|V_{td}|^2 - |V_{ub}|^2 \; , \nonumber \\
{\cal A}_2 & \equiv & |V_{us}|^2 - |V_{cb}|^2 \; =\;
|V_{cd}|^2 - |V_{ts}|^2 \; =\;
|V_{tb}|^2 - |V_{ud}|^2 \; .
\end{eqnarray}
Clearly ${\cal A}_1 =0$ (or ${\cal A}_2 =0$) would imply that
the flavor mixing matrix were symmetric about the
$V_{ud}$-$V_{cs}$-$V_{tb}$ (or $V_{ub}$-$V_{cs}$-$V_{td}$)
axis. Geometrically this corresponds to the congruence between
two unitarity triangles, i.e.,
\begin{eqnarray}
{\cal A}_1 = 0 \; ~~~ \Longrightarrow ~~~ \; \triangle_u & \cong &
\triangle_d \; , ~~~
\triangle_c \; \cong \; \triangle_s \; , ~~~
\triangle_t \; \cong \; \triangle_b \; ; \nonumber \\
{\cal A}_2 = 0 \; ~~~ \Longrightarrow ~~~ \; \triangle_u & \cong &
\triangle_b \; , ~~~
\triangle_c \; \cong \; \triangle_s \; , ~~~
\triangle_t \; \cong \; \triangle_d \; .
\end{eqnarray}
In reality, however, both ${\cal A}_1$ and ${\cal A}_2$
are nonvanishing. In terms of the Wolfenstein parameters of
$V$ \cite{Wolfenstein83}, 
the three measurables ${\cal J}$, ${\cal A}_1$ and ${\cal A}_2$ can be
expressed as follows:
\begin{eqnarray}
{\cal J} & \approx & A^2 \lambda^6 \eta \; ,
\nonumber \\
{\cal A}_1 & \approx & A^2 \lambda^6 (1 -2\rho) \; ,
\nonumber \\
{\cal A}_2 & \approx & \lambda^2 (1 - A^2\lambda^2 )
\; .
\end{eqnarray}
It is clear that ${\cal A}_2 \gg {\cal A}_1 \sim {\cal J}
\sim 10^{-5}$.
Therefore $V$ deviates only a little bit from an exactly
symmetric matrix (about the $V_{ud}$-$V_{cs}$-$V_{tb}$ axis).
With ${\cal A}_1 >0$ one arrives at a clear hierarchy among nine
elements of $V$:
\begin{eqnarray}
|V_{tb}| > |V_{ud}| > |V_{cs}| & \gg &
|V_{us}| > |V_{cd}| \nonumber \\
& \gg & |V_{cb}| > |V_{ts}| \nonumber \\
& \gg & |V_{td}| > |V_{ub}| \; > \; 0 \; .
\end{eqnarray}
Indeed $|V_{ub}| \neq 0$ is a necessary condition for $CP$ violation
in the standard model.

It is worth remarking that the triangle $\Delta_s$ is,
among six unitarity triangles, of particular interest for the study
of $B$-meson physics and $CP$ violation. Indeed three inner angles
of $\triangle_s$, usually denoted as 
\begin{eqnarray}
\alpha & = & \arg \left ( - \frac{V^*_{tb}V_{td}}{V^*_{ub}V_{ud}}
\right ) \;\; , \nonumber \\
\beta & = & \arg \left ( -\frac{V^*_{cb}V_{cd}}{V^*_{tb}V_{td}}
\right ) \;\; , \nonumber \\
\gamma & = & \arg \left ( -\frac{V^*_{ub}V_{ud}}{V^*_{cb}V_{cd}}
\right ) \;\; ,
\end{eqnarray}
can directly be measured from $CP$ asymmetries in specific $B$ decay
modes. On the other hand, these rephasing-invariant observables
can be predicted from specific models of quark mass matrices.
Therefore the unitarity triangle $\triangle_s$ may serve as a 
phenomenological link between the theoretical understanding 
and the experimental measurement of quark flavor mixing and 
$CP$ violation.

\subsection{Unitarity triangles of lepton mixing}

A similar geometric description as that discussed above for 
quark mixing can be introduced for the $3\times 3$ lepton
flavor mixing matrix $V_l$.
The six leptonic unitarity triangles, which have 
eighteen different sides, nine different inner angles and 
the identical area (amounting to ${\cal J}_l/2$), are illustrated
in Fig. 3.2. Our present knowledge on lepton flavor mixing
is rather poor, hence it remains difficult to quantify
the magnitude of leptonic $CP$ violation as well as the
off-diagonal asymmetries of $V_l$. Useful information about
the sides and angles of the leptonic unitarity triangles
are likely to be obtained, in the near future, from
the long-baseline neutrino experiments.
In particular one might determine the three inner angles of the
triangle $\triangle_3$, i.e.,
\begin{eqnarray}
\alpha^{~}_l & = & \arg \left ( - \frac{V^*_{e 1}V_{e 2}}
{V^*_{\mu 1}V_{\mu 2}} \right ) \;\; , 
\nonumber \\
\beta_l & = & \arg \left ( - \frac{V^*_{\mu 1}V_{\mu 2}}
{V^*_{\tau 1}V_{\tau 2}} \right ) \;\; ,
\nonumber \\
\gamma^{~}_l & = & \arg \left ( - \frac{V^*_{\tau 1}V_{\tau 2}}
{V^*_{e 1}V_{e 2}} \right ) \;\; ;
\end{eqnarray}
and then test the self-consistency of the lepton flavor mixing
and of $CP$ violation (e.g., $\alpha^{~}_l + \beta_l + \gamma^{~}_l
\neq \pi$ would imply the presence of new physics which violates
the unitarity of the $3\times 3$ lepton flavor mixing matrix).
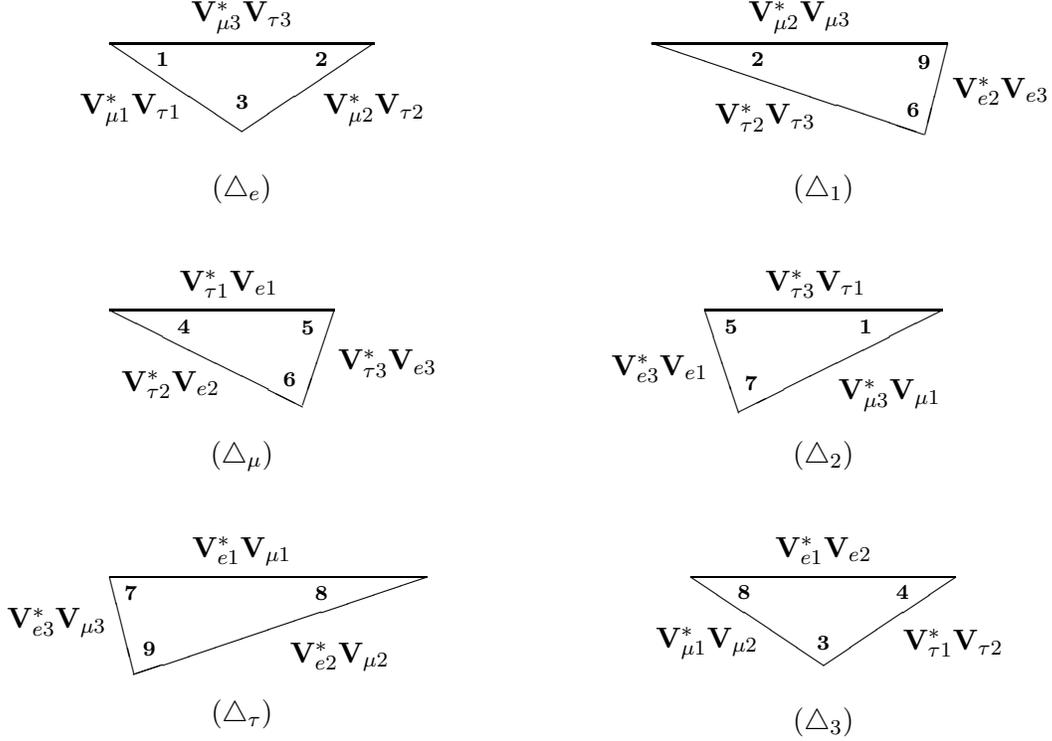
\begin{figure}[t]
\begin{picture}(400,315)(20,0)
\put(70,300){\line(1,0){100}}
\put(120,310){\makebox(0,0){${\bf V}^*_{\mu 3}{\bf V}_{\tau 3}$}}
\put(70,300){\line(3,-2){50}}
\put(78,275){\makebox(0,0){${\bf V}^*_{\mu 1}{\bf V}_{\tau 1}$}}
\put(170,300){\line(-3,-2){50}}
\put(170,275){\makebox(0,0){${\bf V}^*_{\mu 2}{\bf V}_{\tau 2}$}}
\put(90,294){\makebox(0,0){\scriptsize\bf 1}}
\put(150,294){\makebox(0,0){\scriptsize\bf 2}}
\put(120,278){\makebox(0,0){\scriptsize\bf 3}}
\put(120,245){\makebox(0,0){($\triangle_e$)}}
\put(275,300){\line(1,0){112}}
\put(331,310){\makebox(0,0){${\bf V}^*_{\mu 2}{\bf V}_{\mu 3}$}}
\put(275,300){\line(3,-1){103}}
\put(318,272){\makebox(0,0){${\bf V}^*_{\tau 2}{\bf V}_{\tau 3}$}}
\put(387,300){\line(-1,-4){8.6}}
\put(407,282){\makebox(0,0){${\bf V}^*_{e 2}{\bf V}_{e 3}$}}
\put(315,294){\makebox(0,0){\scriptsize\bf 2}}
\put(378,293){\makebox(0,0){\scriptsize\bf 9}}
\put(374,275){\makebox(0,0){\scriptsize\bf 6}}
\put(340,245){\makebox(0,0){($\triangle_1$)}}
\end{picture}

\begin{picture}(400,100)(20,0)
\put(70,300){\line(1,0){85}}
\put(115,310){\makebox(0,0){${\bf V}^*_{\tau 1}{\bf V}_{e 1}$}}
\put(70,300){\line(2,-1){73}}
\put(93,272){\makebox(0,0){${\bf V}^*_{\tau 2}{\bf V}_{e 2}$}}
\put(155,300){\line(-1,-3){12}}
\put(175,280){\makebox(0,0){${\bf V}^*_{\tau 3}{\bf V}_{e 3}$}}
\put(98,294){\makebox(0,0){\scriptsize\bf 4}}
\put(145,293.5){\makebox(0,0){\scriptsize\bf 5}}
\put(138,274){\makebox(0,0){\scriptsize\bf 6}}
\put(120,245){\makebox(0,0){($\triangle_\mu$)}}
\put(295,300){\line(1,0){90}}
\put(337,310){\makebox(0,0){${\bf V}^*_{\tau 3}{\bf V}_{\tau 1}$}}
\put(295,300){\line(1,-3){13}}
\put(278,278){\makebox(0,0){${\bf V}^*_{e 3}{\bf V}_{e 1}$}}
\put(385,300){\line(-2,-1){77}}
\put(365,268){\makebox(0,0){${\bf V}^*_{\mu 3}{\bf V}_{\mu 1}$}}
\put(305,294){\makebox(0,0){\scriptsize\bf 5}}
\put(356,294){\makebox(0,0){\scriptsize\bf 1}}
\put(312.5,272.5){\makebox(0,0){\scriptsize\bf 7}}
\put(340,245){\makebox(0,0){($\triangle_2$)}}
\end{picture}

\begin{picture}(400,100)(20,0)
\put(70,300){\line(1,0){120}}
\put(120,310){\makebox(0,0){${\bf V}^*_{e 1}{\bf V}_{\mu 1}$}}
\put(70,300){\line(1,-4){9.2}}
\put(50,283){\makebox(0,0){${\bf V}^*_{e 3}{\bf V}_{\mu 3}$}}
\put(190,300){\line(-3,-1){111}}
\put(157,270){\makebox(0,0){${\bf V}^*_{e 2}{\bf V}_{\mu 2}$}}
\put(78,294){\makebox(0,0){\scriptsize\bf 7}}
\put(150,294){\makebox(0,0){\scriptsize\bf 8}}
\put(85,273){\makebox(0,0){\scriptsize\bf 9}}
\put(120,248){\makebox(0,0){($\triangle_\tau$)}}
\put(290,300){\line(1,0){100}}
\put(340,310){\makebox(0,0){${\bf V}^*_{e 1}{\bf V}_{e 2}$}}
\put(290,300){\line(3,-2){50}}
\put(296,275){\makebox(0,0){${\bf V}^*_{\mu 1}{\bf V}_{\mu 2}$}}
\put(390,300){\line(-3,-2){50}}
\put(389,275){\makebox(0,0){${\bf V}^*_{\tau 1}{\bf V}_{\tau 2}$}}
\put(310,294.5){\makebox(0,0){\scriptsize\bf 8}}
\put(370,294.5){\makebox(0,0){\scriptsize\bf 4}}
\put(340,275){\makebox(0,0){\scriptsize\bf 3}}
\put(340,245){\makebox(0,0){($\triangle_3$)}}
\end{picture}
\vspace{-8.5cm}
\caption{Unitarity triangles of the lepton flavor mixing matrix 
in the complex plane. Each triangle is named by 
the flavor index that does not manifest in its three sides.}
\end{figure}

To be more specific, we write down the transition probabilities
of three different neutrino flavors in the vacuum:
\begin{equation}
P (\nu_\alpha \rightarrow \nu_\beta ) \; =\;
-4 \sum_{i<j} \left [ {\rm Re} \left ( V_{\alpha i}
V_{\beta j} V^*_{\alpha j} V^*_{\beta i} \right )
\sin^2 F_{ij} \right ] ~ + ~
8 {\cal J}_l \prod_{i<j} \left (\sin F_{ij} \right ) \;
\end{equation}
with $(\alpha, \beta) = (e, \mu)$, $(\mu, \tau)$ or
$(\tau, e)$; $i, j = 1, 2, 3$; and 
$F_{ij} = 1.27 \Delta m^2_{ij} L/E$, where $L$ is the
distance between the neutrino source and the detector
(in unit of km) and $E$ denotes the neutrino beam energy
(in unit of GeV). Note that the $CP$-violating term
($\propto {\cal J}_l$) in $P(\nu_\alpha \rightarrow \nu_\beta)$
is independent of the neutrino flavor indices $\alpha$ and $\beta$.
As the negative neutrino mass-squared differences
$\Delta m^2_{12}$, $\Delta m^2_{13}$ and $\Delta m^2_{23}$ 
can be related to the positive solar and atmospheric neutrino mass-squared
differences given in (2.21), 
we simplify (3.22) to
\begin{eqnarray}
P (\nu_e \rightarrow \nu_\mu) & = & 4 |V_{e3}|^2 |V_{\mu 3}|^2
\sin^2 F_{\rm atm} ~ - ~ 4 {\rm Re} \left (V_{e 1} V_{\mu 2}
V^*_{e 2} V^*_{\mu 1} \right ) \sin^2 F_{\rm sun}
\nonumber \\ 
&& ~~~~~~~~~~~~~~~~~~~~~~~~~~~~~ - ~ 8 {\cal J}_l 
\sin F_{\rm sun} \sin^2 F_{\rm atm} \; ,
\nonumber \\ 
P (\nu_\mu \rightarrow \nu_\tau) & = & 4 |V_{\mu 3}|^2 |V_{\tau 3}|^2
\sin^2 F_{\rm atm} ~ - ~ 4 {\rm Re} \left ( V_{\mu 1} V_{\tau 2}
V^*_{\mu 2} V^*_{\tau 1} \right ) \sin^2 F_{\rm sun}
\nonumber \\
&& ~~~~~~~~~~~~~~~~~~~~~~~~~~~~~ - ~ 8 {\cal J}_l 
\sin F_{\rm sun} \sin^2 F_{\rm atm} \; ,
\nonumber \\ 
P (\nu_\tau \rightarrow \nu_e) & = & 4 |V_{\tau 3}|^2 |V_{e 3}|^2
\sin^2 F_{\rm atm} ~ - ~ 4 {\rm Re} \left ( V_{\tau 1} V_{e 2}
V^*_{\tau 2} V^*_{e 1} \right ) \sin^2 F_{\rm sun}
\nonumber \\
&& ~~~~~~~~~~~~~~~~~~~~~~~~~~~~~ - ~ 8 {\cal J}_l 
\sin F_{\rm sun} \sin^2 F_{\rm atm} \; ,
\end{eqnarray}
where $F_{\rm atm} = F_{32} \approx F_{31}$ and 
$F_{\rm sun} = F_{21}$ measure the oscillation frequencies
of atmospheric and solar neutrinos, respectively.
If the length of the baseline satisfies the condition
$L \sim E/\Delta m^2_{\rm sun}$ (i.e., $F_{\rm sun} \sim 1$),
which is only possible for the large-angle MSW solution to
the solar neutrino problem, the $CP$-conserving quantities
${\rm Re}(V_{\alpha 1} V_{\beta 2} V^*_{\alpha 2} V^*_{\beta 1})$
and the $CP$-violating parameter ${\cal J}_l$ could (in principle)
be determined from (3.23) for changing values of the neutrino
beam energy $E$. In this case we arrive at
\begin{eqnarray}
\tan\alpha^{~}_l & = & - \frac{{\cal J}_l}{{\rm Re}(V_{e1} V_{\mu 2}
V^*_{e 2}V^*_{\mu 1})} \;\; ,
\nonumber \\
\tan\beta_l & = & - \frac{{\cal J}_l}{{\rm Re}(V_{\mu 1} V_{\tau 2}
V^*_{\mu 2} V^*_{\tau 1})} \;\; ,
\nonumber \\
\tan\gamma^{~}_l & = & - \frac{{\cal J}_l}{{\rm Re}(V_{\tau 1} V_{e 2}
V^*_{\tau 2} V^*_{e 1})} \;\; .
\end{eqnarray}
As the $CP$-violating effect is universal in the three different
neutrino transitions, one needs only to measure ${\cal J}_l$
from the probability asymmetry between $\nu_\mu \rightarrow \nu_e$
and $\bar{\nu}_\mu \rightarrow \bar{\nu}_e$ 
(or $\nu_e \rightarrow \nu_\mu$) \cite{FX99CP}.
In practice, however, the earth-induced matter effects, which are
possible to fake the genuine $CP$-violating signals in the 
long-baseline neutrino experiments, have to be taken
into account \cite{CPmatter}. A more detailed discussion about leptonic $CP$
violation will be given in section 5.6.

Finally it is worth mentioning that $|V_{e3}|$, $|V_{\mu 3}|$
and $|V_{\tau 3}|$ can well be determined in the first-round
long-baseline neutrino experiments, in which the scale of
the neutrino mass-squared differences is set to $\Delta m^2_{\rm atm}$
(i.e., the terms proportional to $\sin^2 F_{\rm sun}$ and ${\cal J}_l$
in (3.23) are safely negligible).
This will allow a check of the unitarity of $V_l$, i.e.,
$|V_{e3}|^2 + |V_{\mu 3}|^2 + |V_{\tau 3}|^2 =1$.
If $|V_{e1}|$ and $|V_{e2}|$ are measured in the solar
neutrino experiment, one will be able to test the
unitarity condition $|V_{e1}|^2 + |V_{e2}|^2 + |V_{e3}|^2 =1$.
At this stage all three mixing angles of $V_l$ can
be determined. To isolate the $CP$-violating phase and
make a full test of the unitarity, however, more delicate
long-baseline neutrino experiments are needed.

\subsection{Classification of different parametrizations}

The $3\times 3$ flavor mixing matrix $V$ can be expressed in terms
of four independent parameters, which are usually taken as 
three rotation angles and one $CP$-violating phase angle. 
A number of parametrizations, different from the original
Kobayashi-Maskawa form \cite{KM}, 
have been proposed in the literature \cite{Wolfenstein83,Para}. 
To adopt a specific parametrization
is of course somehow arbitrary, since from a mathematical point of
view the different parametrizations are equivalent.
Nevertheless it is quite likely that the underlying physics
responsible for flavor mixing and $CP$ violation becomes
more transparent in a particular parametrization than in
the others \cite{FX97a}. For this reason, we find
it useful to give a classification of 
all possible parametrizations \cite{FX97b}
(in terms of 
three angles and one phase) and point out the most appropriate 
one for the phenomenology of mass matrices, flavor mixing and
$CP$ violation.

If the flavor mixing matrix $V$ is first assumed to be a real 
orthogonal matrix, it can
in general be written as a product of three matrices $R_{12}$,
$R_{23}$ and $R_{31}$, which describe simple rotations in the (1,2),
(2,3) and (3,1) planes:
\begin{equation}
\begin{tabular}{ccc} 
$R_{12}(\theta)$        & $R_{23}(\sigma)$      & $R_{31}(\tau)$
\\ \hline \\
$\left ( \matrix{
c^{~}_{\theta}  & s^{~}_{\theta}        & 0 \cr
- s^{~}_{\theta}        & c^{~}_{\theta}        & 0 \cr
0       & 0     & 1 \cr} \right )$      ~ & ~
$\left ( \matrix{
1       & 0     & 0 \cr
0       & c_{\sigma}    & s_{\sigma} \cr
0       & - s_{\sigma}  & c_{\sigma} \cr} \right )$     ~ & ~
$\left ( \matrix{
c_{\tau}        & 0     & s_{\tau} \cr
0       & 1     & 0 \cr
- s_{\tau}      & 0     & c_{\tau} \cr} \right )$
\end{tabular}
\end{equation}
where $s^{~}_{\theta} \equiv \sin \theta$, $c^{~}_{\theta} \equiv \cos
\theta$, and so on. 
Clearly these rotation matrices do not commute with one another.
There exist twelve different ways to arrange products of these
matrices such that the most general orthogonal matrix $R$ can be
obtained. 
Note that the matrix $R^{-1}_{ij} (\omega) $ plays an equivalent role
as $R_{ij} (\omega) $ in constructing $R$, because of $R^{-1}_{ij}(\omega) =
R_{ij}(-\omega)$. Note also that $R_{ij} (\omega) R_{ij}
(\omega^{\prime}) = R_{ij} (\omega + \omega^{\prime})$ holds, thus 
$R_{ij}(\omega) R_{ij}(\omega^{\prime})
R_{kl}(\omega^{\prime\prime})$ or $R_{kl}(\omega^{\prime\prime})
R_{ij}(\omega) R_{ij}(\omega^{\prime})$ cannot cover the whole space
of a $3\times 3$ orthogonal matrix and should be excluded.
Explicitly six of the twelve different forms of $R$ belong to
the type
\begin{equation}
R \; = \; R_{ij}(\theta) ~ R_{kl}(\sigma) ~ R_{ij}(\theta^{\prime})
\; ,
\end{equation}
in which the rotation in the $(i,j)$ plane occurs twice;
and the other six belong to the type
\begin{equation}
R \; = \; R_{ij}(\theta) ~ R_{kl}(\sigma) ~ R_{mn}(\tau) \; ,
\end{equation}
where rotations take place in three different planes.
Although all the twelve combinations represent the most general 
orthogonal matrices, only nine of them are structurally different.
The reason is that the products $R_{ij} R_{kl} R_{ij}$ and 
$R_{ij} R_{mn} R_{ij}$ (with $ij\neq kl\neq mn$) in (3.26) 
are correlated with each other, leading
essentially to the same form for $R$ \cite{FX97b}. 
Thus only three of $R$ in (3.26) need be treated as
independent choices. We then draw the conclusion that
there exist {\it nine} different forms for the orthogonal matrix $R$
(see Table 3.1).
\small
\begin{table}
\caption{Classification of different parametrizations for the flavor mixing
matrix.}
\vspace{0.cm}
\begin{center}
\begin{tabular}{ccc} \\ \hline\hline 
Parametrization     & ~~~~ & Useful relations 
\\  \hline 
\underline{{\it P1:} ~ $V \; = \; R_{12}(\theta) ~ R_{23}(\sigma, \varphi)
~ R^{-1}_{12}(\theta^{\prime})$}         
&& ${\cal J} = s^{~}_{\theta} c^{~}_{\theta} s^{~}_{\theta^{\prime}}
c^{~}_{\theta^{\prime}} s^2_{\sigma} c_{\sigma} \sin\varphi$ \\ 
$\left ( \matrix{
s^{~}_{\theta} s^{~}_{\theta^{\prime}} c_{\sigma} + c^{~}_{\theta} c^{~}_{\theta^{\prime}} e^{-{\rm i}\varphi}  & 
s^{~}_{\theta} c^{~}_{\theta^{\prime}} c_{\sigma} - c^{~}_{\theta}
s^{~}_{\theta^{\prime}} e^{-{\rm i}\varphi}     & s^{~}_{\theta} s_{\sigma} \cr
c^{~}_{\theta} s^{~}_{\theta^{\prime}} c_{\sigma} - s^{~}_{\theta} c^{~}_{\theta^{\prime}} e^{-{\rm i}\varphi}  &
c^{~}_{\theta} c^{~}_{\theta^{\prime}} c_{\sigma} + s^{~}_{\theta}
s^{~}_{\theta^{\prime}} e^{-{\rm i}\varphi}     & c^{~}_{\theta} s_{\sigma} \cr
- s^{~}_{\theta^{\prime}} s_{\sigma}    & - c^{~}_{\theta^{\prime}}
s_{\sigma}      & c_{\sigma} \cr} \right )
$
&& $\matrix{
\tan\theta = |V_{ub}/V_{cb}| \cr
\tan\theta^{\prime} = |V_{td}/V_{ts}| \cr
\cos\sigma = |V_{tb}| \cr} $ \\ \\
\underline{{\it P2:} ~ $V \; = \; R_{23}(\sigma) ~ R_{12}(\theta, \varphi)
~ R^{-1}_{23}(\sigma^{\prime})$}         
&& ${\cal J} = s^2_{\theta} c^{~}_{\theta} s_{\sigma} c_{\sigma} s_{\sigma^{\prime}} c_{\sigma^{\prime}} \sin\varphi$ \\ 
$\left ( \matrix{
c^{~}_{\theta}  & s^{~}_{\theta} c_{\sigma^{\prime}}    & -s^{~}_{\theta} s_{\sigma^{\prime}} \cr 
-s^{~}_{\theta} c_{\sigma}      & c^{~}_{\theta} c_{\sigma} c_{\sigma^{\prime}} + s_{\sigma} s_{\sigma^{\prime}} e^{-{\rm i}\varphi}    
& -c^{~}_{\theta} c_{\sigma} s_{\sigma^{\prime}} + s_{\sigma} c_{\sigma^{\prime}} e^{-{\rm i}\varphi} \cr
s^{~}_{\theta} s_{\sigma}       & -c^{~}_{\theta} s_{\sigma} c_{\sigma^{\prime}} + c_{\sigma} s_{\sigma^{\prime}} e^{-{\rm i}\varphi}   
& c^{~}_{\theta} s_{\sigma} s_{\sigma^{\prime}} + c_{\sigma} c_{\sigma^{\prime}} e^{-{\rm i}\varphi} \cr} \right )
$
&& $\matrix{
\cos\theta = |V_{ud}| \cr
\tan\sigma = |V_{td}/V_{cd}| \cr
\tan\sigma^{\prime} = |V_{ub}/V_{us}| \cr} $ \\ \\
\underline{{\it P3:} ~ $V \; = \; R_{23}(\sigma) ~ R_{31}(\tau, \varphi)
~ R_{12}(\theta)$}
&& ${\cal J} = s^{~}_{\theta} c^{~}_{\theta} s_{\sigma} c_{\sigma} s_{\tau} c^2_{\tau} \sin\varphi$
\\ 
$\left ( \matrix{
c^{~}_{\theta} c_{\tau}         & s^{~}_{\theta} c_{\tau}       & s_{\tau} \cr 
-c^{~}_{\theta} s_{\sigma} s_{\tau} - s^{~}_{\theta} c_{\sigma} e^{-{\rm i}\varphi}     
& -s^{~}_{\theta} s_{\sigma} s_{\tau} + c^{~}_{\theta} c_{\sigma} e^{-{\rm i}\varphi}   & s_{\sigma} c_{\tau} \cr
-c^{~}_{\theta} c_{\sigma} s_{\tau} + s^{~}_{\theta} s_{\sigma} e^{-{\rm i}\varphi}     
& -s^{~}_{\theta} c_{\sigma} s_{\tau} - c^{~}_{\theta} s_{\sigma} e^{-{\rm i}\varphi}   & c_{\sigma} c_{\tau} \cr} \right )
$
&& $\matrix{
\tan\theta = |V_{us}/V_{ud}| \cr
\tan\sigma = |V_{cb}/V_{tb}| \cr
\sin\tau = |V_{ub}| \cr} $ \\ \\
\underline{{\it P4:} ~ $V \; = \; R_{12}(\theta) ~ R_{31}(\tau, \varphi)
~ R^{-1}_{23}(\sigma)$}  
&& ${\cal J} = s^{~}_{\theta} c^{~}_{\theta} s_{\sigma} c_{\sigma} s_{\tau} c^2_{\tau} \sin\varphi$ 
\\ 
$\left ( \matrix{
c^{~}_{\theta} c_{\tau}         & c^{~}_{\theta} s_{\sigma} s_{\tau} + s^{~}_{\theta} c_{\sigma} e^{-{\rm i}\varphi}    
& c^{~}_{\theta} c_{\sigma} s_{\tau} - s^{~}_{\theta} s_{\sigma} e^{-{\rm i}\varphi} \cr
-s^{~}_{\theta} c_{\tau}        & -s^{~}_{\theta} s_{\sigma} s_{\tau} + c^{~}_{\theta} c_{\sigma} e^{-{\rm i}\varphi}   
& -s^{~}_{\theta} c_{\sigma} s_{\tau} - c^{~}_{\theta} s_{\sigma} e^{-{\rm i}\varphi} \cr
-s_{\tau}       & s_{\sigma} c_{\tau}   & c_{\sigma} c_{\tau} \cr} \right )
$
&& $\matrix{
\tan\theta = |V_{cd}/V_{ud}| \cr
\tan\sigma = |V_{ts}/V_{tb}| \cr
\sin\tau = |V_{td}| \cr} $ \\ \\
\underline{{\it P5:} ~ $V \; = \; R_{31}(\tau) ~ R_{12}(\theta, \varphi)
~ R^{-1}_{31}(\tau^{\prime})$}   
&& ${\cal J} = s^2_{\theta} c^{~}_{\theta} s_{\tau} c_{\tau} s_{\tau^{\prime}} c_{\tau^{\prime}} \sin\varphi$ 
\\ 
$\left ( \matrix{
c^{~}_{\theta} c_{\tau} c_{\tau^{\prime}} + s_{\tau} s_{\tau^{\prime}} e^{-{\rm i}\varphi}      & s^{~}_{\theta} c_{\tau}
& -c^{~}_{\theta} c_{\tau} s_{\tau^{\prime}} + s_{\tau} c_{\tau^{\prime}} e^{-{\rm i}\varphi} \cr
-s^{~}_{\theta} c_{\tau^{\prime}}       & c^{~}_{\theta}        & s^{~}_{\theta} s_{\tau^{\prime}} \cr 
-c^{~}_{\theta} s_{\tau} c_{\tau^{\prime}} + c_{\tau} s_{\tau^{\prime}} e^{-{\rm i}\varphi}     & -s^{~}_{\theta} s_{\tau}
& c^{~}_{\theta} s_{\tau} s_{\tau^{\prime}} + c_{\tau} c_{\tau^{\prime}} e^{-{\rm i}\varphi} \cr} \right )
$
&& $\matrix{
\cos\theta = |V_{cs}| \cr
\tan\tau = |V_{ts}/V_{us}| \cr
\tan\tau^{\prime} = |V_{cb}/V_{cd}| \cr} $ \\ \\
\underline{{\it P6:} ~ $V \; = \; R_{12}(\theta) ~ R_{23}(\sigma, \varphi)
~ R_{31}(\tau)$}         
&& ${\cal J} = s^{~}_{\theta} c^{~}_{\theta} s_{\sigma} c^2_{\sigma} s_{\tau} c_{\tau} \sin\varphi$ 
\\ 
$\left ( \matrix{
-s^{~}_{\theta} s_{\sigma} s_{\tau} + c^{~}_{\theta} c_{\tau} e^{-{\rm i}\varphi}       & s^{~}_{\theta} c_{\sigma}     &
s^{~}_{\theta} s_{\sigma} c_{\tau} + c^{~}_{\theta} s_{\tau} e^{-{\rm i}\varphi} \cr
-c^{~}_{\theta} s_{\sigma} s_{\tau} - s^{~}_{\theta} c_{\tau} e^{-{\rm i}\varphi}       & c^{~}_{\theta} c_{\sigma}     &
c^{~}_{\theta} s_{\sigma} c_{\tau} - s^{~}_{\theta} s_{\tau} e^{-{\rm i}\varphi} \cr
-c_{\sigma} s_{\tau}    & -s_{\sigma}           & c_{\sigma} c_{\tau} \cr} \right )
$
&& $\matrix{
\tan\theta = |V_{us}/V_{cs}| \cr
\sin\sigma = |V_{ts}| \cr
\tan\tau = |V_{td}/V_{tb}| \cr} $ \\ \\
\underline{{\it P7:} ~ $V \; = \; R_{23}(\sigma) ~ R_{12}(\theta, \varphi)
~ R^{-1}_{31}(\tau)$}    
&& ${\cal J} = s^{~}_{\theta} c^2_{\theta} s_{\sigma} c_{\sigma} s_{\tau} c_{\tau} \sin\varphi$ 
\\ 
$\left ( \matrix{
c^{~}_{\theta} c_{\tau}         & s^{~}_{\theta}        & -c^{~}_{\theta} s_{\tau} \cr 
-s^{~}_{\theta} c_{\sigma} c_{\tau} + s_{\sigma} s_{\tau} e^{-{\rm i}\varphi}   & c^{~}_{\theta} c_{\sigma}
& s^{~}_{\theta} c_{\sigma} s_{\tau} + s_{\sigma} c_{\tau} e^{-{\rm i}\varphi} \cr
s^{~}_{\theta} s_{\sigma} c_{\tau} + c_{\sigma} s_{\tau} e^{-{\rm i}\varphi}    & -c^{~}_{\theta} s_{\sigma}    
& -s^{~}_{\theta} s_{\sigma} s_{\tau} + c_{\sigma} c_{\tau} e^{-{\rm i}\varphi} \cr} \right )
$
&& $\matrix{
\sin\theta = |V_{us}| \cr
\tan\sigma = |V_{ts}/V_{cs}| \cr
\tan\tau = |V_{ub}/V_{ud}| \cr} $ \\ \\
\underline{{\it P8:} ~ $V \; = \; R_{31}(\tau) ~ R_{12}(\theta, \varphi)
~ R_{23}(\sigma)$}       
&& ${\cal J} = s^{~}_{\theta} c^2_{\theta} s_{\sigma} c_{\sigma} s_{\tau} c_{\tau} \sin\varphi$ 
\\ 
$\left ( \matrix{
c^{~}_{\theta} c_{\tau}         & s^{~}_{\theta} c_{\sigma} c_{\tau} - s_{\sigma} s_{\tau} e^{-{\rm i}\varphi}  
& s^{~}_{\theta} s_{\sigma} c_{\tau} + c_{\sigma} s_{\tau} e^{-{\rm i}\varphi} \cr
-s^{~}_{\theta} & c^{~}_{\theta} c_{\sigma}     & c^{~}_{\theta} s_{\sigma} \cr
-c^{~}_{\theta} s_{\tau}        & -s^{~}_{\theta} c_{\sigma} s_{\tau} - s_{\sigma} c_{\tau} e^{-{\rm i}\varphi} 
& -s^{~}_{\theta} s_{\sigma} s_{\tau} + c_{\sigma} c_{\tau} e^{-{\rm i}\varphi} \cr} \right )
$
&& $\matrix{
\sin\theta = |V_{cd}| \cr
\tan\sigma = |V_{cb}/V_{cs}| \cr
\tan\tau = |V_{td}/V_{ud}| \cr} $ \\ \\
\underline{{\it P9:} ~ $V \; = \; R_{31}(\tau) ~ R_{23}(\sigma, \varphi)
~ R^{-1}_{12}(\theta)$}  
&& ${\cal J} = s^{~}_{\theta} c^{~}_{\theta} s_{\sigma} c^2_{\sigma} s_{\tau} c_{\tau} \sin\varphi$ 
\\ 
$\left ( \matrix{
-s^{~}_{\theta} s_{\sigma} s_{\tau} + c^{~}_{\theta} c_{\tau} e^{-{\rm i}\varphi}       
& -c^{~}_{\theta} s_{\sigma} s_{\tau} - s^{~}_{\theta} c_{\tau} e^{-{\rm i}\varphi}     & c_{\sigma} s_{\tau} \cr
s^{~}_{\theta} c_{\sigma}       & c^{~}_{\theta} c_{\sigma}     & s_{\sigma} \cr 
-s^{~}_{\theta} s_{\sigma} c_{\tau} - c^{~}_{\theta} s_{\tau} e^{-{\rm i}\varphi}       
& -c^{~}_{\theta} s_{\sigma} c_{\tau} + s^{~}_{\theta} s_{\tau} e^{-{\rm i}\varphi}     & c_{\sigma} c_{\tau} \cr} \right )
$
&& $\matrix{
\tan\theta = |V_{cd}/V_{cs}| \cr
\sin\sigma = |V_{cb}| \cr
\tan\tau = |V_{ub}/V_{tb}| \cr} $ \\ 
\hline\hline
\end{tabular}
\end{center}
\end{table}
\normalsize

We proceed to introduce the $CP$-violating phase, denoted by $\varphi$.
The complex rotation matrices should be
unitary, such that a unitary flavor mixing matrix results.
There are several different ways for
$\varphi$ to enter $R_{12}$, $R_{23}$ or
$R_{31}$. Then the flavor mixing matrix $V$ can be constructed, as a product of
three rotation matrices, by use of one complex $R_{ij}$ and two real ones. 
Note that the location of the $CP$-violating phase in $V$ can be arranged by
redefining the quark field phases, thus it does not play an essential role in 
classifying different parametrizations. We find that it is always
possible to locate the phase parameter $\varphi$ in a $2\times 2$ submatrix of
$V$, in which each element is a sum of two terms with the relative
phase $\varphi$. The remaining five elements of $V$ are real in such a 
phase assignment. Accordingly we arrive at nine distinctive
parametrizations of the flavor mixing  matrix $V$ as listed in Table 3.1, where
the complex rotation matrices $R_{12}(\theta, \varphi)$,
$R_{23}(\sigma, \varphi)$ and $R_{31}(\tau, \varphi)$ are obtained
directly from the real ones in (3.25) with the replacement $1
\rightarrow e^{-{\rm i}\varphi}$. 

Some instructive relations of each parametrization, together
with the expression for the ${\cal J}$ parameter defined in (3.11), 
are also given in Table 3.1.
One can see that the parametrizations {\it P2} and {\it P3}
correspond to the Kobayashi-Maskawa \cite{KM} and 
Maiani \cite{Para} parametrizations, 
although different
notations for the $CP$-violating phase and three mixing angles are adopted
here. The latter is indeed equivalent to the 
``standard'' parametrization advocated by the Particle Data Group \cite{PDG}.
This can be seen clearly if one makes 
three transformations of quark-field phases: 
$c \rightarrow c ~ e^{-{\rm i} \varphi}$, $t
\rightarrow t ~ e^{-{\rm i} \varphi}$, and $b \rightarrow 
b ~ e^{-{\rm i} \varphi}$. In addition, the parametrization
{\it P1} is just the one proposed recently by the authors \cite{FX97a}.
The specific advantages of this parametrization for the study of
quark mass matrices and $B$-meson physics will become obvious
in section 3.6 and section 4.

Besides the aforementioned parametrizations, which take advantage of
three rotation angles and one $CP$-violating phase, there are 
some other ways to parametrize the $3\times 3$ flavor mixing matrix. 
A representation based on a hierarchical expansion was proposed by 
Wolfenstein \cite{Wolfenstein83}:
\begin{equation}
V \; \approx \; \left (\matrix{
1 - \frac{1}{2} \lambda^2       & \lambda       &
A \lambda^3 (\rho - {\rm i}\eta ) \cr
-\lambda        & 1 -\frac{1}{2} \lambda^2      &
A\lambda^2 \cr
A \lambda^3 (1-\rho -{\rm i}\eta )      & -A\lambda^2
& 1 \cr} \right ) \; ,
\end{equation}
in which $\lambda \approx 0.22$, $A\approx 0.83$, and the
magnitudes of $\rho$ and $\eta$ are smaller than one.
Note that at the accuracy level
of (3.28) the $CP$-violating parameter $\eta$
appears only in two smallest elements ($V_{ub}$ and
$V_{td}$), whose magnitudes are suppressed by $\lambda^3$.
If the three sides of the unitarity triangle 
$\triangle_s$ are rescaled by $V_{cd}V^*_{cb}$, then the
resultant triangle has three vertices $(0,0)$, $(1,0)$ and 
$(\rho, \eta)$ in the complex plane \cite{PDG}. 
To discuss more precise experimental data on quark
mixing and $CP$ violation, modified versions of the
Wolfenstein parametrization with higher accuracy have 
been proposed in the literature \cite{NewWolf}. 

Furthermore one can parametrize $V$ in terms of the moduli of four 
independent matrix elements, in terms of four independent 
angles of the unitarity triangles, or in terms of four
characteristic quantities of $V$ \cite{Bjorken}.
The choice of any four
parameters is arbitrary, and different choices may be mathematically
convenient for different physical purposes. 

\subsection{A unique description of flavor mixing}

{\rm From} a mathematical point of view, all different parametrizations
of the flavor mixing matrix are equivalent. However, this is not the 
case if we take the hierarchical structure of the quark mass spectrum and its
implications on the flavor mixing phenomenon into account. 
It is well known that both the observed
quark mass spectrum and the observed values of the flavor mixing
parameters exhibit a striking hierarchical structure. The latter can
be understood in a natural way as the consequence of a specific
pattern of chiral symmetries, whose breaking causes the towers of
different masses to appear step by step.
Such a chiral evolution of the
mass matrices leads, as argued in Ref. \cite{FX97a}, to a
specific way to describe the flavor mixing (i.e.,
the parametrization {\it P1} in Table 3.1):
\begin{eqnarray}
V & = & \left ( \matrix{
c_{\rm u}       & s_{\rm u}     & 0 \cr
-s_{\rm u}      & c_{\rm u}     & 0 \cr
0       & 0     & 1 \cr } \right )  \left ( \matrix{
e^{-{\rm i}\varphi}     & 0     & 0 \cr
0       & c     & s \cr
0       & -s    & c \cr } \right )  \left ( \matrix{
c_{\rm d}       & -s_{\rm d}    & 0 \cr
s_{\rm d}       & c_{\rm d}     & 0 \cr
0       & 0     & 1 \cr } \right )  \nonumber \\ \nonumber \\
& = & \left ( \matrix{
s_{\rm u} s_{\rm d} c + c_{\rm u} c_{\rm d} e^{-{\rm i}\varphi} &
s_{\rm u} c_{\rm d} c - c_{\rm u} s_{\rm d} e^{-{\rm i}\varphi} &
s_{\rm u} s \cr
c_{\rm u} s_{\rm d} c - s_{\rm u} c_{\rm d} e^{-{\rm i}\varphi} &
c_{\rm u} c_{\rm d} c + s_{\rm u} s_{\rm d} e^{-{\rm i}\varphi}   &
c_{\rm u} s \cr
- s_{\rm d} s   & - c_{\rm d} s & c \cr } \right ) \; ,
\end{eqnarray}
where $s_{\rm u} \equiv \sin\theta_{\rm u}$, $s_{\rm d} \equiv
\sin\theta_{\rm d}$, $c\equiv \cos\theta$, and so on. The three mixing
angles may all be arranged to lie in the first quadrant, and the phase
parameter $\varphi$ generally takes values from $0$ to $2\pi$. 
This parametrization follows automatically from the generic Hermitian 
mass matrices $M_{\rm u}$ and $M_{\rm d}$, 
if one imposes the constraints from the chiral
symmetries and the quark mass hierarchy on them (see section 4.1 for 
detailed discussions). Therefore it is particularly useful
for the study of realistic quark mass matrices.
We shall see later on
that this representation also makes the properties of flavor
mixing and $CP$ violation more transparent and proves to be
very convenient for the phenomenology of $B$-meson decays.

The three mixing angles $\theta$, $\theta_{\rm u}$ and 
$\theta_{\rm d}$ have direct physical meanings. 
The angle $\theta$
describes the mixing between the second and third families.
We shall refer to this mixing (involving $t$ and $b$ quarks)
as the ``heavy quark mixing''.
The angle $\theta_{\rm u}$ primarily describes the $u$-$c$ 
mixing and will be denoted as the ``u-channel mixing''.
The angle $\theta_{\rm d}$ primarily describes 
the $d$-$s$ mixing, and will be denoted as the ``d-channel mixing''. 
Clearly there exists an asymmetry between the mixing of the first and
second families and that of the second and third families,
which in our view reflects interesting details of the underlying 
dynamics of flavor mixing. 
The heavy quark mixing is a combined effect, involving both charge
$+2/3$ and charge $-1/3$ quarks, while the u- or d-channel mixing
proceeds 
primarily in the charge $+2/3$ or charge $-1/3$ sector. Therefore an
experimental determination of $\theta_{\rm u}$ and $\theta_{\rm d}$
would give us useful information about the underlying patterns of 
up and down mass matrices.

The angles $\theta$, $\theta_{\rm u}$ and $\theta_{\rm d}$
are related in a very simple way to observable quantities of $B$-meson 
physics. 
For example, $\theta$ is related to 
the rate of the semileptonic decay $B\rightarrow D^*l\nu^{~}_l$; 
$\theta_{\rm u}$ is associated with the ratio of the decay rate of
$B\rightarrow (\pi, \rho) l \nu^{~}_l$ to that of $B\rightarrow 
D^* l\nu^{~}_l$; and $\theta_{\rm d}$ can be determined from the ratio of
the mass difference between two $B_d$ mass eigenstates to that between
two $B_s$ mass eigenstates. From (3.29) one can find the following exact
relations \cite{FX97a}:
\begin{eqnarray}
\tan\theta_{\rm u} & = & \left | \frac{V_{ub}}{V_{cb}} \right | \; ,
\nonumber \\
\tan\theta_{\rm d} & = & \left | \frac{V_{td}}{V_{ts}} \right | \; ,
\nonumber \\
\sin \theta & = & \sqrt{|V_{ub}|^2 + |V_{cb}|^2} \; .
\end{eqnarray}
These results show that the parametrization (3.29) is particularly
useful for $B$-meson physics. A global analysis
of current experimental data on $|\epsilon^{~}_K|$, $|V_{ub}/V_{cb}|$,
$\Delta M_{B_d}$ and $\Delta M_{B_s}$, as already done in Ref. \cite{Stocchi},
yields $\theta = 2.30^{\circ} \pm 0.09^{\circ}$,
$\theta_{\rm u} = 4.87^{\circ} \pm 0.86^{\circ}$,
$\theta_{\rm d} \; = \; 11.71^{\circ} \pm 1.09^{\circ}$,
and $\varphi = 91.1^{\circ} \pm 11.8^{\circ}$. The smallness
of $\theta$ or $\theta_{\rm u}$ implies the approximate decoupling 
between two different flavor mixing channels, while the fact $\varphi
\approx \pi/2$ might have a deeper meaning for $CP$
violation in the standard model (see section 4).
 
The dynamics of flavor mixing can easily be understood by
considering certain limiting cases in (3.29). In the limit $\theta
\rightarrow 0$ the
flavor mixing exists only between the first and
second families, described by a single (overall) mixing angle \cite{Cabibbo}
-- the Cabibbo angle $\theta_{\rm C}$ (i.e.,
$s^{~}_{\rm C} \equiv \sin \theta_{\rm C} = |V_{us}| = |V_{cd}|$):
\begin{equation}
s^{~}_{\rm C} \; =\; \left | s_{\rm u} c_{\rm d} ~ - ~
c_{\rm u} s_{\rm d} e^{-{\rm i}\varphi} \right | \; .
\end{equation}
Note that the Cabibbo angle
(or the matrix element $V_{us}$ or $V_{cd}$) is
indeed a superposition of two terms including a phase. This feature
comes up naturally in the hierarchical approach of quark mass
matrices \cite{Fr77,Fr78}, since the limit 
$\theta \rightarrow 0$ is essentially
equivalent to the heavy quark limit $m_t \rightarrow \infty$,
$m_b \rightarrow \infty$. Note also that 
(3.31) defines a triangle in the complex plane, 
denoted as the light-quark triangle (see Fig. 3.3(a)
for illustration). The limit $\theta \rightarrow 0$, in which a variety of
interesting patterns of quark mass matrices predict \cite{FX99} 
\begin{eqnarray}
\tan \theta_{\rm u} & = & \sqrt{\frac{m_u}{m_c}} \;\; ,
\nonumber \\
\tan \theta_{\rm d} & = & \sqrt{\frac{m_d}{m_s}} \;\; ,
\end{eqnarray}
is very close to the reality (see section 4.3). It assures that 
the light-quark triangle is approximately congruent with the
rescaled unitarity triangle $\triangle_s$, as to be shown later.
\begin{figure}[t]
\hspace*{-1.8cm}\begin{picture}(400,160)(-10,210)
\put(80,300){\line(1,0){150}}
\put(80,300.5){\line(1,0){150}}
\put(150,285.5){\makebox(0,0){$s^{~}_{\rm C}$}}
\put(80,300){\line(1,3){21.5}}
\put(80,300.5){\line(1,3){21.5}}
\put(80,299.5){\line(1,3){21.5}}
\put(71,333){\makebox(0,0){$s_{\rm u}c_{\rm d}$}}
\put(230,300){\line(-2,1){128}}
\put(230,300.5){\line(-2,1){128}}
\put(178,343.5){\makebox(0,0){$c_{\rm u}s_{\rm d}$}}
\put(109,350){\makebox(0,0){$\varphi$}}
\put(150,260){\makebox(0,0){(a)}}
\hspace*{-0.5cm}
\put(300,300){\line(1,0){150}}
\put(300,300.5){\line(1,0){150}}
\put(370,285.5){\makebox(0,0){$s^{~}_{\rm C}$}}
\put(300,300){\line(1,3){21.5}}
\put(300,300.5){\line(1,3){21.5}}
\put(300,299.5){\line(1,3){21.5}}
\put(287,333){\makebox(0,0){$s_{\rm u} c_{\rm d}$}}
\put(450,300){\line(-2,1){128}}
\put(450,300.5){\line(-2,1){128}}
\put(395,343.5){\makebox(0,0){$s_{\rm d}$}}
\put(315,310){\makebox(0,0){$\gamma$}}
\put(408,309){\makebox(0,0){$\beta$}}
\put(329,350){\makebox(0,0){$\alpha$}}
\put(370,260){\makebox(0,0){(b)}}
\end{picture}
\vspace{-1.5cm}
\caption{The light-quark triangle (a) and the rescaled 
unitarity triangle (b) in the complex plane.}
\end{figure}
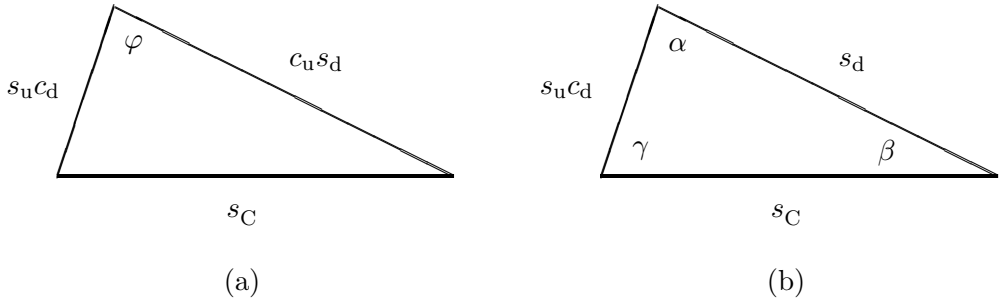

As we have pointed out, the unitarity triangle 
$\triangle_s$ is of particular interest for $B$-meson physics
and $CP$ violation. 
Rescaling the three sides of $\triangle_s$ by $V^*_{cb}$, we
obtain 
\begin{equation}
|V^*_{ub}V_{ud}| ~ : ~ |V^*_{cb}V_{cd}| ~ : ~ |V^*_{tb}V_{td}| 
\; \approx \; s_{\rm u} c_{\rm d} ~ : ~ s^{~}_{\rm C} ~ : ~ s_{\rm d} \; 
\end{equation}
as a good approximation (with an error of a few percent).
The rescaled unitarity triangle $\triangle_s$ is illustrated in Fig. 3.3(b).
Comparing this
triangle with the light-quark triangle shown 
in Fig. 3.3(a), we find that they are 
indeed congruent with each other to a high degree of 
accuracy \cite{FX97a,FX95}
\footnote{Strictly, the mixing angles $\theta_{\rm u}$ and
$\theta_{\rm d}$ in the light-quark triangle should be redefined
in accordance with the limit $\theta \rightarrow 0$. We have omitted
this redefinition, because corrections from $\theta \neq 0$ are
tiny. This point will be seen more clearly in sections 4.2 and 4.3, where we
derive both the light-quark triangle and the unitarity triangle
from a realistic texture of quark mass matrices.}.
The congruent relation between these two triangles is particularly
interesting, since the light-quark
triangle is essentially a feature of the physics
of the first two quark families, while the unitarity triangle is
linked to all three families and describes $CP$ violation.
As a straightforward result, 
$\alpha \approx \varphi$ holds. The other two angles
of the unitarity triangle (i.e., $\beta$ and $\gamma$) can then be
predicted in terms of $\varphi$ and the sides of the light-quark
triangles, which may be simple functions of light quark mass
ratios as indicated in (3.32). 
Note that both the light-quark triangle and the rescaled unitarity
triangle are approximately independent of the renormalization-group 
effects \cite{FX99,FX95}.
Hence information about the $CP$-violating angles $\alpha$,
$\beta$ and $\gamma$ obtained at
low-energy scales can directly be extended to an arbitrary high-energy
scale, at which a dynamical understanding of quark mass generation and
$CP$ violation might be available.

Taking the relationship in (3.33) and
the approximate congruency between the unitarity
triangles $\triangle_c$ and $\triangle_s$ (see Fig. 3.1) into
account, we find that the parametrization (3.29)
takes an instructive leading-order form:
\begin{equation}
V \; \approx \; \left (\matrix{
e^{-{\rm i}\alpha}      & s^{~}_{\rm C} e^{{\rm i}\gamma}       & s_{\rm u} s \cr
s^{~}_{\rm C} e^{{\rm i}\beta}  & 1     & s \cr
-s_{\rm d} s    & -s    & 1 \cr} \right ) \; .
\end{equation}
The values of $\alpha$, $\beta$ and $\gamma$ can
therefore be given in terms of $s_{\rm u}$, $s_{\rm d}$ and 
$s^{~}_{\rm C}$.

Finally it is worth pointing out that there exist simple relations
between the mixing parameters in (3.29) and those in (3.28).
In particular $\varphi = 90^\circ$ implies 
$\eta^2 \approx \rho (1-\rho)$, 
on the condition $0 < \rho < 1$ \cite{FX97a}.
In this interesting case, of course, the flavor mixing matrix
can fully be described in terms of only three independent parameters.

In summary, the parametrization (3.29)
has many distinctions and advantages compared
with other representations. We therefore
highlight its usefulness in the
description of flavor mixing and $CP$ violation, in particular, 
for the studies of quark mass matrices and $B$-meson physics.

\section{Realistic schemes of quark mass matrices}
\setcounter{equation}{0}
\setcounter{figure}{0}
\setcounter{table}{0}

\subsection{Generic Hermitian mass matrices}

For a deeper understanding of the quark flavor mixing and $CP$-violating
phenomena, it is desirable to study the properties of the quark mass
matrices $M_{\rm u, d}$, 
whose textures are completely unknown within the standard electroweak
model. A theory more fundamental than the standard model should
be able to determine $M_{\rm u}$ and $M_{\rm d}$ exclusively,
such that the associated physical parameters (six quark masses,
three flavor mixing angles, and one $CP$-violating phase) could be
predicted. Attempts in this direction, e.g., those
starting from supersymmetric grand unified theories and from 
superstring theories, have so far proved not to be very 
successful \cite{Raby95}.
Phenomenologically the natural approach is to look for the 
simplest pattern of $M_{\rm u}$ and $M_{\rm d}$, which can 
result in self-consistent and experimentally-favored 
relations between quark masses and flavor mixing parameters.
The discrete flavor symmetries hidden in such mass matrix
textures might finally provide useful hints towards
the dynamics of quark mass generation and $CP$ violation in a
more fundamental theoretical framework.

Before discussing a variety of specific patterns,
we first make some analyses of the general quark mass
matrices and their consequences for flavor mixing and $CP$ violation.
It is worth remarking that the mass matrices $M_{\rm u}$
and $M_{\rm d}$ can be taken to be Hermitian, without loss of generality,
in the standard model or its extensions which have no flavor-changing
right-handed currents. The essential point is that the 
right-handed fields in the Lagrangians of such models
are SU(2) singlets. Therefore it is always possible to choose a
suitable basis of the right-handed quarks, so that the resultant
up- and down-type mass matrices become Hermitian. Such a redefinition
of the right-handed fields does not alter the rest of the
Lagrangians, i.e., it keeps the generality of the
relevant physics of the model. For this reason we shall simply assume
$M_{\rm u}$ and $M_{\rm d}$ to be Hermitian mass matrices in the
following.

A general Hermitian mass matrix $M_{\rm q}$ (for $\rm q =u$ or 
$\rm d$) can be written as 
\begin{equation}
M_{\rm q} \; =\; \left ( \matrix{
E_{\rm q}       & D_{\rm q}     & F_{\rm q} \cr
D^*_{\rm q}     & C_{\rm q}     & B_{\rm q} \cr
F^*_{\rm q}     & B^*_{\rm q}   & A_{\rm q} \cr } \right ) \; . 
\end{equation}
Note that physics is invariant under 
a common unitary transformation of $M_{\rm u}$ and $M_{\rm d}$,
i.e., $M_{\rm q} \rightarrow S M_{\rm q} S^{\dagger}$, where
$S$ is an arbitrary unitary matrix. Using this freedom,
one can always arrange $F_{\rm u}$ and $F_{\rm d}$ to vanish
and obtain the following form of $M_{\rm q}$ \cite{FX97a}:
\begin{equation}
M_{\rm q} \; =\; \left ( \matrix{
E_{\rm q}       & D_{\rm q}     & {\bf 0} \cr
D^*_{\rm q}     & C_{\rm q}     & B_{\rm q} \cr
{\bf 0} & B^*_{\rm q}   & A_{\rm q} \cr } \right ) \; .
\end{equation}
The condition $F_{\rm u} = F_{\rm d} =0$, which imposes
four constraints on $S$, can always be fulfilled, as $S$
generally consists of nine parameters (three mixing angles
and six complex phases). This implies that by going from (4.1) to (4.2) 
the generality of the physics remains. 
The basis in which the mass matrices take the form (4.2) is a 
particularly interesting basis of the quark flavor space.
The texture zeros of $M_{\rm u}$ and $M_{\rm d}$ in this basis
allows the absence of direct mixing between the heavy quark ($t$
or $b$) and the light quark ($u$ or $d$). Such a feature seems
quite natural from the phenomenological point of view, but it
does not imply any special relations among
mass eigenvalues and flavor mixing parameters. 
Note that $M_{\rm u}$ and 
$M_{\rm d}$ in (4.2) totally have twelve nontrivial parameters
(ten moduli and two phase differences \cite{Shrock}), 
thus no definite prediction 
can be obtained for the flavor mixing matrix.

It is worth remarking that the specific description of flavor mixing
given in (3.29) can naturally be obtained from the diagonalization of 
$M_{\rm u,d}$ in (4.2). 
The latter can in the absence 
of complex phases be diagonalized by a $3\times 3$
orthogonal matrix, described only by two rotation
angles in the hierarchy limit of quark masses \cite{Fr79}.
First, the off-diagonal element $B_{\rm q}$ is rotated away 
by a rotation matrix $R_{23}$
between the second and third families.
Then the element $D_{\rm q}$ is rotated away by a
transformation $R_{12}$ between the first and second families. 
No rotation between the first and third families is
necessary in either the limit $m_u\rightarrow 0$,
$m_d\rightarrow 0$ or the limit $m_t\rightarrow
\infty$, $m_b\rightarrow \infty$. In reality
one needs an additional transformation $R_{31}$ with a tiny rotation angle 
to fully diagonalize $M_{\rm q}$. Note, however, that the rotation sequence
$(R^{\rm u}_{12} R^{\rm u}_{23}) (R^{\rm d }_{12}
R^{\rm d }_{23})^{\rm T}$ is enough to describe the $3\times 3$ real
flavor mixing matrix, as the effects of $R^{\rm u}_{31}$
and $R^{\rm d}_{31}$ can always be absorbed into
this sequence through redefining the relevant rotation
angles. By introducing a complex phase angle into 
the rotation combination $(R^{\rm u}_{23}) (R^{\rm d}_{23})^{\rm T}$,
we are then able to arrive at the representation of
quark flavor mixing in (3.29).
Although we have derived this particular
parametrization in a heuristic way from the
hierarchical mass matrices in (4.2), we should like to emphasize 
that (3.29) is a model-independent description of any $3\times 3$
flavor mixing matrix.

Now we start from the Hermitian mass matrices (4.2) to 
derive the flavor mixing angles ($\theta_{\rm u}$, 
$\theta_{\rm d}$, $\theta$) and the $CP$-violating phase ($\varphi$).
Note that $M_{\rm q}$ can be decomposed 
into $M_{\rm q} = P^{\dagger}_{\rm q} \overline{M}_{\rm q}
P_{\rm q}$, where
\begin{equation}
\overline{M}_{\rm q} \; =\; \left ( \matrix{
E_{\rm q}       & |D_{\rm q}|   & {\bf 0} \cr
|D_{\rm q}|     & C_{\rm q}     & |B_{\rm q}| \cr
{\bf 0}       & |B_{\rm q}|   & A_{\rm q} \cr} \right ) \; 
\end{equation}
is a real symmetric matrix, and 
$P_{\rm q} = {\rm Diag} \{ 1, e^{{\rm i}\phi_{D_{\rm q}}}, 
e^{{\rm i} (\phi_{B_{\rm q}} + \phi_{D_{\rm q}})} \}$ 
with $\phi_{D_{\rm q}} = \arg (D_{\rm q})$ and 
$\phi_{B_{\rm q}} = \arg (B_{\rm q})$
is a diagonal phase matrix. For simplicity we shall neglect
the subscript ``q'' in the following,
when it is unnecessary to distinguish between
the up and down quark sectors.
$\overline{M}$ can be diagonalized by use of the orthogonal transformation 
$O^{\rm T} \overline{M} O = {\rm Diag} \{ \lambda_1, \lambda_2, \lambda_3 \}$,
where $\lambda_i$ ($i=1,2,3$) are quark mass eigenvalues and may be either
positive or negative.
As a result, we have
\begin{eqnarray}
\sum_{i=1}^3 \lambda_i & = & A + C + E \; ,
\nonumber \\
\prod_{i=1}^3 \lambda_i & = & ACE - A |D|^2 - E |B|^2  \; ,
\nonumber \\
\sum_{i=1}^3 \lambda^2_i & = & A^2 + 2|B|^2 + C^2 + 2|D|^2 + E^2 \; .
\end{eqnarray}
It is a simple exercise to solve the nine matrix elements of
$O$ in terms of the parameters of quark mass matrices. Explicitly,
three diagonal elements of $O$ read \cite{FX99}
\footnote{Here and hereafter, the off-diagonal elements $B$ and $D$ are
both taken to be nonvanishing. The relevant calculations will somehow
be simplified if one of them vanishes.} :
\begin{eqnarray}
O_{11} & = & \left [ 1 + \left ( \frac{\lambda_1 - E}{|D|} \right )^2 + \left
( \frac{|B|}{|D|} \cdot \frac{\lambda_1 - E}{\lambda_1 - A} \right )^2 
\right ]^{-1/2} \; , \nonumber \\
O_{22} & = & \left [ 1 + \left ( \frac{|D|}{\lambda_2 - E} \right )^2 + \left
( \frac{|B|}{\lambda_2 - A} \right )^2 \right ]^{-1/2} \; , \nonumber \\
O_{33} & = & \left [ 1 + \left ( \frac{\lambda_3 - A}{|B|} \right )^2 + \left
( \frac{|D|}{|B|} \cdot \frac{\lambda_3 - A}{\lambda_3 - E} \right )^2 
\right ]^{-1/2} \; ;
\end{eqnarray}
and then six off-diagonal elements of $O$ can be obtained from the
relations
\begin{eqnarray}
O_{2i} & = & \frac{\lambda_i - E}{|D|} O_{1i} \; , \nonumber \\
O_{3i} & = & \frac{|B|}{\lambda_i - A}O_{2i} \; .
\end{eqnarray}
The flavor mixing matrix turns out to be $V \equiv O^{\rm T}_{\rm u} (P_{\rm u}
P_{\rm d}^{\dagger}) O_{\rm d}$. More specifically, we have
\begin{equation}
V_{i\alpha} \; =\; O^{\rm u}_{1i} O^{\rm d}_{1\alpha} ~ + ~
O^{\rm u}_{2i} O^{\rm d}_{2\alpha} e^{{\rm i}\phi_1} ~ + ~
O^{\rm u}_{3i} O^{\rm d}_{3\alpha} e^{{\rm i}(\phi_1 + \phi_2)} \; ,
\end{equation}
where $\phi_1 \equiv \phi_{D_{\rm u}} - \phi_{D_{\rm d}}$
and $\phi_2  \equiv  \phi_{B_{\rm u}} - \phi_{B_{\rm d}}$;
and the Latin subscript $i$ and the Greek subscript
$\alpha$ run over $(u,c,t)$ and $(d,s,b)$, respectively.

To link the parameters of flavor mixing with those of
quark mass matrices in an analytically concise way, 
we first define four dimensionless quantities:
\begin{eqnarray}
X_{\rm u} & \equiv & \left | \frac{|D_{\rm u}|}{\lambda^{\rm u}_1 - 
E_{\rm u}} \cdot 
\frac{|D_{\rm d}| \left (\lambda^{\rm d}_3 - A_{\rm d} \right )}
{|B_{\rm d}| \left
(\lambda^{\rm d}_3 - E_{\rm d} \right )} ~ + ~ \frac{\lambda^{\rm d}_3 
- A_{\rm d}}{|B_{\rm d}|}
e^{{\rm i}\phi_1} ~ + ~ \frac{|B_{\rm u}|}{\lambda^{\rm u}_1 - A_{\rm u}}
e^{{\rm i}(\phi_1 + \phi_2)}
\right | \; , \nonumber \\
Y_{\rm u} & \equiv & \left | \frac{|D_{\rm u}|}{\lambda^{\rm u}_2 - 
E_{\rm u}} \cdot 
\frac{|D_{\rm d}| \left (\lambda^{\rm d}_3 - A_{\rm d} \right )}
{|B_{\rm d}| \left
(\lambda^{\rm d}_3 - E_{\rm d} \right )} ~ + ~ 
\frac{\lambda^{\rm d}_3 - A_{\rm d}}{|B_{\rm d}|}
e^{{\rm i}\phi_1} ~ + ~ \frac{|B_{\rm u}|}{\lambda^{\rm u}_2 - A_{\rm u}}
e^{{\rm i}(\phi_1 + \phi_2)}
\right | \; ;
\end{eqnarray}
and $(X_{\rm d}, Y_{\rm d})$ can directly be obtained from $(X_{\rm
u}, Y_{\rm u})$ through the subscript exchange ${\rm u}
\leftrightarrow {\rm d}$ in (4.8). 
After a lengthy but straightforward calculation, we arrive at
\begin{eqnarray}
\tan\theta_{\rm u} & = & \frac{O^{\rm u}_{21}}{O^{\rm u}_{22}} \cdot
\frac{X_{\rm u}}{Y_{\rm u}} \; , \nonumber \\
\tan\theta_{\rm d} & = & \frac{O^{\rm d}_{21}}{O^{\rm d}_{22}} \cdot
\frac{X_{\rm d}}{Y_{\rm d}} \; ,
\end{eqnarray}
and 
\begin{eqnarray}
\sin\theta & = & \left [ (O^{\rm u}_{21} )^2 X^2_{\rm u} 
~ + ~ (O^{\rm u}_{22} )^2 Y^2_{\rm u} \right ]^{1/2}
O^{\rm d}_{33} \; , \nonumber \\
& = & \left [ (O^{\rm d}_{21} )^2 X^2_{\rm d} 
~ + ~ (O^{\rm d}_{22} )^2 Y^2_{\rm d} \right ]^{1/2}
O^{\rm u}_{33} \; , 
\end{eqnarray}
where $O_{21}$, $O_{22}$ and $O_{33}$ for up and down sectors have been
given in (4.5) and (4.6). Also an indirect relation 
between $\varphi$ and $\phi_{1,2}$ can be obtained as follows:
\begin{equation}
\cos \varphi \; =\; \frac{s^2_{\rm u} c^2_{\rm d} c^2 + c^2_{\rm u}
s^2_{\rm d} - |V_{us}|^2}{2 s_{\rm u} c_{\rm u} s_{\rm d} c_{\rm d} c} 
\; ,
\end{equation}
where
\begin{equation}
|V_{us}| \; =\; O^{\rm u}_{11} O^{\rm d}_{22} \left | \frac{|D_{\rm
d}|}{\lambda^{\rm d}_2 - E_{\rm d}} + \frac{\lambda^{\rm u}_1 - 
E_{\rm u}}{|D_{\rm u}|}
e^{{\rm i} \phi_1} \left ( 1 + 
\frac{|B_{\rm u}|}{\lambda^{\rm u}_1 - A_{\rm u}} \cdot \frac{|B_{\rm
d}|}{\lambda^{\rm d}_2 - A_{\rm d}} e^{{\rm i} \phi_2} \right ) \right | \; .
\end{equation}
If the hierarchies of the matrix elements and mass eigenvalues of
$M_{\rm u, d}$ are taken into account, one can see that the
effect of $\phi_2$ on $|V_{us}|$ is strongly suppressed and 
thus negligible.
Fitting $|V_{us}|$ with current data should essentially determine the
magnitude of $\phi_1$. Note also that the terms associated with $\phi_1$ and
$\phi_2$ may primarily be cancelled in the ratios $X_{\rm u}/Y_{\rm u}$
and $X_{\rm d}/Y_{\rm d}$ due to the hierarchical structures of $M_{\rm u}$
and $M_{\rm d}$, hence the dependence of $\theta_{\rm u}$ and $\theta_{\rm d}$
on $\phi_{1,2}$ could be negligible in the leading order approximation. 
Although the mixing angle $\theta$ may be sensitive to $\phi_1$ and $\phi_2$
(or one of them), its smallness indicated by current data makes the 
factor $\cos\theta$
in the denominator of $\cos\varphi$ completely negligible. As a result, 
(4.11) and (4.12)
imply that the $CP$-violating phase $\varphi$ depends dominantly on $\phi_1$
through $|V_{us}|$, unless the magnitude of $\phi_1$ is very small.
Without fine-tuning, we find that a delicate numerical
analysis does support the argument made here, i.e., 
$\phi_2$ plays a negligible role for $CP$ violation in $V$, 
because of the hierarchy of quark masses. 
The observed $CP$ violation is 
linked primarily to the phases in the (1,2) and
(2,1) elements of the quark mass matrices.

\subsection{Symmetry limits of quark masses}

Given Hermitian mass matrices of the form (4.2), 
one may consider two useful symmetry limits of quark masses 
and analyze their corresponding consequences on the
flavor mixing angles.

\underline{\it The chiral limit of quark masses} ~
In the limit $m_u \rightarrow 0$, $m_d \rightarrow 0$ (``chiral
limit''), where both the up- and down-type quark mass 
matrices have zeros in 
the positions $(1,1)$, $(1,2)$, $(2,1)$, $(1,3)$ and $(3,1)$,
the flavor mixing angles 
$\theta_{\rm u}$ and $\theta_{\rm d}$ vanish. Only the $\theta$ rotation
affecting the heavy quark sector remains, i.e., 
the flavor mixing matrix effectively takes the form
\begin{equation}
\hat{V} \; =\; \left ( \matrix{
\cos\hat{\theta} & \sin\hat{\theta} \cr
- \sin\hat{\theta} & \cos\hat{\theta} \cr } \right ) \; ,
\end{equation}
where $\hat{\theta}$ denotes the value of $\theta$ which one
obtains in 
the limit $\theta_{\rm u} \rightarrow 0$, $\theta_{\rm d} \rightarrow
0$. We see that $\hat{V}$ is a real orthogonal matrix, arising naturally 
from $V$ in the chiral limit.

The flavor mixing angle $\hat{\theta}$ can be derived from Hermitian
quark mass matrices of the following general form (in the limit $m_u
\rightarrow 0$, $m_d \rightarrow 0$):
\begin{equation}
\hat{M}_{\rm q} \; =\; \left ( \matrix{
\hat{C}_{\rm q} & \hat{B}_{\rm q} \cr
\hat{B}^*_{\rm q} & \hat{A}_{\rm q} \cr } \right ) \; ,
\end{equation}
where $|\hat{A}_{\rm q}| \gg |\hat{B}_{\rm q}| , |\hat{C}_{\rm q}|$; 
and q = u (up) or d 
(down). Note that the phase difference between $\hat{B}_{\rm u}$ and
$\hat{B}_{\rm d}$, denoted as $\kappa \equiv \arg (\hat{B}_{\rm u})
- \arg (\hat{B}_{\rm d})$, has no effect on $CP$ symmetry 
in the chiral limit, but it may affect the magnitude of
$\hat{\theta}$. It is known that current data on the top-quark mass
and the $B$-meson lifetime disfavor the special case $\hat{C}_{\rm u}
= \hat{C}_{\rm d} =0$ for $\hat{M}_{\rm u}$ and 
$\hat{M}_{\rm d}$,
hence we take $\hat{C}_{\rm q} \neq 0$ and define a ratio $\hat{r}_{\rm q}
\equiv |\hat{B}_{\rm q}|/\hat{C}_{\rm q}$ for convenience. Then we 
can obtain the flavor mixing angle $\hat{\theta}$, in terms of the
quark mass ratios $m_c/m_t$, $m_s/m_b$ and the parameters
$\hat{r}_{\rm u}$, $\hat{r}_{\rm d}$, by diagonalizing the mass
matrices in (4.14). In the next-to-leading order approximation,
$\sin\hat{\theta}$ reads
\begin{equation}
\sin\hat{\theta} \; = \; \left | \hat{r}_{\rm d} \frac{m_s}{m_b} \left (1 -
\hat{\delta}_{\rm d} \right ) ~ - ~ \hat{r}_{\rm u} \frac{m_c}{m_t} \left (1 -
\hat{\delta}_{\rm u} \right ) e^{{\rm i} \kappa} \right | \; ,
\end{equation}
where two correction terms are given by
\begin{eqnarray}
\hat{\delta}_{\rm u} & = & \left (1 + \hat{r}^2_{\rm u} 
\right ) \frac{m_c}{m_t} \; , \nonumber \\
\hat{\delta}_{\rm d} & = & \left (1 + \hat{r}^2_{\rm d} \right ) 
\frac{m_s}{m_b}
\; .
\end{eqnarray}
In view of the fact $m_s/m_b \sim {\cal O}(10) ~  m_c/m_t$ from current 
data \cite{PDG}, we find that the
flavor mixing angle $\hat{\theta}$ is primarily linked to $m_s/m_b$
provided $|\hat{r}_{\rm u}| \approx |\hat{r}_{\rm d}|$. Note that in specific 
models, e.g., those describing the mixing between the second and third
families as an effect related to the breaking of an underlying ``democratic
symmetry'' \cite{Democracy}, 
the ratios $\hat{r}_{\rm u}$ and $\hat{r}_{\rm d}$ 
are purely algebraic numbers (such as
$|\hat{r}_{\rm u}| = |\hat{r}_{\rm d}| = 1/\sqrt{2}$ 
or $\sqrt{2}$).

For illustration, we take $\hat{r}_{\rm u} = \hat{r}_{\rm d} \equiv
\hat{r}$ to fit the experimental result $\sin\hat{\theta} = 0.040 \pm 0.002$
with the typical inputs $m_b/m_s = 26 - 36$ and $m_t/m_c \sim 250$. 
It is found that the favored value of $|\hat{r}|$ varies in the range
1.0 -- 2.5, dependent weakly on the phase parameter $\kappa$. 

Note that both $m_s/m_b$ and $m_c/m_t$ evolve with the
energy scale (see section 4.7 for a detailed discussion),
therefore $\hat{\theta}$ itself is also scale-dependent. 
 
\underline{\it The heavy quark limit} ~
The limit $m_t \rightarrow \infty$, $m_b \rightarrow \infty$ is
subsequently referred to as the ``heavy quark limit''. In this limit,
in which the $(3,3)$ elements of the up- and down-type  mass matrices formally
approach infinity but
all other matrix elements are fixed, the angle $\theta$ vanishes.
The flavor mixing matrix,
which is nontrivial only in the light quark sector, takes the form:
\begin{eqnarray}
\tilde{V} & = & \left ( \matrix{
\tilde{c}_{\rm u}       & \tilde{s}_{\rm u}     \cr
-\tilde{s}_{\rm u}      & \tilde{c}_{\rm u}     \cr } \right )  
\left ( \matrix{
e^{-{\rm i}\tilde{\varphi}}     & 0     \cr
0       & 1 \cr } \right )  \left ( \matrix{
\tilde{c}_{\rm d}       & -\tilde{s}_{\rm d}    \cr
\tilde{s}_{\rm d}       & \tilde{c}_{\rm d}     \cr } \right )  
\nonumber \\ \nonumber \\
& = & \left ( \matrix{
\tilde{s}_{\rm u} \tilde{s}_{\rm d} + \tilde{c}_{\rm u} \tilde{c}_{\rm d} 
e^{-{\rm i}\tilde{\varphi}} &
\tilde{s}_{\rm u} \tilde{c}_{\rm d} - \tilde{c}_{\rm u} \tilde{s}_{\rm d} 
e^{-{\rm i}\tilde{\varphi}} \cr
\tilde{c}_{\rm u} \tilde{s}_{\rm d} - \tilde{s}_{\rm u} \tilde{c}_{\rm d} 
e^{-{\rm i}\tilde{\varphi}} &
\tilde{c}_{\rm u} \tilde{c}_{\rm d} + \tilde{s}_{\rm u} \tilde{s}_{\rm
d} e^{-{\rm i}\tilde{\varphi}} \cr } \right ) \; .
\end{eqnarray}
where $\tilde{s}_{\rm u} = {\rm sin} \tilde{\theta }_{\rm u},
\tilde{c}_{\rm u} = {\rm cos} \tilde{\theta}_{\rm u}$, and so on. The angles
$\tilde{\theta}_{\rm u}$ and $\tilde{\theta}_{\rm d}$ are the 
values for $\theta_{\rm u}$ and $\theta_{\rm d}$
obtained in the heavy quark limit. Since the $(t, b)$ system is decoupled from
the $(c, s)$ and $(u, d)$ systems, the flavor mixing can be described as in
the case of two families. Therefore the mixing matrix $\tilde{V}$ is
effectively given in terms of only a single rotation angle, 
the Cabbibo angle $\theta_{\rm C}$:
\begin{equation}
\sin \theta_{\rm C} = \mid \tilde{s}_{\rm u} \tilde{c}_{\rm d}
~ - ~ \tilde{c}_{\rm u} \tilde{s}_{\rm d} ~ e^{-{\rm i}
\tilde{\varphi}} \mid \; .
\end{equation}
Of course $\tilde{V}(\theta_{\rm C})$ 
is essentially a real matrix, because its complex
phases can always be rotated away by redefining the quark fields.

We stress that the heavy quark limit, which carries
the flavor mixing matrix $V$ to its simplified form $\tilde{V}$, is not far
from the reality, since $1 - c \approx 0.1 \%$ holds \cite{FX97a}.
Therefore $\theta_{\rm u}$, 
$\theta_{\rm d}$ and $\varphi $ are expected to
approach $\tilde{\theta}_{\rm u}$, $\tilde{\theta}_{\rm d}$ and 
$\tilde{\varphi}$ rapidly,
as $\theta \rightarrow 0$, corresponding to
$m_t \rightarrow \infty $ and $m_b \rightarrow \infty$. However, the concrete
limiting behavior depends on the specific algebraic 
structure of the up- and down-type mass matrices.
If two Hermitian mass matrices have the parallel hierarchy with texture zeros
in the (1,1) (2,2), (1,3) and (3,1) elements \cite{Fr78}, 
for example, the magnitude of $\theta $ is suppressed by the terms 
proportional to $1/\sqrt{m_t}$ and $1/\sqrt{m_b}$; and 
if the (2,2) elements are kept nonvanishing and comparable in magnitude 
with the (2,3) and (3,2) elements \cite{FX99}, then $\theta $ is dependent 
on $1/m_t$ and $1/m_b$.

The angles $\tilde{\theta}_{\rm u}$ and 
$\tilde{\theta}_{\rm d}$ as well as the phase
$\tilde{\varphi }$ are well-defined quantities in the heavy quark limit. The
physical meaning of these quantities can be seen more clearly, if we take
the up and down mass matrices to be of the 
following specific and realistic structure in the symmetry limit 
(i.e., $m_t\rightarrow \infty$ and $m_b\rightarrow \infty$) \cite{Fr77}:
\begin{equation}
\tilde{M}_{\rm q} \; = \; \left( \matrix{
{\bf 0} & \tilde{B}_{\rm q} \cr
\tilde{B}^*_{\rm q} & \tilde{A}_{\rm q} \cr } \right) \; .
\end{equation}
The diagonalization of $\tilde{M}_{\rm u}$ and
$\tilde{M}_{\rm d}$ leads to the Cabibbo-type mixing
\begin{equation}
\sin \theta_{\rm C} \; = \; \mid R_{\rm u} ~ - ~ R_{\rm d} ~ e^{-{\rm i}
\psi} \mid \; ,
\end{equation}
where
\begin{eqnarray}
R_{\rm u} & = & \sqrt{\frac{m_u}{m_u + m_c}} \, \sqrt{\frac{m_s}{m_d + m_s}}
\;\; , \nonumber \\
R_{\rm d} & = & \sqrt{\frac{m_c}{m_u + m_c}} \, \sqrt{\frac{m_d}{m_d +
m_s}} \; \; ,
\end{eqnarray}
and $\psi \equiv \arg (\tilde{B}_{\rm u}) - \arg (\tilde{B}_{\rm d})$ 
denotes the relative phase between the off-diagonal elements
$\tilde{B}_{\rm u}$ and $\tilde{B}_{\rm d}$ 
(in the limit $m_u \rightarrow 0$ this phase can be absorbed through a
redifinition of the quark fields). We find that the same structure for the
Cabibbo-type mixing matrix has been 
obtained as in the decoupling limit discussed
above. If we set
\begin{eqnarray}
\tan \tilde{\theta}_{\rm u} & = & \sqrt{\frac{m_u}{m_c}} \; , \nonumber \\
\tan \tilde{\theta}_{\rm d} & = & \sqrt{\frac{m_d}{m_s}} \; ,
\end{eqnarray}
and $\tilde{\varphi} = \psi$ for (4.18), then the result in (4.20)
and (4.21) can exactly be reproduced.
\begin{figure}[t]
\begin{picture}(400,160)(-45,210)
\put(80,300){\line(1,0){150}}
\put(80,300.5){\line(1,0){150}}
\put(150,285.5){\makebox(0,0){$\sin\theta_{\rm C}$}}
\put(80,300){\line(1,3){21.5}}
\put(80,300.5){\line(1,3){21.5}}
\put(80,299.5){\line(1,3){21.5}}
\put(71,333){\makebox(0,0){$R_{\rm u}$}}
\put(230,300){\line(-2,1){128}}
\put(230,300.5){\line(-2,1){128}}
\put(178,343.5){\makebox(0,0){$R_{\rm d}$}}
\put(107.5,348.5){\makebox(0,0){$\tilde{\varphi}$}}
\end{picture}
\vspace{-2.6cm}
\caption{The light-quark triangle (LT) in the complex plane.}
\end{figure}
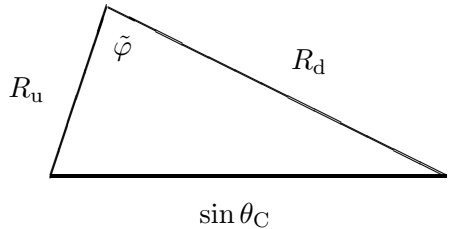

Indeed the relation in (4.18) or (4.20), analogous to that in (3.31),  
defines a triangle in the complex plane
(see Fig. 4.1 and Fig. 3.3 for illustration), 
which has been named as the ``light-quark 
triangle''(LT) in section 3.6. Taking into account the central values of the
Cabibbo angle ($\sin \theta_{\rm C} = |V_{us}| = 0.2205$ \cite{PDG}) 
and the light quark mass ratios ($m_s / m_d = 18.8$ and
$m_c /m_u = 265$ \cite{Leutwyler82,Leutwyler96}, we
can calculate the phase parameter from (4.20) and obtain 
$\tilde{\varphi} = \psi \approx 79^{\circ}$. If we allow the mass ratios 
and $\theta_{\rm C}$ to vary in
their ranges given above, then $\tilde{\varphi}$ may vary in the range
$38^{\circ}$ to $115^{\circ}$. We find that $\tilde{\varphi}$ has a good
chance to be around $90^{\circ}$.
The case $\tilde{\varphi} \approx 90^{\circ}$ (i.e., the LT is
rectangular) is of special interest, as we shall see later, since it
implies that the area of the unitarity triangle of flavor mixing 
takes its maximum value for the fixed quark mass ratios -- in this
sense, the $CP$ symmetry of weak interactions would be maximally violated.

It is worth remarking that the quark mass ratios
$m_d/m_s$ and $m_u/m_c$ are essentially
independent of the renormalization-group effect from the weak scale
to a superhigh scale (or vice versa), so is the Cabibbo angle
$\theta_{\rm C}$. As a result the sides and angles of the LT
are to a very good degree of accuracy
scale-independent. This interesting feature of the light
quark sector implies
that the prediction for $\tilde{\theta}_{\rm u}$
and $\tilde{\theta}_{\rm d}$ from quark mass matrices at any
high energy scale (e.g., $\mu \sim 10^{16}$ GeV) can directly
be confronted with the low-scale experimental data.

The two symmetry limits discussed above are both not far from
the reality, in which the strong hierarchy of quark masses
($m_u\ll m_c \ll m_t$ and $m_d\ll m_s\ll m_b$) has been
observed. They will serve as a guide in the subsequent 
discussions about specific quark mass matrices and their 
consequences on flavor mixing. In particular one expects that
a realistic pattern of $3\times 3$ quark mass matrices should naturally
reproduce the $2\times 2$ textures (4.14) and (4.19), respectively, in the 
chiral limit and in the heavy quark limit. This observation implies that
a specific $3\times 3$ texture of the form (4.2) with $E_{\rm q} =0$  
might be the best candidate for realistic quark mass matrices, which
is able to accommodate the experimental data on quark 
masses, flavor mixing and $CP$ violation. A detailed analysis of this
texture will be given later on.

\subsection{A Hermitian scheme with four texture zeros}

In order to get predictions for the flavor mixing angles and $CP$ 
violation, we proceed to specify the general Hermitian mass matrices
in (4.2) by taking $E_{\rm u} = E_{\rm d} = 0$:
\begin{eqnarray}
M_{\rm u} & = & \left ( \matrix{
{\bf 0} & D_{\rm u} & {\bf 0} \cr
D^*_{\rm u} & C_{\rm u} & B_{\rm u} \cr
{\bf 0} & B^*_{\rm u} & A_{\rm u} \cr } \right ) \; , 
\nonumber \\
M_{\rm d} & = & \left ( \matrix{
{\bf 0} & D_{\rm d}     & {\bf 0} \cr
D^*_{\rm d}     & C_{\rm d}     & B_{\rm d} \cr
{\bf 0} & B^*_{\rm d}   & A_{\rm d} \cr } \right ) .
\end{eqnarray}
Then $M_{\rm u}$ and $M_{\rm d}$ totally have four texture zeros
(here a pair of off-diagonal texture zeros are counted, due to
the Hermiticity, as one zero).
As remarked above, the texture zeros
in (1,3) and (3,1) positions of a general Hermitian mass
matrix can always be arranged. 
Thus the physical constraint is as follows: In the
flavor basis in which (1,3) and (3,1) elements of $M_{\rm u,d}$
vanish, the (1,1) element of $M_{\rm u,d}$ vanishes as well.
This may be true only at a particular energy scale. The vanishing of
the (1,1) element can be viewed as a result of 
an underlying flavor symmetry, which
may either be discrete or continuous. In the literature a number of such
possibilities have been discussed 
(see, e.g., Refs. \cite{FX95}--\cite{4zero}).
Here we shall not discuss further details in this respect, 
but concentrate on the phenomenological
consequences of $M_{\rm u}$ and $M_{\rm d}$ on
flavor mixing and $CP$ violation. 

\underline{\it Flavor mixing angles} ~
Now we start from (4.23) to calculate the flavor mixing
matrix. Here again we take $C_{\rm q} \neq 0$ and define 
$|B_{\rm q}|/C_{\rm q} \equiv r^{~}_{\rm q}$ for each
quark sector. The magnitude of $r^{~}_{\rm q}$ 
is expected to be of ${\cal O}(1)$.
The parameters $A_{\rm q}$, $|B_{\rm q}|$, $C_{\rm q}$ 
and $|D_{\rm q}|$ of $M_{\rm q}$ can be expressed in terms of 
the quark mass eigenvalues and $r^{~}_{\rm q}$. 
Applying such results to the general formulas listed in (4.9)
and (4.10), we get three mixing angles of $V$ as follows:
\begin{eqnarray}
\tan\theta_{\rm u} & = & \sqrt{\frac{m_u}{m_c}} ~ \left ( 1 + \Delta_{\rm u}
\right ) \; , \nonumber \\
\tan\theta_{\rm d} & = & \sqrt{\frac{m_d}{m_s}} ~ \left ( 1 + \Delta_{\rm d}
\right ) \; , 
\end{eqnarray}
and
\begin{equation}
\sin\theta \; = \; \left | r_{\rm d} \frac{m_s}{m_b} \left (1 -
\delta_{\rm d} \right ) ~ - ~ r_{\rm u} \frac{m_c}{m_t} \left ( 1 -
\delta_{\rm u} \right ) e^{{\rm i}\phi_2} \right | \; ,
\end{equation}
where the next-to-leading order corrections read
\begin{eqnarray}
\Delta_{\rm u} & = & \sqrt{\frac{m_c m_d}{m_u m_s}} ~ \frac{m_s}{m_b} 
~ \left | {\rm Re} \left [ e^{{\rm i}\phi_1} ~ - ~ \frac{r_{\rm
u}}{r_{\rm d}} \cdot \frac{m_c m_b}{m_t m_s} 
e^{{\rm i}(\phi_1 + \phi_2)} \right
]^{-1} \right | \; , \nonumber \\
\Delta_{\rm d} & = & \sqrt{\frac{m_u m_s}{m_c m_d}} ~ \frac{m_c}{m_t} 
~ \left | {\rm Re} \left [ e^{{\rm i}\phi_1} ~ - ~ \frac{r_{\rm
d}}{r_{\rm u}} \cdot \frac{m_t m_s}{m_c m_b} 
e^{{\rm i}(\phi_1 + \phi_2)} \right
]^{-1} \right | \; ;
\end{eqnarray}
as well as 
\begin{eqnarray}
\delta_{\rm u} & = & \frac{m_u}{m_c} ~ + 
\left ( 1 + r^2_{\rm u} \right ) \frac{m_c}{m_t}
\; , \nonumber \\
\delta_{\rm d} & = & \frac{m_d}{m_s} ~ + \left ( 1 + r^2_{\rm d} 
\right ) \frac{m_s}{m_b}
\; .
\end{eqnarray}
Clearly the result for $\hat{\delta}_{\rm u,d}$ in (4.16) can be
reproduced from $\delta_{\rm u,d}$ in (4.27) if one takes the chiral 
limit $m_u \rightarrow 0$, $m_d \rightarrow 0$. 
From (4.25) we also observe that the phase $\phi_2$ is 
only associated with the small quantity $m_c/m_t$ in $\sin\theta$. 
To get the relationship between $\varphi$ and $\phi_1$ or $\phi_2$,
we first calculate $|V_{us}|$ from quark mass matrices by use of
(4.12). We obtain 
\begin{equation}
|V_{us}| \; = \; \left (1 -\frac{1}{2} \frac{m_u}{m_c} - \frac{1}{2}
\frac{m_d}{m_s} \right ) \left | \sqrt{\frac{m_d}{m_s}} ~ - ~
\sqrt{\frac{m_u}{m_c}} ~ e^{{\rm i}\phi_1} \right | \; , 
\end{equation}
in the next-to-leading order approximation. Note that this result can
also be achieved from (4.20) and (4.21), which were obtained in the
heavy quark limit. Confronting (4.28) with current data on
$|V_{us}|$ leads to the result $\phi_1 \sim
90^{\circ}$, as we have pointed out before. Therefore $\cos\phi_1$ is
expected to be a small quantity.
Then we use (4.11) together with (4.24) and (4.28) 
to calculate $\cos\varphi$. In the same order approximation, we arrive at
\begin{equation}
\cos\varphi \; =\; \sqrt{\frac{m_u m_s}{m_c m_d}} ~ \Delta_{\rm u} ~ + ~
\sqrt{\frac{m_c m_d}{m_u m_s}} ~ \Delta_{\rm d} ~ + ~ \left (1 -
\Delta_{\rm u} - \Delta_{\rm d} \right ) \cos\phi_1 \; .
\end{equation}
The contribution of $\phi_2$ to $\varphi$ is substantially suppressed at
this level of accuracy. 

For simplicity, we proceed by taking $r_{\rm u} = r_{\rm d} \equiv r$,
which holds in some aforementioned models with natural flavor
symmetries \cite{Democracy}. Then
$\sin\theta$ becomes proportional to a universal parameter $|r|$.
In view of the fact $m_s/m_b \sim {\cal O}(10) ~ m_c/m_t$, 
we find that the result in (4.26) can be simplified as
\begin{eqnarray}
\Delta_{\rm u} & = & \sqrt{\frac{m_c m_d}{m_u m_s}} ~ \frac{m_s}{m_b} 
~ \cos\phi_1 \; , \nonumber \\
\Delta_{\rm d} & = & 0 \; . 
\end{eqnarray}
Also the relation between $\varphi$ and $\phi_1$ in (4.29) is simplified to
\begin{equation}
\cos\varphi \; =\; \left (1 + \frac{m_s}{m_b} \right ) \cos\phi_1 \; .
\end{equation}
As $m_s/m_b \sim 4\%$, it becomes apparent that 
$\varphi \approx \phi_1$ is a good approximation.
With these results one may further calculate the
rephasing-invariant $CP$-violating parameter $\cal J$
defined in (3.11).
We confirm that the magnitude of $\cal J$
is dominated by the $\sin\phi_1$ term and receives
one-order smaller corrections from the $\sin (\phi_1 \pm \phi_2)$
terms. As a result, 
\begin{equation}
{\cal J} \; \approx \; |r|^2 \sqrt{\frac{m_u}{m_c}}
\sqrt{\frac{m_d}{m_s}} \left (\frac{m_s}{m_b}\right )^2
\sin\phi_1 \; 
\end{equation}
holds to a good degree of accuracy. Clearly 
${\cal J} \sim {\cal O}(10^{-5}) \times \sin\phi_1$ with $\sin\phi_1 \sim 1$
is favored by the current experimental data.

The result of $\cal J$ in (4.32) might give one an impression that
$CP$ violation is absent if either $m_u$ or $m_d$ vanishes. This
is not exactly true, however. If we set $m_u=0$,
$\cal J$ is not zero, but it becomes dependent on $\sin \phi_2$
with a factor which is about two orders of magnitude
smaller (i.e., of order $10^{-7}$):
\begin{equation}
{\cal J} \; \approx \; |r|^2 ~ \frac{m_c}{m_t} \cdot
\frac{m_d}{m_s} \left (\frac{m_s}{m_b} \right )^2
\sin\phi_2 \; .
\end{equation}
Certainly this possibility has been ruled out by the present
experimental data
(i.e., both $m_u$ and $m_d$ must be different from zero).

Note also that the texture of $M_{\rm u}$ and $M_{\rm d}$ in (4.23) predicts 
\begin{eqnarray}
\tan\theta_{\rm u} \; =\; 
\left | \frac{V_{ub}}{V_{cb}} \right | & = &
\sqrt{\frac{m_u}{m_c}} ~ \left (1 + \Delta_{\rm u} \right ) \; ,
\nonumber \\
\tan\theta_{\rm d} \; =\;
\left | \frac{V_{td}}{V_{ts}} \right | & = &
\sqrt{\frac{m_d}{m_s}} ~ \left (1 + \Delta_{\rm d} \right ) \; .
\end{eqnarray}
In $B$-meson physics, $|V_{ub}/V_{cb}|$ can be determined
from the ratio of the decay rate of $B\rightarrow 
(\pi, \rho) l \nu^{~}_l$ to that of $B\rightarrow D^* l\nu^{~}_l$;
and $|V_{td}/V_{ts}|$ can be extracted from the ratio of
the rate of $B^0_d$-$\bar{B}^0_d$ mixing to that of 
$B^0_s$-$\bar{B}^0_s$ mixing \cite{FX97a}. 
\begin{figure}[t]
\begin{picture}(400,160)(170,210)
\put(300,300){\line(1,0){150}} 
\put(300,300.5){\line(1,0){150}}
\put(370,285.5){\makebox(0,0){$|V_{cd}|$}}
\put(300,300){\line(1,3){21.5}}
\put(300,300.5){\line(1,3){21.5}}
\put(300,299.5){\line(1,3){21.5}}
\put(292,333){\makebox(0,0){$S_{\rm u}$}}
\put(450,300){\line(-2,1){128}}
\put(450,300.5){\line(-2,1){128}}
\put(395,343.5){\makebox(0,0){$S_{\rm d}$}}
\put(315,310){\makebox(0,0){$\gamma$}}
\put(408,309){\makebox(0,0){$\beta$}}
\put(328,350){\makebox(0,0){$\alpha$}}
\end{picture}
\vspace{-2.6cm}
\caption{The rescaled unitarity triangle (UT) in the complex plane.}
\end{figure}
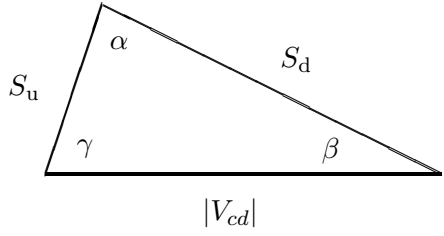

\underline{\it The unitarity triangle} ~
We are now in a position to calculate the unitarity triangle (UT) of
quark flavor mixing defined by the orthogonality relation
\begin{equation}
V^*_{ub}V_{ud} + V^*_{cb}V_{cd} + V^*_{tb}V_{td} \; =\; 0 \; ,
\end{equation}
i.e., the triangle $\triangle_s$ in Fig. 3.1. 
The three inner angles of this triangle are
denoted as $\alpha$, $\beta$ and $\gamma$ in (3.20). Three
sides of $\Delta_s$ can be rescaled by $V_{cb}^*$ (see
Fig. 3.3(b) and Fig. 4.2 for illustration). The resultant triangle reads
\begin{equation}
|V_{cd}| \; =\; \left | S_{\rm d} ~ -~ S_{\rm u} ~ e^{-{\rm i}\alpha}
\right | \; ,
\end{equation}
where $S_{\rm u} = |V_{ub}V_{ud}/V_{cb}|$ and $S_{\rm d} =
|V_{tb}V_{td}/V_{cb}|$. After some calculations we obtain
$S_{\rm u}$, $S_{\rm d}$ and $\alpha$ from $M_{\rm u}$ and $M_{\rm d}$
in the next-to-leading order approximation \cite{FX99}:
\begin{eqnarray}
S_{\rm u} & = & \sqrt{\frac{m_u}{m_c}} \left ( 1 - \frac{1}{2}
\frac{m_u}{m_c} - \frac{1}{2} \frac{m_d}{m_s} + \sqrt{\frac{m_c
m_d}{m_u m_s}} ~ \frac{m_s}{m_b} ~ \cos\phi_1 + \sqrt{\frac{m_u
m_d}{m_c m_s}} ~ \cos\phi_1 \right ) \; , \nonumber \\ S_{\rm d} & = &
\sqrt{\frac{m_d}{m_s}} \left ( 1 + \frac{1}{2} \frac{m_u}{m_c} -
\frac{1}{2} \frac{m_d}{m_s} \right ) \; ;
\end{eqnarray}
and
\begin{equation}
\sin\alpha \; = \; \left ( 1 - \sqrt{\frac{m_u m_d}{m_c m_s}} ~
\cos\phi_1 \right ) \sin\phi_1 \; .
\end{equation}
A comparison of the rescaled UT in Fig. 4.2 with the LT in Fig. 4.1, which
is obtained in the heavy quark limit, is interesting. We find
\begin{eqnarray}
\frac{S_{\rm u} - R_{\rm u}}{R_{\rm u}} & = & \left ( 1 + \frac{m_c
m_s}{m_u m_b} \right ) \sqrt{\frac{m_u m_d}{m_c m_s}} ~
\cos\tilde{\varphi} \; , \nonumber \\ \frac{S_{\rm d} - R_{\rm
d}}{R_{\rm d}} & = & \frac{m_u}{m_c} \; , \nonumber \\
\frac{\sin\alpha - \sin\tilde{\varphi}}{\sin\tilde{\varphi}} & = & -
\sqrt{\frac{m_u m_d}{m_c m_s}} ~ \cos\tilde{\varphi} \; ,
\end{eqnarray}
which are of order $15\% \cos\tilde{\varphi}$, $0.4\%$ and $1.4\%
\cos\tilde{\varphi}$, respectively. Obviously $R_{\rm d} \approx
S_{\rm d}$ is an excellent approximation, and $\alpha \approx
\tilde{\varphi} \approx \varphi$ is a good approximation. As $\varphi$
(or $\tilde{\varphi}$) is expected to be close to $90^{\circ}$,
$R_{\rm u} \approx S_{\rm u}$ should also be accurate enough in the
next-to-leading order estimation.  Therefore the light-quark triangle
is essentially {\it congruent with} the rescaled unitarity triangle!
This result has two straightforward implications: first, $CP$
violation is an effect arising primarily from the light quark sector;
second, the $CP$-violating observables ($\alpha$, $\beta$, $\gamma$)
can be predicted in terms of the light quark masses and the phase
difference between up and down mass matrices \cite{FX95}.  If we use
the value of $|V_{cd}|$, which is expected to be equal to $|V_{us}|$ within
the $0.1\%$ error bar \cite{Xing96}, then all three angles of the
unitarity triangle can be calculated in terms of $m_u/m_c$, $m_d/m_s$
and $|V_{cd}|$ to a good degree of accuracy.

The three inner angles of the UT ($\alpha$, $\beta$ and $\gamma$) will be
determined at $B$-meson factories,
e.g., from the $CP$
asymmetries in $B_d\rightarrow \pi^+\pi^-$, $B_d\rightarrow J/\psi
K_S$ and $B^{\pm}_u\rightarrow (D^0, \bar{D}^0) + K^{(*)\pm}$
decays \cite{CDF99}.
The characteristic measurable quantities are $\sin
(2\alpha)$, $\sin (2\beta)$ and $\sin^2\gamma$, respectively. For
the purpose of illustration, we typically take 
$|V_{us}| = |V_{cd}| =0.22$,
$m_u/m_c =0.0056$, $m_d/m_s = 0.045$ and $m_s/m_b = 0.033$ to
calculate these three $CP$-violating parameters from 
the LT and from the rescaled UT separately. 
Both approaches lead to
$\alpha \approx 90^{\circ}$,
$\beta \approx 20^{\circ}$ and
$\gamma \approx 70^{\circ}$,
which are in good agreement with the results obtained from
the standard 
analysis of current data on $|V_{ub}/V_{cb}|$, $\epsilon^{~}_K$,
$B^0_d$-$\bar{B}^0_d$ mixing and $B^0_s$-$\bar{B}^0_s$ 
mixing \cite{Stocchi}. 
Note that among three $CP$-violating observables
only $\sin (2\beta)$ is remarkably sensitive to the value of
$m_u/m_c$, which involves quite large uncertainty (e.g., $\sin
(2\beta)$ may change from $0.4$ to $0.8$ if $m_u/m_c$ varies in the
range $0.002$ to $0.01$). For this reason 
we emphasize again that the numbers given above can only serve
as an illustration.
A more reliable determination of the 
quark mass values is crucial, in order to test the patterns
of quark mass matrices in a numerically decisive way \cite{Barbieri99}.

It is also worth mentioning that
the result $\tan\theta_{\rm d} = \sqrt{m_d/m_s}$ is particularly 
interesting for the mixing rates of 
$B^0_d$-$\bar{B}^0_d$ and $B^0_s$-$\bar{B}^0_s$
systems, measured by $x_{\rm d}$ and $x_{\rm s}$ respectively.
The ratio $x_{\rm s} /x_{\rm d}$ amounts to $|V_{ts}/V_{td}|^2 =
\tan^{-2} \theta_{\rm d}$ multiplied by a factor $\chi_{\rm su(3)} = 1.45
\pm 0.13$, which reflects the $\rm SU(3)_{\rm flavor}$ symmetry breaking
effects \cite{Martinelli}. 
As $x_{\rm d} = 0.723 \pm 0.032$ has been well 
determined \cite{PDG}, the prediction for the value of $x_{\rm s}$ is
\begin{equation}
x_{\rm s} \; =\; x_{\rm d} ~ \chi_{\rm su(3)} ~ \frac{m_s}{m_d} \; =\;
19.8 \pm 3.5 \; ,
\end{equation}
where $m_s/m_d = 18.9 \pm 0.8$, obtained from the chiral perturbation
theory \cite{Leutwyler96}, 
has been used. This result is certainly consistent
with the present experimental bound on $x_{\rm s}$, i.e.,
$x_{\rm s} > 14.0$ at the $95\%$ confidence level \cite{PDG}.
A measurement 
of $x_{\rm s} \sim 20$ may be realized at the forthcoming
HERA-$B$ and LHC-$B$ experiments.

\underline{\it Further discussions} ~
Finally let us make some further remarks on the quark mass matrices
(4.23), its phenomenological hints and its theoretical
prospects. 

(1) Naively one might not expect any prediction from the 
four-texture-zero mass matrices in (4.23), 
since they totally consist of ten free parameters
(two of them are the phase differences between $M_{\rm u}$ and $M_{\rm d}$).
This is not true, however, as we have seen. 
We find that two predictions,
$\tan\theta_{\rm u} \approx \sqrt{m_u/m_c}$ and $\tan\theta_{\rm d} \approx 
\sqrt{m_d/m_s}$ , can be obtained in the leading order approximation. 
In some cases the latter may even hold in the next-to-leading order 
approximation, as shown in (4.30). Note again that these two relations,
as a consequence of the hierarchy and the texture zeros of 
quark mass matrices, are essentially independent of the renormalization-group 
effects. This interesting scale-independent
feature has also manifested itself in the LT and the rescaled
UT, as well as their inner angles $(\alpha, \beta, \gamma)$. 

(2) It remains to be seen whether the interesting possibility
$\varphi \approx \phi_1 \approx 90^{\circ}$, indicated by current
data of quark masses and flavor mixing, could arise from an
underlying flavor symmetry or a dynamical
symmetry breaking scheme. Some speculations about this problem 
have been made 
(see, e.g., Refs. \cite{FX99,FX95} and Refs. \cite{Shin,Weyers}).
However, no final conclusion has been reached thus far.
It is remarkable, nevertheless, that we have at least observed a
useful relation between the area of the UT 
(${\cal A}_{\rm UT}$) and that of the LT
(${\cal A}_{\rm LT}$) to a good degree of accuracy \cite{FX99}:
\begin{equation}
{\cal A}_{\rm UT} \; \approx \; |V_{cb}|^2 {\cal A}_{\rm LT}
\; \approx \; \sin^2\theta ~ {\cal A}_{\rm LT} \; .
\end{equation}
Since ${\cal A}_{\rm UT} = {\cal J}/2$ measures the magnitude
of $CP$ violation in the standard model, we conclude that 
$CP$ violation is primarily linked to the 
light quark sector. This is a natural consequence of the
strong hierarchy between the heavy and light quark masses,
which is on the other hand responsible for the smallness of 
${\cal J}$ or ${\cal A}_{\rm UT}$. 

(3) Is it possible to derive the quark mass matrix (4.23) in an
underlying theoretical framework? To answer this question
we first specify the hierarchical structure of $M_{\rm q}$ in 
terms of the mixing angle $\theta_{\rm q}$ (for q = d or s). 
Adopting the radiant
unit for the mixing angles (i.e., $\theta_{\rm u}
\approx 0.085$, $\theta_{\rm d} \approx 0.204$ and
$\theta \approx 0.040$), we have
\begin{eqnarray}
\frac{m_u}{m_c} & \sim & \frac{m_c}{m_t} \;
\sim \; \theta^2_{\rm u} \; ,  \nonumber \\
\frac{m_d}{m_s} & \sim & \frac{m_s}{m_b} \;
\sim \; \theta^2_{\rm d} \; .
\end{eqnarray}
Then the mass matrices $M_{\rm u}$ and $M_{\rm d}$,
which have the mass  scales $m_t$  and $m_b$ 
respectively, take the following {\it parallel} hierarchies:
\begin{eqnarray}
M_{\rm u} & \sim & m_t \left ( \matrix{
{\bf 0} & \theta^3_{\rm u}      & {\bf 0} \cr
\theta^3_{\rm u} & \theta^2_{\rm u} & \theta^2_{\rm u} \cr
{\bf 0} & \theta^2_{\rm u}      & {\bf 1} \cr}
\right ) \; \; , \nonumber \\
\nonumber \\
M_{\rm d} & \sim & m_b \left ( \matrix{
{\bf 0} & \theta^3_{\rm d}      & {\bf 0} \cr
\theta^3_{\rm d} & \theta^2_{\rm d} & \theta^2_{\rm d} \cr
{\bf 0} & \theta^2_{\rm d}      & {\bf 1} \cr}
\right ) \; \; ,
\end{eqnarray}
where the relevant complex phases have been neglected.
Clearly all three flavor mixing angles can properly
be reproduced from (4.43), once one takes $\theta \approx
\theta^2_{\rm d} \gg \theta^2_{\rm u}$ into account. 
The $CP$-violating phase $\varphi$ in $V$ comes
essentially from the phase difference between the
$\theta^3_{\rm u}$ and $\theta^3_{\rm d}$ terms.

Of course $\theta_{\rm u}$ and $\theta_{\rm d}$, which
are more fundamental than the Cabibbo angle $\theta_{\rm C}$
in our point of view, denote perturbative corrections to the
rank-one limits of $M_{\rm u}$ and $M_{\rm d}$
respectively. They are responsible for the generation 
of light quark masses as well as the flavor mixing.
They might also be responsible for $CP$ violation in
a specific theoretical framework (e.g., a pure
real $\theta_{\rm u}$ and a pure imaginary $\theta_{\rm d}$
will lead to a phase difference of about $90^{\circ}$
between $M_{\rm u}$ and $M_{\rm d}$, which is just the
source of $CP$ violation favored by current data).
The small parameter $\theta_{\rm q}$ could get its
physical meaning in the Yukawa coupling of
an underlying superstring theory: 
$\theta_{\rm q} = \langle \Theta_{\rm q} \rangle /\Omega_{\rm q}$,
where $\langle \Theta_{\rm q} \rangle $ denotes the
vacuum expectation value of the singlet field $\Theta_{\rm q}$,
and $\Omega_{\rm q}$ represents  the unification (or string)
mass scale which governs higher  dimension operators
(see, e.g., Refs. \cite{Froggatt79,Ross94,Xing97}). The quark mass
matrices of the form (4.43) could then be obtained 
by introducing an extra (horizontal) U(1) gauge symmetry
or assigning the matter fields appropriately.

It is concluded that
the four texture zeros and parallel hierarchies of up and down
quark mass matrices do imply specific symmetries, perhaps 
at a superhigh scale, and have instructive consequences
on flavor mixing and $CP$-violating phenomena.
The new parametrization of the flavor mixing matrix
that we advocated in (3.29) is particularly useful in studying
the quark mass generation, flavor mixing 
and $CP$ violation. 

\subsection{Hermitian schemes with five texture zeros}

The quark mass matrices $M_{\rm u}$ and $M_{\rm d}$ given in (4.23)
have parallel structures with four texture zeros.
Giving up the parallelism between the
structures of $M_{\rm u}$ and $M_{\rm d}$, Ramond, Roberts and Ross
(RRR) have found that there exist five phenomenologically allowed
patterns of Hermitian quark mass matrices -- each of them has five texture 
zeros \cite{RRR}, as listed in Table 4.1. 
The RRR patterns I, II and IV can be formally regarded
as a special case of the four-texture-zero pattern (4.23), with
$B_{\rm u} =0$, $C_{\rm u} =0$ and $B_{\rm d} =0$, respectively. Note that
$M_{\rm u}$ of the RRR pattern III or V has nonvanishing (1,3) and (3,1)
elements, therefore these two patterns are essentially different 
from the mass matrices prescribed in (4.2) or (4.23). 
As a comparison, here 
we make some brief comments on consequences of the RRR patterns on the
flavor mixing angles $\theta$, $\theta_{\rm u}$ and $\theta_{\rm d}$.
\small
\begin{table}[t]
\caption{Five RRR patterns of Hermitian quark mass matrices.}
\vspace{0.2cm}
\begin{center}
\begin{tabular}{ccccc}\hline\hline 
Pattern &~~~~~~~~~~& $M_{\rm u}$        & ~~~~~~~~~~ & $M_{\rm d}$ \\ \hline 
I       && $\left ( \matrix{ {\bf 0} & D_{\rm u} & {\bf 0} \cr
D^*_{\rm u} & C_{\rm u} & {\bf 0} \cr
{\bf 0} & {\bf 0} & A_{\rm u} \cr } \right )$ 
&       & $\left ( \matrix{ {\bf 0} & D_{\rm d} & {\bf 0} \cr
D^*_{\rm d} & C_{\rm d} & B_{\rm d} \cr
{\bf 0} & B^*_{\rm d} & A_{\rm d} \cr } \right )$ \\ \\ 
II      && $\left ( \matrix{ {\bf 0} & D_{\rm u} & {\bf 0} \cr
D^*_{\rm u} & {\bf 0} & B_{\rm u} \cr
{\bf 0} & B^*_{\rm u} & A_{\rm u} \cr } \right )$
&       & $\left ( \matrix{ {\bf 0} & D_{\rm d} & {\bf 0} \cr
D^*_{\rm d} & C_{\rm d} & B_{\rm d} \cr 
{\bf 0} & B^*_{\rm d} & A_{\rm d} \cr } \right )$ \\ \\
III      && $\left ( \matrix{ {\bf 0} & {\bf 0} & F_{\rm u} \cr
{\bf 0} & C_{\rm u} & {\bf 0} \cr
F^*_{\rm u} & {\bf 0} & A_{\rm u} \cr } \right )$ 
&       & $\left ( \matrix{ {\bf 0} & D_{\rm d} & {\bf 0} \cr
D^*_{\rm d} & C_{\rm d} & B_{\rm d} \cr
{\bf 0} & B^*_{\rm d} & A_{\rm d} \cr } \right )$ \\ \\
IV      && $\left ( \matrix{ {\bf 0} & D_{\rm u} & {\bf 0} \cr
D^*_{\rm u} & C_{\rm u} & B_{\rm u} \cr
{\bf 0} & B^*_{\rm u} & A_{\rm u} \cr } \right )$
&       & $\left ( \matrix{ {\bf 0} & D_{\rm d} & {\bf 0} \cr
D^*_{\rm d} & C_{\rm d} & {\bf 0} \cr 
{\bf 0} & {\bf 0} & A_{\rm d} \cr } \right )$ \\ \\
V       && $\left ( \matrix{ {\bf 0} & {\bf 0} & F_{\rm u} \cr
{\bf 0} & C_{\rm u} & B_{\rm u} \cr
F^*_{\rm u} & B^*_{\rm u} & A_{\rm u} \cr } \right )$
&       & $\left ( \matrix{ {\bf 0} & D_{\rm d} & {\bf 0} \cr
D^*_{\rm d} & C_{\rm d} & {\bf 0} \cr
{\bf 0} & {\bf 0} & A_{\rm d} \cr } \right )$ \\  \hline \hline
\end{tabular}
\end{center}
\end{table}
\normalsize
\small
\begin{table}[t]
\caption{Approximate forms of real RRR-type quark mass matrices.}
\vspace{0.2cm}
\begin{center}
\begin{tabular}{ccccc}\hline\hline 
Pattern &~~~~~~~~~~& $M_{\rm u}$        & ~~~~~~~~~~ & $M_{\rm d}$ \\ \hline  
I       && $ m_t \left ( \matrix{ {\bf 0} & \theta^3_{\rm u} & {\bf 0} \cr
\theta^3_{\rm u} & \theta^2_{\rm u} & {\bf 0} \cr
{\bf 0} & {\bf 0} & {\bf 1} \cr } \right )$ 
&       & $ m_b \left ( \matrix{ {\bf 0} & \theta^3_{\rm d} & {\bf 0} \cr
\theta^2_{\rm d} & \theta^2_{\rm d} & \theta^2_{\rm d} \cr
{\bf 0} & \theta^2_{\rm d} & {\bf 1} \cr } \right )$ \\ \\ 
II      && $ m_t \left ( \matrix{ {\bf 0} & \theta^3_{\rm u} & {\bf 0} \cr
\theta^3_{\rm u} & {\bf 0} & \theta_{\rm u} \cr
{\bf 0} & \theta_{\rm u} & {\bf 1} \cr } \right )$
&       & $ m_b \left ( \matrix{ {\bf 0} & \theta^3_{\rm d} & {\bf 0} \cr
\theta^3_{\rm d} & \theta^2_{\rm d} & \theta^2_{\rm d} \cr 
{\bf 0} & \theta^2_{\rm d} & {\bf 1} \cr } \right )$ \\ \\
III      && $ m_t \left ( \matrix{ {\bf 0} & {\bf 0} & \theta^2_{\rm u} \cr
{\bf 0} & \theta^2_{\rm u} & {\bf 0} \cr
\theta^2_{\rm u} & {\bf 0} & {\bf 1} \cr } \right )$ 
&       & $ m_b \left ( \matrix{ {\bf 0} & \theta^3_{\rm d} & {\bf 0} \cr
\theta^3_{\rm d} & \theta^2_{\rm d} & \theta^2_{\rm d} \cr
{\bf 0} & \theta^2_{\rm d} & {\bf 1} \cr } \right )$ \\ \\
IV      && $ m_t \left ( \matrix{ {\bf 0} & \theta^3_{\rm u} & {\bf 0} \cr
\theta^3_{\rm u} & \theta^2_{\rm u} & \theta_{\rm u} \cr
{\bf 0} & \theta_{\rm u} & {\bf 1} \cr } \right )$
&       & $ m_b \left ( \matrix{ {\bf 0} & \theta^3_{\rm d} & {\bf 0} \cr
\theta^3_{\rm d} & \theta^2_{\rm d} & {\bf 0} \cr 
{\bf 0} & {\bf 0} & {\bf 1} \cr } \right )$ \\ \\
V       && $ m_t \left ( \matrix{ {\bf 0} & {\bf 0} & \theta^2_{\rm u} \cr
{\bf 0} & \theta^2_{\rm u} & \theta_{\rm u} \cr
\theta^2_{\rm u} & \theta_{\rm u} & {\bf 1} \cr } \right )$
&       & $ m_b \left ( \matrix{ {\bf 0} & \theta^3_{\rm d} & {\bf 0} \cr
\theta^3_{\rm d} & \theta^2_{\rm d} & {\bf 0} \cr
{\bf 0} & {\bf 0} & {\bf 1} \cr } \right )$ \\  \hline \hline
\end{tabular}
\end{center}
\end{table}
\normalsize

(a) For the RRR pattern I, the magnitude of $\sin\theta$ is governed by
the (2,3) and (3,2) elements of $M_{\rm d}$. Therfore $|B_{\rm d}| \sim
|C_{\rm d}|$ is expected, to result in $\sin\theta \sim m_s/m_b$. In
the leading order approximation, $\tan\theta_{\rm u} = \sqrt{m_u/m_c}$
and $\tan\theta_{\rm d} = \sqrt{m_d/m_s}$ hold. The next-to-leading 
order corrections to these two quantities are almost indistinguishable 
from those obtained in (4.26) and (4.30) for the four-texture-zero
ansatz. 

(b) The mass matrix $M_{\rm u}$ of the RRR pattern II takes the
well-known form suggested originally in Ref. \cite{Fr78}.
To reproduce the experimental
value of $\sin\theta$, the possibility $|B_{\rm d}| \gg |C_{\rm d}|$
has to be abandoned and the condition $|B_{\rm d}| \sim |C_{\rm d}|$
is required. However, significant cancellation between the term
proportional to $\sqrt{m_c/m_t}$ (from $M_{\rm u}$) 
and that proportional to $m_s/m_b$ (from $M_{\rm d}$) in 
$\sin\theta$ may take place, if the phase difference between
$B_{\rm u}$ and $B_{\rm d}$ is vanishing or very small. The leading 
order results $\tan\theta_{\rm u} = \sqrt{m_u/m_c}$ and
$\tan\theta_{\rm d} = \sqrt{m_d/m_s}$ can still be obtained here,
but their next-to-leading order corrections may deviate somehow
from those obtained in section 4.3.

(c) From the RRR pattern IV one can arrive at $\sin\theta \sim \sqrt{m_c/m_t}$
with the necessary condition $|C_{\rm u}| \ll |B_{\rm u}|$, since the 
(2,3) and (3,2) elements of $M_{\rm d}$ vanish. The results for 
$\tan\theta_{\rm u}$ and $\tan\theta_{\rm d}$ are similar to those
obtained from the RRR pattern I.

(d) The nonvanishing (1,3) and (3,1) elements of $M_{\rm u}$ in the
RRR pattern III make its prediction for the mixing angles $\theta_{\rm u}$
and $\theta_{\rm d}$ quite different from all patterns discussed above.
Analytically one can find $\tan\theta_{\rm u} \sim (m_b/m_s) \sqrt{m_u/m_t}$ ,
while $\tan\theta_{\rm d}$ is a complicated combination of the
terms $\sqrt{m_d/m_s}$ and $(m_b/m_s) \sqrt{m_u/m_t}$ with a relative
phase. In addition, $\sin\theta \sim m_s/m_b$ holds under the condition
$|B_{\rm d}| \sim |C_{\rm d}|$, similar to the RRR pattern I.

(e) For the RRR pattern V, the necessary condition $|B_{\rm u}| \gg
|C_{\rm u}|$ is required in order to reproduce 
$\sin\theta \sim \sqrt{m_c/m_t}$ .
Here again the nonvanishing (1,3) and (3,1) elements of $M_{\rm u}$
result in very complicated expressions for $\tan\theta_{\rm u}$ and
$\tan\theta_{\rm d}$ 
(even more complicated than those in the RRR pattern III \cite{Xing97}).

For reasons of naturalness and simplicity, we argue that
the RRR patterns III and V are unlikely to be good candidates for the
quark mass matrices in an underlying theory of fermion mass generation.
In analogy with (4.43), we illustrate different magnitudes 
of the elements of five
RRR-type mass matrices approximately in terms of two basic
flavor mixing parameters $\theta_{\rm u}$ and $\theta_{\rm d}$. 
The results are listed in Table 4.2, where the relevant complex phases of
$M_{\rm u}$ and $M_{\rm d}$ are neglected. It is clear that the
hierarchy of either (up or down) mass matrix can be governed by a single
small parameter, whose physical mixing is apparently related to 
the smallness of the flavor mixing angles.

A numerical illustration of the consequences of five RRR 
ans$\rm\ddot{a}$tze on flavor mixing and $CP$ violation
has been given in Ref. \cite{RRR}. It is worth remarking
that with the same input values of quark mass ratios only one
or two of them can be in good agreement with current experimental
data. In other words, the five textures cannot simultaneously
survive.

\subsection{Mass matrices from flavor democracy breaking}

Towards an understanding of the hierarchy of quark masses and
flavor mixing angles, it has been speculated by a number of authors
that the realistic quark mass matrices like that in (4.23)
might arise from the flavor democracy symmetry and its 
explicit breaking \cite{Democracy}. 
Let us describe the basic idea of this approach and give a simple
example for illustration.

Under exact $\rm S(3)_L \times S(3)_R$ symmetry
(i.e., the flavor democracy), the mass spectrum for either
up or down quark sector consists of only two levels:
one is of two-fold degeneracy with vanishing mass eigenvalues,
and the other is nondegenerate and massive. The mass matrix
in this ``democratic'' basis is of rank one and takes the form
\begin{equation}
M^{\rm (0)}_{\rm q} \; =\; \frac{\chi_{\rm q}}{3}
\left ( \matrix{
1       & ~ 1 ~     & 1 \cr
1       & ~ 1 ~     & 1 \cr
1       & ~ 1 ~     & 1 \cr } \right ) \; 
\end{equation}
with $\chi_{\rm u} =m_t$ and $\chi_{\rm d} =m_b$. Through the
orthogonal transformation $U_0M^{\rm (0)}_{\rm q}U^{\dagger}_0$, where
\begin{equation}
U_0 \; =\; \left ( \matrix{
\frac{1}{\sqrt{2}}      & -\frac{1}{\sqrt{2}}   & 0 \cr
\frac{1}{\sqrt{6}}      & \frac{1}{\sqrt{6}}    & -\frac{2}{\sqrt{6}} \cr
\frac{1}{\sqrt{3}}      & \frac{1}{\sqrt{3}}    & \frac{1}{\sqrt{3}} \cr }
\right ) \; ,
\end{equation}
one arrives at another rank-one matrix in the ``hierarchical''
basis:
\begin{equation}
M^{\rm (H)}_{\rm q} \; =\; \chi_{\rm q} \left ( \matrix{
0       & ~ 0 ~     & 0 \cr
0       & ~ 0 ~     & 0 \cr
0       & ~ 0 ~     & 1 \cr } \right ) \; ,
\end{equation}
like that in (2.10).
Clearly quarks of the first two families are degenerate
and massless. The mass matrices $M^{\rm (H)}_{\rm q}$ and
$M^{\rm (0)}_{\rm q}$ have equivalent physical significance
and either of them can be used as a starting symmetry limit
to construct the full quark mass matrix. However, we should like 
to remark that the democratic pattern (4.44) plays
an important role in some other aspects of physics, where
the mass-gap phenomena have also been observed. For example,
in the BCS theory of superconductivity, the energy gap
is related to a democratic matrix in the Hilbert space
of the Cooper pairs \cite{Nambu}; 
the pairing force in nuclear physics,
which is introduced to explain large mass gaps in nuclear
energy levels, has the property that the associated Hamiltonian
in the space of nucleon pairs possesses equal mass elements,
i.e., it has the structure like $M^{(0)}_{\rm q}$ \cite{F96}; 
the mass pattern of the $(\pi^0, \eta, \eta')$
pseudoscalar mesons in QCD can be 
described by $M^{(0)}_{\rm q}$ in the chiral limit
$m_u = m_d =0$, where $\pi^0$ and $\eta$ are massless 
Goldstone bosons but $\eta'$ has a nonvanishing mass as a consequence of
the gluon anomaly \cite{Fr77b}. Such examples, together with the 
observed hierarchy in quark
and charged lepton mass spectra, might imply that the mass-gap 
phenomena in physics come in general from analogous 
dynamical mechanisms with the underlying democracy symmetry.

The $\rm S(3)_L \times S(3)_R$ symmetry of 
$M^{\rm (0)}_{\rm q}$ need be broken to the
$\rm S(2)_L \times S(2)_R$ symmetry, in order to
generate masses of the second-family quarks. 
For this purpose we consider the following (real)
symmetry-breaking correction to $M^{\rm (0)}_{\rm q}$:
\begin{equation}
\Delta M^{(1)}_{\rm q} \; =\; \frac{\chi_{\rm q}}{3} \left (\matrix{
0       & 0     & \delta_{\rm q} \cr 
0       & 0     & \delta_{\rm q} \cr 
\delta_{\rm q}  & \delta_{\rm q}        & \varepsilon^{~}_{\rm q} 
\cr } \right ) \; ,
\end{equation}
where $|\delta_{\rm q}| \ll 1$ and $|\varepsilon_{\rm q}| \ll 1$.
Diagonalizing $M^{\rm (0)}_{\rm q} + \Delta M^{(1)}_{\rm q}$ 
through the orthogonal transformation which depends on a single
rotation angle, one can obtain quark mass eigenvalues of the second 
and third families. Explicitly,
\begin{eqnarray}
m_c & = & \frac{\chi_{\rm u}}{6} \left [ 3 + \varepsilon^{~}_{\rm u} 
- \sqrt{ (1-\varepsilon^{~}_{\rm u})^2 + 8 (1+\delta_{\rm u})^2}
\right ] \; , \nonumber \\
m_t & = & \frac{\chi_{\rm u}}{6} \left [ 3 + \varepsilon^{~}_{\rm u} 
+ \sqrt{ (1-\varepsilon^{~}_{\rm u})^2 + 8 (1+\delta_{\rm u})^2}
\right ] \; ;
\end{eqnarray}
and $m_s$, $m_b$ can similarly be given in terms of $\varepsilon^{~}_{\rm d}$
and $\delta_{\rm d}$.  
The flavor mixing angle between the second and third quark families, 
denoted as $\hat{\theta}$, reads as
\begin{eqnarray}
\sin\hat{\theta} & = & \frac{\sqrt{2}}{9} \left | ~ \left (\varepsilon_{\rm u}
- \varepsilon_{\rm d} \right ) \left [ 1+ \frac{1}{9} 
\left ( \varepsilon_{\rm u} + \varepsilon_{\rm d} \right ) \right ] 
+ \left ( \delta_{\rm u} - \delta_{\rm d} \right )
\left [ 1 - \frac{8}{9} \left ( \delta_{\rm u} + \delta_{\rm d} \right )
\right ] \right . \nonumber \\
&  & \left . - \frac{7}{9} \left ( \varepsilon_{\rm u} \delta_{\rm u} - 
\varepsilon_{\rm d}\delta_{\rm d} \right ) \right | \; 
\end{eqnarray}
in the next-to-leading order approximation.
We see that the magnitude of this mixing parameter
(i.e., $|V_{cb}|$ or $|V_{ts}|$) is dominated by the linear difference
between $\varepsilon_{\rm u}$ and $\varepsilon_{\rm d}$ and (or) that between
$\delta_{\rm u}$ and $\delta_{\rm d}$.
Two instructive examples are in order:

(a) $\delta_{\rm q} = 0$ yielding 
\begin{equation}
\sin\hat{\theta} \; = \; \frac{1}{\sqrt{2}} \left ( \frac{m_s}{m_b} -
\frac{m_c}{m_t} \right ) \left [ 1 + \frac{3}{2} \left ( \frac{m_s}{m_b} +
\frac{m_c}{m_t} \right ) \right ] \; ; 
\end{equation}

(b) $\delta_{\rm q} = \varepsilon_{\rm q}$ yielding
\begin{equation}
\sin\hat{\theta} \; = \; 
\sqrt{2} ~ \left ( \frac{m_s}{m_b} - \frac{m_c}{m_t} \right )
\left [ 1 + 3 \left (\frac{m_s}{m_b} + \frac{m_c}{m_t} \right ) \right ] \; .
\end{equation}
These two cases have been discussed in Refs. \cite{Fr94} and \cite{XingD}, 
respectively.
The latter is more favored by current data on the magnitude of $V_{cb}$.
\footnotesize
\begin{table}
\caption{Translations of quark or lepton mass matrices from 
the ``democratic'' basis ($M_{\rm D}$) into the ``hierarchical'' 
basis ($M_{\rm H}$), or vice versa.}
\vspace{0.3cm}
\begin{center}
\begin{tabular}{ccccc}\hline\hline 
Democratic pattern $M_{\rm D}$  &~~~~&  Hierarchical pattern $M_{\rm H}$        
&&      Flavor symmetry of $M_{\rm D}$
\\ \hline 
$c \left ( \matrix{
1       &       1       &       1 \cr
1       &       1       &       1 \cr
1       &       1       &       1 \cr} \right )$
& $\Longrightarrow$     & 
$3c \left ( \matrix{
0       & ~~~~ 0 ~~~~   &       0 \cr
0       & ~~~~ 0 ~~~~   &       0 \cr
0       & ~~~~ 0 ~~~~   &       1 \cr} \right )$
&& $\rm S(3)_L \times S(3)_R$ \\ \\
$c \left ( \matrix{
0       &       0       &       0 \cr
0       &       0       &       0 \cr
0       &       0       &       \epsilon \cr} \right )$
& $\Longrightarrow$     & 
$\displaystyle\frac{c ~\epsilon}{3} \left ( \matrix{
0       &       0       &       0 \cr
0       &       2       &       -\sqrt{2} \cr
0       &       -\sqrt{2}       &       1 \cr}
\right )$
&& $\rm S(2)_L \times S(2)_R$ \\ \\
$c \left ( \matrix{
0       &       0       &       0 \cr
0       &       \epsilon        &       0 \cr
0       &       0       &       0 \cr} \right )$
& $\Longrightarrow$     & 
$\displaystyle\frac{c ~\epsilon}{6} \left ( \matrix{
3       &       -\sqrt{3}       &       -\sqrt{6} \cr
-\sqrt{3}       &       1       &       \sqrt{2} \cr
-\sqrt{6}       &       \sqrt{2}        &       2 \cr} \right )$
&& $\rm S(2)_L \times S(2)_R$ \\ \\
$c \left ( \matrix{
\epsilon        &       0       &       0 \cr
0       &       0       &       0 \cr
0       &       0       &       0 \cr} \right )$
& $\Longrightarrow$     & 
$\displaystyle\frac{c ~\epsilon}{6} \left ( \matrix{
3       &       \sqrt{3}        &       \sqrt{6} \cr
\sqrt{3}        &       1       &       \sqrt{2} \cr
\sqrt{6}        &       \sqrt{2}        &       2 \cr} \right )$
&& $\rm S(2)_L \times S(2)_R$ \\ \\
$c \left ( \matrix{
0       &       0       &       0 \cr
0       &       0       &       \epsilon \cr
0       &       \epsilon        &       0 \cr} \right )$
& $\Longrightarrow$     &  
$\displaystyle\frac{c ~\epsilon}{3\sqrt{2}} \left ( \matrix{
0       &       \sqrt{6}        &       -\sqrt{3} \cr
\sqrt{6}        &       -2\sqrt{2}      &       -1 \cr
-\sqrt{3}       &       -1      &       2\sqrt{2} \cr} \right )$
&& $\rm S(1)_L \times S(1)_R$ \\ \\
$c \left ( \matrix{
0       &       0       &       \epsilon \cr
0       &       0       &       0 \cr
\epsilon        &       0       &       0 \cr} \right )$
& $\Longrightarrow$     & 
$\displaystyle\frac{c ~\epsilon}{3\sqrt{2}} \left ( \matrix{
0       &       -\sqrt{6}       &       \sqrt{3} \cr
-\sqrt{6}       &       -2\sqrt{2}      &       -1 \cr
\sqrt{3}        &       -1      &       2\sqrt{2} \cr} \right )$
&& $\rm S(1)_L \times S(1)_R$ \\ \\
$c \left ( \matrix{
0       &       \epsilon        &       0 \cr
\epsilon        &       0       &       0 \cr
0       &       0       &       0 \cr} \right )$
& $\Longrightarrow$     & 
$\displaystyle\frac{c ~\epsilon}{3} \left ( \matrix{
- 3     &       0       &       0 \cr
0       &       1       &       \sqrt{2} \cr
0       &       \sqrt{2}        &       2 \cr} \right )$
&& $\rm S(1)_L \times S(1)_R$ \\ 
\hline\hline
Hierarchical pattern $M_{\rm H}$        &~~~~~~~&       
Democratic pattern $M_{\rm D}$  &~~&    Flavor symmetry of $M_{\rm D}$
\\ \hline 
$A \left ( \matrix{
0       &       0       &       0 \cr
0       &       0       &       0 \cr
0       &       0       &       1 \cr} \right )$
& $\Longrightarrow$     & 
$\displaystyle\frac{A}{3} \left ( \matrix{
1       & ~~~ 1 ~~~     &       1 \cr
1       & ~~~ 1 ~~~     &       1 \cr
1       & ~~~ 1 ~~~     &       1 \cr} \right )$
&& $\rm S(3)_L \times S(3)_R$ \\ \\
$B \left ( \matrix{
0       &       0       &       0 \cr
0       &       0       &       1 \cr
0       &       1       &       0 \cr} \right )$
& $\Longrightarrow$     & 
$\displaystyle\frac{B}{3\sqrt{2}} \left ( \matrix{
2       &       2       &       -1 \cr
2       &       2       &       -1 \cr
-1      &       -1      &       -4 \cr} \right )$
&& $\rm S(2)_L \times S(2)_R$ \\ \\
$C \left ( \matrix{
0       &       0       &       0 \cr
0       &       1       &       0 \cr
0       &       0       &       0 \cr} \right )$
& $\Longrightarrow$     & 
$\displaystyle\frac{C}{6} \left ( \matrix{
1       &       1       &       -2 \cr
1       &       1       &       -2 \cr
-2      &       -2      &       4 \cr} \right )$
&& $\rm S(2)_L \times S(2)_R$ \\ \\
$D \left ( \matrix{
0       &       1       &       0 \cr
1       &       0       &       0 \cr
0       &       0       &       0 \cr} \right )$
& $\Longrightarrow$     & 
$\displaystyle\frac{D}{\sqrt{3}} \left ( \matrix{
1       &       0       &       -1 \cr
0       &       -1      &       1 \cr
-1      &       1       &       0 \cr} \right )$
&& $\rm S(1)_L \times S(1)_R$ \\ \\
$E \left ( \matrix{
1       &       0       &       0 \cr
0       &       0       &       0 \cr
0       &       0       &       0 \cr} \right )$
& $\Longrightarrow$     & 
$\displaystyle\frac{E}{2} \left ( \matrix{
1       &       -1      &       0 \cr
-1      &       1       &       0 \cr
0       &       0       &       0 \cr} \right )$
&& $\rm S(1)_L \times S(1)_R$ \\ \\
$F \left ( \matrix{
0       &       0       &       1 \cr
0       &       0       &       0 \cr
1       &       0       &       0 \cr} \right )$
& $\Longrightarrow$     & 
$\displaystyle\frac{F}{\sqrt{6}} \left ( \matrix{
2       &       0       &       1 \cr
0       &       -2      &       -1 \cr
1       &       -1      &       0 \cr} \right )$
&& $\rm S(1)_L \times S(1)_R$ \\ \hline\hline
\end{tabular}
\end{center}
\end{table}
\normalsize

To generate masses of the first-family quarks and $CP$ violation,
a further (complex) perturbation need be introduced to
$M^{\rm (0)}_{\rm q}$. Then the $\rm S(3)_L \times S(3)_R$
symmetry of $M^{\rm (0)}_{\rm q}$ will be broken to
$\rm S(1)_L \times S(1)_R$. A simple symmetry-breaking
pattern of this category is
\begin{equation}
\Delta M^{(2)}_{\rm q} \; =\; \frac{\chi_{\rm q}}{3} \left [
\sigma_{\rm q} \cos\phi_{\rm q} \left ( \matrix{
1       &       0       &       -1 \cr
0       &       -1      &       1 \cr
-1      &       1       &       0 \cr } \right )
~ + ~ {\rm i} \sigma_{\rm q} \sin\phi_{\rm q} \left ( \matrix{
0       &       1       &       -1 \cr
-1      &       0       &       1  \cr
1       &       -1      &       0  \cr } \right )
\right ] \; ,
\end{equation}
where $|\sigma_{\rm q}| \ll 1$.
Transforming $M^{\rm (0)}_{\rm q} + \Delta M^{(1)}_{\rm q} + \Delta 
M^{(2)}_{\rm q}$ into the heavy basis, we obtain (for the case
$\delta_{\rm q} = \varepsilon_{\rm q}$) the following mass matrices:
\begin{equation}
M_{\rm q} \; = \; \chi_{\rm q} \left ( \matrix{
{\bf 0}         & \frac{1}{\sqrt{3}} \sigma_{\rm q} 
e^{+{\rm i}\phi_{\rm q}} & {\bf 0} \cr
\frac{1}{\sqrt{3}} \sigma_{\rm q} e^{-{\rm i}\phi_{\rm q}}      
& -\frac{2}{9}\varepsilon_{\rm q}            
& -\frac{2\sqrt{2}}{9}\varepsilon_{\rm q} \cr
{\bf 0}         & -\frac{2\sqrt{2}}{9}\varepsilon_{\rm q}    
& 1 + \frac{5}{9}\varepsilon_{\rm q} \cr } \right ) \; .
\end{equation}
We see that $M_{\rm u}$ and $M_{\rm d}$ totally have four
texture zeros, just like the more general pattern given in (4.23).
This implies that consequences of the mass matrices in (4.53) 
on flavor mixing and $CP$
violation can essentially be obtained with the help of the relevant formulas 
derived from (4.23) in section 4.3; i.e., the former 
may result from the latter by taking
$\phi_1 = \phi_{\rm u} - \phi_{\rm d}$, $\phi_2 = 0$ and
$|r_{\rm u}| = |r_{\rm d}| = \sqrt{2}$ \cite{XingD}.
We find that the results for $\theta_{\rm u}$,
$\theta_{\rm d}$ and angles of the unitarity triangles
are almost insensitive to the explicit values of
$\phi_2$ and $|r_{\rm u}| = |r_{\rm d}|$, which mainly
affect the mixing angle $\theta$.
The point is simply that the hierarchy of quark masses
and the texture of quark mass matrices are enough to
determine, at least partly, some features of the flavor 
mixing. In this sense we emphasize again that the scheme (4.23)
definitely has some predictive power for the flavor mixing
angles, although it totally involves ten parameters.  

Of course there are many other possibilities to break the
symmetry of flavor democracy for $M^{\rm (0)}_{\rm q}$. The
corresponding forms of quark mass matrices in the hierarchical
basis can be obtained in a straightforward way (see Table 4.3, in which only
real perturbations to $M^{\rm (0)}_{\rm q}$ are considered, for
illustration). Conversely one may
transform a general (Hermitian) mass matrix from the
hierarchical basis to the democratic basis, as also shown by the
translation ``dictionary'' in Table 4.3.

\subsection{Non-Hermitian textures of mass matrices}

The textures of quark mass matrices discussed above all 
possess the Hermitian feature. In general, however,
hermiticity is not a necessary constraint on
the up- or down-type mass matrix, or both of them,
for any theory of fermion mass generation. Some 
attention has recently been paid to the non-Hermitian
schemes of quark mass matrices. In the following we 
illustrate three typical ans$\rm\ddot{a}$tze of
this nature.

\underline{\it Nearest-neighbor mixing pattern} ~
In the standard electroweak model or its extensions which do not have
flavor-changing right-handed currents, it is always
possible to find a specific flavor basis in which the arbitrary $3\times 3$
quark mass matrices take the nearest-neighbor mixing form \cite{Branco89}:
\begin{eqnarray}
M_{\rm u} & = & \left ( \matrix{
{\bf 0} & x_{\rm u}     & {\bf 0} \cr
x'_{\rm u}      & {\bf 0}       & y_{\rm u} \cr
{\bf 0} & y'_{\rm u}    & z_{\rm u} \cr} \right ) \; ,
\nonumber \\
M_{\rm d} & = & \left ( \matrix{
{\bf 0} & x_{\rm d}     & {\bf 0} \cr 
x'_{\rm d}      & {\bf 0}       & y_{\rm d} \cr
{\bf 0} & y'_{\rm d}    & z_{\rm d} \cr} \right ) \; ,
\end{eqnarray}
where $\arg (x'_{\rm q}) = \arg (x^*_{\rm q})$ and
$\arg (y'_{\rm q}) = \arg (y^*_{\rm q})$ for q = u and d.
The texture of $M_{\rm u}$ and $M_{\rm d}$ can be
regarded as a non-Hermitian extension of the six-texture-zero
quark mass matrices prescribed first in Ref. \cite{Fr78}. 
To reproduce the measured
magnitude of the flavor mixing matrix element $V_{cb}$ 
(or $V_{ts}$), the restriction $|y_{\rm q}| < |y'_{\rm q}|$
is generally required. If the exotic
hierarchy $|y_{\rm q}| \ll |y'_{\rm q}| \sim |z_{\rm q}|$
is assumed, one obtains $|V_{cb}| \approx |V_{ts}|
\sim {\cal O} (m_s/m_b)$ \cite{NNI}, which is consistent with the
experimental measurement. On the other hand, 
the Cabibbo mixing angle (i.e., $|V_{us}|$ and $|V_{cd}|$)
can properly be reproduced from (4.54) even in the 
assumption of $|x'_{\rm q}| = |x_{\rm q}|$. 

As the mass matrices $M_{\rm u}$ and $M_{\rm d}$
consist totally of twelve independent
parameters (two of them are the phase differences between 
$M_{\rm u}$ and $M_{\rm d}$), 
it is not difficult to find out the 
parameter space allowed by current data of flavor mixing and
$CP$ violation. The number of free parameters could 
further be reduced to ten, e.g., by taking
$\arg (y_{\rm u}) = \arg (y_{\rm d})$
and $x'_{\rm q} = x_{\rm q}$ (for q = u and d).

\underline{\it Triangular mass matrices} ~
It has been shown in Ref. \cite{Scheck} that the arbitrary up- and
down-type mass matrices can always be transformed into a 
triangular form by proper rotations of the right-handed
quark fields. If one of the
two triangular mass matrices is diagonalized, 
the other can again be cast into
the triangular form without loss of any generality. There 
totally exist four types of triangular textures with three
zeros in the upper-left, upper-right, lower-left and lower-right
corners of the mass matrix, depending on the chosen basis of
flavor space. While the first two types do not yield simple
relations between quark masses and flavor mixing angles, the
latter two do \cite{Scheck,WuGH}. 
For example one may take 
\begin{equation}
M_{\rm q} \; =\; \left ( \matrix{
C_{\rm q}       & C'_{\rm q}    & C''_{\rm q} \cr
{\bf 0} & B_{\rm q}     & B'_{\rm q} \cr
{\bf 0} & {\bf 0}       & A_{\rm q} \cr} \right ) \; ,
\end{equation}
where q = u and d, and all non-zero elements are generally
complex. 

One may diagonalize $M_{\rm q}$ with the help of
the bi-unitary transformations, as shown in (3.1), but the
flavor mixing matrix $V$ depends only upon the unitary 
matrices used to rotate the left-handed quark fields. 
In the flavor basis where the up-type mass matrix is diagonal (i.e.,
$M_{\rm u} = {\rm Diag} \{ m_u, m_c, m_t \}$), the 
triangular down-type mass matrix $M_{\rm d}$ can simply be
expressed, in terms of the quark masses $(m_d, m_s, m_b)$ and the
matrix elements of $V$, as follows:
\begin{equation}
M_{\rm d} \; \approx \; \left ( \matrix{
m_d/ V^*_{ud}      & m_s V_{us}    & m_b V_{ub} \cr
{\bf 0}         & m_s V_{cs}    & m_b V_{cb} \cr
{\bf 0}         & {\bf 0}       & m_b V_{tb} \cr} \right ) \; .
\end{equation}
This interesting pattern, valid up to the accuracy of 
${\cal O}(\lambda^4)$ for $\lambda = \sin \theta^{~}_{\rm C}
\approx 0.22$ \cite{WuGH},  
reflects the hierarchy of quark
masses and flavor mixing angles vividly. In a similar way
and to a similar accuracy,
we may choose the flavor basis in which $M_{\rm d}$ is 
diagonal and $M_{\rm u}$ takes the following 
triangular form:
\begin{equation}
M_{\rm u} \; \approx \; \left ( \matrix{
m_u/ V_{ud}    & m_c V^*_{cd}  & m_t V^*_{td} \cr
{\bf 0}         & m_c V^*_{cs}  & m_t V^*_{ts} \cr
{\bf 0}         & {\bf 0}       & m_t V^*_{tb} \cr} \right ) \; .
\end{equation}
As $m_d/m_s \sim m_s/m_b \sim \lambda^2$ and
$m_u/m_c \sim m_c/m_t \sim \lambda^4$ hold, 
the approximate relations between flavor mixing angles
and quark mass ratios can straightforwardly be obtained
from diagonalizing either $M_{\rm d}$ in (4.56) or
$M_{\rm u}$ in (4.57). 

\underline{\it Pure phase mass matrices} ~
The pure phase pattern of quark mass matrices, given by
\begin{equation}
M_{\rm q} \; =\; \frac{\chi^{~}_{\rm q}}{3} \left ( \matrix{
e^{{\rm i}\phi^{\rm q}_{11}}    & e^{{\rm i}\phi^{\rm q}_{12}}
& e^{{\rm i}\phi^{\rm q}_{13}} \cr
e^{{\rm i}\phi^{\rm q}_{21}}    & e^{{\rm i}\phi^{\rm q}_{22}}
& e^{{\rm i}\phi^{\rm q}_{23}} \cr
e^{{\rm i}\phi^{\rm q}_{31}}    & e^{{\rm i}\phi^{\rm q}_{32}}
& e^{{\rm i}\phi^{\rm q}_{33}} \cr} \right ) \; 
\end{equation}
with q = u and d, 
comes from
the phenomenological conjecture that all the Yukawa
couplings of quarks have the identical moduli but 
their phases may be different \cite{Phase}. 
This idea, usually 
referred to as the hypothesis of the universal strength 
for Yukawa couplings, is a complex variation of the
democratic form of quark mass matrices in (4.44). 
Again the freedom of right-handed quark fields
can be used to recast the pure phase mass matrices.
One has verified that the Hermiticity of
$M_{\rm u}$ and $M_{\rm d}$ in (4.58) must be abandoned, in order
to accommodate the observed values of quark masses and
flavor mixing parameters.

Further assumptions need be made for the generic pattern
of pure phase mass matrices, otherwise there is little
calculability to reproduce the measured spectrum of 
quark masses and flavor mixing angles. An interesting
ansatz is to assume $M_{\rm u}$ and $M_{\rm d}$
to be symmetric and have the following parallel 
structures \cite{Silva}:
\begin{eqnarray}
M_{\rm u} & = & \frac{\chi^{~}_{\rm u}}{3} \left ( \matrix{
1       & 1     & e^{{\rm i}\sigma_{\rm u}} \cr
1       & 1     & e^{{\rm i}\rho^{~}_{\rm u}} \cr
e^{{\rm i} \sigma_{\rm u}}      & e^{{\rm i}\rho^{~}_{\rm u}}
& e^{{\rm i}\rho^{~}_{\rm u}} \cr} \right ) \; ,
\nonumber \\
M_{\rm d} & = & \frac{\chi^{~}_{\rm d}}{3} \left ( \matrix{
1       & 1     & e^{{\rm i}\sigma_{\rm d}} \cr
1       & 1     & e^{{\rm i}\rho^{~}_{\rm d}} \cr
e^{{\rm i} \sigma_{\rm d}}      & e^{{\rm i}\rho^{~}_{\rm d}}
& e^{{\rm i}\rho^{~}_{\rm d}} \cr} \right ) \; .
\end{eqnarray}
The six free parameters in $M_{\rm u}$ and $M_{\rm d}$ can be determined 
by six known quark masses. Therefore the three flavor mixing
angles can be derived from (4.59) in terms of only four 
quark mass ratios (i.e., without additional free parameters).  
The problem associated with this ansatz or its 
variations is that the measured strength of $CP$ violation 
in quark mixing cannot correctly be reproduced. To remedy
this defect one has to slightly relax the universality constraint on
the moduli of $M_{\rm u}$ or $M_{\rm d}$, or both of them.

\subsection{Mass matrices at superhigh energy scales}

The phenomenological schemes of quark mass matrices can be 
studied not only at the experimentally accessible scales
(e.g., $\mu \leq 10^2$ GeV) but also at the scales of string or
grand unified theories (e.g., $\mu \geq 10^{15}$ GeV). For the
latter case the running of quark masses and flavor mixing parameters
from the superhigh scale to the weak-interaction scale must be
investigated with the help of renormalization-group equations.
It is in general expected that a particular scheme of mass matrices
at one scale may be significantly changed at another scale, e.g.,
the original texture zeros become nonvanishing as a 
straightforward consequence
of the running effect \cite{XingD,Lindner88}.  
In this section we first point out 
some generic running behaviors of quark mass ratios and flavor
mixing angles. Then we take into account a specific 
ansatz of fermion mass matrices in the context of 
supersymmetric grand unified theories for illustration.

The one-loop evolution equations for the Yukawa coupling matrices
of quarks and charged leptons, in the assumption of no threshold effect
at the intermediate or superhigh scales, have been presented in
Ref. \cite{Babu}. They can considerably be simplified if the hierarchy
of fermion masses and that of the CKM matrix elements are
taken into account. In the leading order approximation one finds
that the running behaviors of quark mass ratios and flavor mixing
angles are governed only by the third-family Yukawa coupling 
eigenvalues $f_t$, $f_b$ and $f_\tau$. From a superhigh scale
$\mu = M_X$ to the weak scale $\mu =M_Z$, the relevant evolution
functions can be defined as
\begin{equation}
\xi_i \; = \; \exp \left [ -\frac{1}{16\pi^2} 
\int^{\ln (M_X/M_Z)} _0 f^2_i (\chi) ~ {\rm d}\chi \right ] \; 
\end{equation}
with $i=t$, $b$ or $\tau$ and $\chi = \ln (\mu/M_Z)$, whose
values depend on the specific model of spontaneous symmetry breaking
below the scale $M_X$.
We then arrive at the following results, which are valid to a good
degree of accuracy:
\begin{eqnarray}
\left [ \frac{m_u}{m_c} \right ]_{M_Z} & = & 
\left [ \frac{m_u}{m_c} \right ]_{M_X} \; ,
\nonumber \\
\left [ \frac{m_d}{m_s} \right ]_{M_Z} & = &
\left [ \frac{m_d}{m_s} \right ]_{M_X} \; ,
\nonumber \\
\left [ \frac{m_c}{m_t} \right ]_{M_Z} & = & 
\left [ \frac{m_c}{m_t} \right ]_{M_X} 
\left (\xi^x_t ~ \xi^y_b \right ) \; ,
\nonumber \\
\left [ \frac{m_s}{m_b} \right ]_{M_Z} & = & 
\left [ \frac{m_s}{m_b} \right ]_{M_X} 
\left (\xi^y_t ~ \xi^x_b \right ) \; ;
\end{eqnarray}
and
\begin{eqnarray}
\left [ \tan\theta_{\rm u} \right ]_{M_Z} & = &
\left [ \tan\theta_{\rm u} \right ]_{M_X} \; ,
\nonumber \\
\left [ \tan\theta_{\rm d} \right ]_{M_Z} & = &
\left [ \tan\theta_{\rm d} \right ]_{M_X} \; , 
\nonumber \\
\left [ \sin\theta \right ]_{M_Z} & = & 
\left [ \sin\theta \right ]_{M_X} \left (\xi_t ~ \xi_b \right )^y \; ,
\nonumber \\
\left [ {\cal J} \right ]_{M_Z} & = &
\left [ {\cal J} \right ]_{M_X} \left (\xi_t ~ \xi_b \right )^{2y} \; ,
\end{eqnarray}
where the coefficients $(x, y) = (-1.5, 1.5)$,
$(-1.5, -0.5)$ and $(-3, -1)$ in the standard model, the 
two-Higgs doublet model and the minimal supersymmetric model,
respectively. Some remarks are in order: (a) the evolution
of $m_u/m_c$ (or $m_d/m_s$) is negligibly small, thus $m_u/m_t$
and $m_c/m_t$ (or $m_d/m_b$ and $m_s/m_b$) have the identical
running effects; (b) among four parameters of the CKM
matrix $V$ (i.e., $\theta_{\rm u}$, $\theta_{\rm d}$, $\theta$ and
$\varphi$), only the mixing angle $\theta$ may have significant 
evolution; (c) the running of the $CP$-violating parameter $\cal J$ 
depends on that of $\sin^2\theta$, but the three angles of the 
unitarity triangle $\Delta_s$ (i.e., $\alpha$, $\beta$ and $\gamma$)
are essentially scale-independent.

\begin{figure}[t]
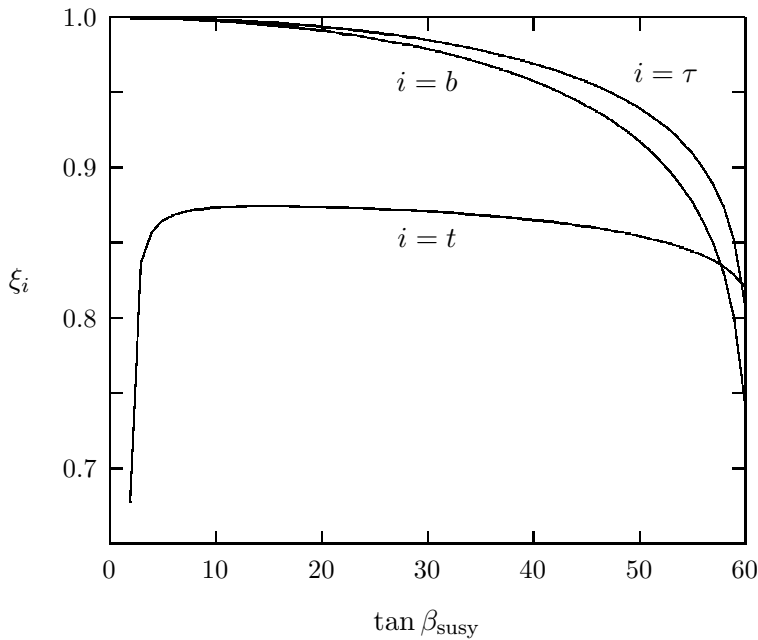

\setlength{\unitlength}{0.240900pt}
\ifx\plotpoint\undefined\newsavebox{\plotpoint}\fi
\sbox{\plotpoint}{\rule[-0.200pt]{0.400pt}{0.400pt}}%

\vspace{0.4cm}
\caption{The magnitudes of $\xi_t$, $\xi_b$ and $\xi_\tau$ changing with
$\tan\beta_{\rm susy}$ in the minimal supersymmetric standard model.}
\end{figure}
For illustration we evaluate the running functions $\xi_t$,
$\xi_b$ and $\xi_\tau$
in the framework of the minimal supersymmetric standard model,
which is best motivated by the present achievement towards a
grand unified theory of all interactions. With the typical inputs
$m_t (M_Z) = 175$ GeV, $m_b (M_Z) = 2.9$ GeV and $m_\tau (M_Z)
= 1.777$ GeV, we calculate $\xi_t$, $\xi_b$ and $\xi_\tau$ for arbitrary
$\tan \beta_{\rm susy}$ (the ratio of Higgs vacuum expectation values)
from $M_X = 10^{16}$ GeV to $M_Z = 91.187$ GeV. The result is 
shown in Fig. 4.3. One can see that $\xi_b \approx \xi_\tau \approx 1$ for 
$\tan\beta_{\rm susy} \leq 20$. Within the perturbatively
allowed region $\xi_b$ and $\xi_\tau$ may be
comparable in magnitude with $\xi_t$ when $\tan\beta_{\rm susy}\geq 50$.
In this case the evolutions of quark masses and flavor mixing
angles are sensitive to both $t$- and $b$-quark contributions,
and to the $\tau$-lepton contribution in a hidden way.

\begin{figure}[t]
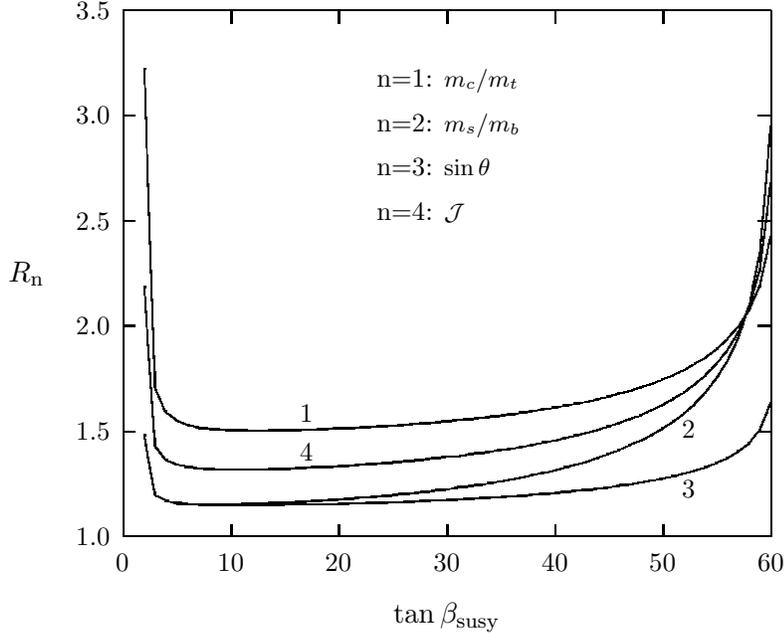

\setlength{\unitlength}{0.240900pt}
\ifx\plotpoint\undefined\newsavebox{\plotpoint}\fi
\sbox{\plotpoint}{\rule[-0.200pt]{0.400pt}{0.400pt}}%

\vspace{0.4cm}
\caption{The running factors $R_{\rm n}$ (from $M_X$ to $M_Z$)
changing with $\tan\beta_{\rm susy}$ in the minimal supersymmetric
standard model.}
\end{figure}

Let us define the running factors of $m_c/m_t$, $m_s/m_b$, $\sin\theta$
and $\cal J$ in (4.61) and (4.62) as $R_1$, $R_2$, $R_3$ and $R_4$
respectively. Their magnitudes can then be obtained from the result
of $\xi_t$ and $\xi_b$ in the minimal supersymmetric standard model
(with $x=-3$ and $y=-1$), as shown in Fig. 4.4. We observe that 
these evolution factors are quite stable for a large range of the
ratio of Higgs vacuum expectation values, i.e.,
$5\leq \tan\beta_{\rm susy} \leq 50$. Such features are essentially
independent of the specific structures of quark mass matrices.

Now we concentrate on specific patterns of quark and charged
lepton mass matrices at the scale of supersymmetric grand 
unified theories ($M_X = 10^{16}$ GeV). In view of the 
success of the Hermitian quark mass matrices with four texture zeros
in accounting for the low-energy flavor mixing phenomena, we 
take the following ansatz at the scale $M_X$:
\begin{eqnarray}
M_{\rm u} & = & \left ( \matrix{
{\bf 0}       & ~ +{\rm i} x            & {\bf 0} \cr
-{\rm i} x             & ~ y             & r y \cr
{\bf 0}       & ~ r y          & z \cr } \right ) \; , 
\nonumber \\
M_{\rm d} & = & \left ( \matrix{
~ {\bf 0}         & ~~~ x'          & ~ {\bf 0} \cr
~ x'              & ~~~ y'         & ~ r y' \cr
~ {\bf 0}         & ~~~ r y'        & ~ z' \cr } \right ) \; ,
\nonumber \\
M_{\rm e} & = & \left ( \matrix{
~ {\bf 0}       & ~ x'          & {\bf 0} \cr
~ x'            & ~ -3 y'       & r y' \cr
~ {\bf 0}       & ~ r y'        & z' \cr } \right ) \; ,
\end{eqnarray}
where $|x|\ll |y| \ll |z|$, $|x'|\ll |y'|\ll |z'|$, and
$r$ is a constant of ${\cal O}(1)$. The texture zeros of $M_{\rm u,d,e}$
as well as the relationship between $M_{\rm d}$
and $M_{\rm e}$ can naturally be obtained in a variety of
grand unified models where the down quarks and the charged
leptons lie in the same multiplet \cite{Fr75}. For example, the coupling
of a Higgs boson in the $\bf 10$ plets of an $\rm SO (10)$ 
model gives certain entries in the Yukawa coupling matrices of the form 
$(M_{\rm d})_{ij} = (M_{\rm e})_{ij}$ (for $i\neq j$), while a Higgs boson
in the $\bf 126$ plets yields $(M_{\rm d})_{22} = -3 (M_{\rm e})_{22}$.
For simplicity we have taken the phase difference between
$(M_{\rm u})_{12}$ and $(M_{\rm d})_{12}$ to be $\pi/2$, a 
value favored by current data on $CP$ violation. The constant $r$
may take values such as $1$, $\sqrt{2}$ or $2$, from the
phenomenological point of view. It is obvious that the ansatz
(4.63) totally involves six free parameters ($x$, $y$, $z$
and $x'$, $y'$, $z'$), which can be determined
from the inputs of three up-type quark
masses and three charged lepton masses. Thus it is able
to give seven predictions at the weak scale $M_Z$, 
three for the down-type quark masses
and four for the parameters of flavor mixing and $CP$ violation.
Of course some of these predictions depend sensitively upon 
the unknown value of $\tan\beta_{\rm susy}$. 
 
The predictions of this ansatz for the down-type quark masses 
read, at the scale $M_X$ and in the next-to-leading order approximation, 
as follows:
\begin{eqnarray}
m_d & = & 3 m_e \left (1 + \frac{4r^2}{9} \cdot\frac{m_\mu}{m_\tau} \right ) \; , 
\nonumber \\
m_s & = & \frac{m_\mu}{3} ~ \left (1 - \frac{4r^2}{9}\cdot \frac{m_\mu}{m_\tau}
\right ) \; , \nonumber \\
m_b & = & m_\tau \; .
\end{eqnarray}
To renormalize these results down to the weak scale $M_Z$, we need
to take into account the evolutions of charged lepton masses 
\begin{eqnarray}
\left [ \frac{m_e}{m_\mu} \right ]_{M_Z} & = &
\left [ \frac{m_e}{m_\mu} \right ]_{M_X} \; ,
\nonumber \\
\left [ \frac{m_\mu}{m_\tau} \right ]_{M_Z} & = &
\left [ \frac{m_\mu}{m_\tau} \right ]_{M_X} \left (\xi^{-3}_\tau 
\right ) \; 
\end{eqnarray}
and make use of the following running function 
in the framework of the minimal supersymmetric standard model:
\begin{equation}
\zeta_{\rm de} \; =\; \exp \left [ +\frac{1}{16\pi^2}
\int^{\ln (M_X/M_Z)}_0 G_{\rm de} (\chi) ~ {\rm d}\chi \right ] \; ,
\end{equation}
where $G_{\rm de} = G_{\rm d} - G_{\rm e}$ with $G_{\rm d}$ and
$G_{\rm e}$ given in terms of the gauge couplings $g^{~}_1$, $g^{~}_2$
and $g^{~}_3$ in Ref. \cite{XingD}. With the inputs $g^2_1 = 0.127$,
$g^2_2 =0.42$ and $g^2_3 =1.44$ at the scale $M_Z$, one
finds $\zeta_{\rm de} = 2.27$. The down-type quark masses 
at the weak scale turn out to be
\begin{eqnarray}
m_d & = & 3m_e \left (1 + \frac{4r^2}{9} \cdot \frac{m_\mu}{m_\tau}
~ \xi^3_\tau \right ) \zeta_{\rm de} \; ,
\nonumber \\
m_s & = & \frac{m_\mu}{3} ~ \left (1 - \frac{4r^2}{9} \cdot
\frac{m_\mu}{m_\tau} ~ \xi^3_\tau \right ) \zeta_{\rm de} \; ,
\nonumber \\
m_b & = & m_\tau \frac{\xi_t \xi^3_b}{\xi^3_\tau} ~ \zeta_{\rm de} \; .
\end{eqnarray}
Taking $r^2 =2$ and $\tan\beta_{\rm susy} =50$ for example, we
obtain $m_d \approx 3.6$ MeV, $m_s \approx 76$ MeV and
$m_b \approx 3.2$ GeV, essentially in agreement with 
the results listed in (2.9).

The predictions of the ansatz (4.63) for flavor mixing and
$CP$ violation at the weak scale $M_Z$ can be obtained 
with the help of the renormalization relations in (4.61) and (4.62).
To the leading order accuracy we arrive at
\begin{eqnarray}
\tan\theta_{\rm u} & = & \sqrt{\frac{m_u}{m_c}} \; , \nonumber \\
\tan\theta_{\rm d} & = & \sqrt{\frac{m_d}{m_s}} \; ;
\end{eqnarray}
and
\begin{eqnarray}
\sin\theta & = & |r| \left (\frac{m_s}{m_b} ~ \xi^2_b 
~ - ~ \frac{m_c}{m_t} ~ \xi^2_t \right ) \; ,
\nonumber \\
{\cal J} & = & r^2 \sqrt{\frac{m_u}{m_c}} \sqrt{\frac{m_d}{m_s}}
\left (\frac{m_s}{m_b} \right )^2 \xi^4_b \; .
\end{eqnarray}
One can see again that the mixing angles $\theta_{\rm u}$ and $\theta_{\rm d}$
are basically scale-independent. The running effects of $\theta$ and $\cal J$ 
depend mainly upon the change of $\xi_b$ from $M_X$ to $M_Z$, which becomes
significant only for $\tan\beta_{\rm susy} \geq 30$.
On this point we conclude that the ansatz (4.63) for fermion mass matrices
has instructive predictions and is favored by current data. It might be
a natural consequence of a complete grand unified model involving
some family symmetries at superhigh scales.

A few similar schemes of fermion mass matrices, based also on the
grand unification ideas, have been proposed in the literature 
(see, e.g., Refs. \cite{Georgi,Hall92}). 
For a brief review of these works, we refer the reader to Ref. \cite{Raby95}.

\section{Lepton mass matrices and neutrino oscillations}
\setcounter{equation}{0}
\setcounter{figure}{0}
\setcounter{table}{0}

\subsection{Classification of $3\times 3$ neutrino mass matrices}

Now we turn to the phenomenological schemes of lepton mass
matrices $M_l$ and $M_\nu$, which can naturally lead to 
hierarchical neutrino mass-squared differences and large
lepton flavor mixing angles as required by current
neutrino oscillation data. Attempts in this direction are of
course restricted by both the preliminary experimental knowledge
and the premature theoretical insight that one has today for
neutrino physics. Hence we shall pay particular attention to
the possible underlying symmetries of lepton flavors, from
which realistic models of lepton mass generation and flavor
mixing could simply be constructed. The possibility of
leptonic $CP$ violation, which is a crucial ingredient of
leptogenesis to explain the cosmological baryon 
asymmetry \cite{Yanagida}, will also be discussed.

Before introducing some phenomenologically favored patterns of
lepton mass matrices, we first give a rough but instructive
classification of $3\times 3$ neutrino mass matrices in the 
flavor basis where the charged lepton mass matrix is diagonal.
To do so the LSND evidence for $\nu_\mu \rightarrow \nu_e$
and $\bar{\nu}_\mu \rightarrow \bar{\nu}_e$ oscillations
is tentatively put aside.

As briefly summarized in section 2.2, the observed neutrino anomalies
can most naturally be interpreted by the hypothesis of neutrino
oscillations, indicating that neutrinos are massive and lepton
flavors are mixed. In the framework of three light neutrinos, to interpret 
current atmospheric and solar neutrino oscillation data requires
$\Delta m^2_{21} = \Delta m^2_{\rm sun} \ll \Delta m^2_{\rm atm}
= \Delta m^2_{32}$ and $|V_{e3}|^2 \ll 1$; i.e., 
the solar and atmospheric neutrino oscillations are respectively
dominated by $\nu_\mu \rightarrow \nu_e$ and
$\nu_\mu \rightarrow \nu_\tau$ transitions, and thus approximately
decoupled. Given the large hierarchy between $\Delta m^2_{21}$
and $\Delta m^2_{32}$, there are three different possibilities
for the spectrum of neutrino masses \cite{Altarelli}:
\begin{eqnarray}
{\rm (a)}:      &~~~& m_1 \; , \; m_2 \; \ll \; m_3 \; ; 
\nonumber \\
{\rm (b)}:      &~~~& m_1 \; \approx \; m_2  \gg  m_3 \; ;
\nonumber \\
{\rm (c)}:      &~~~& m_1 \; \approx \; m_2 \; \approx \; m_3 \; .
\end{eqnarray}
In case (a) the relative magnitude of $m_1$ and $m_2$ is not
restricted; in case (b) we require the inequality
$|m_2 -m_1| \ll m_{1,2}$; and in case (c) the inequality
$|m_2-m_1| \ll |m_3-m_2|$ should be satisfied. Neglecting
possible $CP$-violating phases and taking $|V_{e3}|=0$ 
in the leading order approximation, one can parametrize
the lepton flavor mixing in terms of two rotation angles:
\begin{equation}
V \; \approx \; \left ( \matrix{
c_\odot & -s_\odot    & {\bf 0} \cr
s_\odot c_\bullet     & c_\odot c_\bullet   & -s_\bullet \cr
s_\odot s_\bullet     & c_\odot s_\bullet   & c_\bullet \cr } \right ) \; ,
\end{equation}
where $s_\odot \equiv \sin\theta_{\rm sun}$,
$c_\bullet \equiv \cos\theta_{\rm atm}$, and so on.
Obviously $\theta_{\rm sun}$ and $\theta_{\rm atm}$ correspond to
the mixing angles of solar and atmospheric neutrino oscillations.
The present experimental data favor $\sin^2 2\theta_{\rm atm}
> 0.8$ or 
$\theta_{\rm atm} \sim 32^{\circ}
- 45^{\circ}$; while $\sin^2 2\theta_{\rm sun}$ may be either small
($\theta_{\rm sun} \sim 1^{\circ} - 3^{\circ}$, small-angle
MSW solution) or large ($\theta_{\rm sun} \sim 27^{\circ} 
- 45^{\circ}$, large-angle MSW solution), or nearly
maximal ($\theta_{\rm sun} \sim 45^{\circ}$, vacuum oscillation).

In order to build realistic models of lepton mass generation which 
lead to the flavor mixing pattern (5.2),
it is phenomenologically useful to figure out possible forms
of the neutrino mass matrix that are compatible with 
the mass spectra in (5.1). One should keep in mind that
the flavor mixing matrix $V$ arises from the mismatch between
the diagonalization of 
the charged lepton mass matrix $M_l$ and that of the neutrino mass
matrix $M_\nu$ in a specific flavor basis.
Hence the flavor mixing angles $\theta_{\rm sun}$ and $\theta_{\rm atm}$
depend in general on both the neutrino masses and the
charged lepton masses.
Even though the latter exhibit a strong hierarchy in magnitude
(see (2.1) for illustration), their contributions to
$\theta_{\rm sun}$ and (or) $\theta_{\rm atm}$ may be
non-negligible in some cases.
For instance, the small MSW-type mixing angle $\theta_{\rm sun}$
could be dominated by a contribution of order 
$\arcsin (\sqrt{m_e/m_\mu}) \approx 4^{\circ}$ from the
charged lepton sector, if $M_l$ takes the Hermitian texture with full
nearest-neighbor mixing \cite{Fr78}. The large mixing angle
$\theta_{\rm atm}$ might get a significant contribution of
order $\arcsin (\sqrt{m_\mu/m_\tau}) \approx 14^{\circ}$
(i.e., about $30\%$ of $\theta_{\rm atm} \sim 45^{\circ}$)
in the same scenario of $M_l$. It is therefore improper, 
even misleading, to account for the lepton flavor mixing solely
in terms of the mixing parameters in the neutrino sector.

The observation made above has an important implication: if one
writes the neutrino mass matrix $M_\nu$ taking into account the flavor
mixing matrix $V$ in the basis where the charged lepton mass
matrix $M_l$ is diagonalized,
the resultant $M_\nu$ remains dependent on $M_l$ in a hidden
way (i.e., through unspecified flavor mixing angles).
In this sense the useful information that one can obtain 
about the textures of $M_l$ is limited. Nevertheless
it should still be instructive to find out the possible leading order
forms of $M_\nu$, versus the diagonal $M_l$, at the 
present phenomenological stage. Such forms of
$M_\nu$ might provide useful hints for model buildings, which
could finally shed light on the dynamics of lepton mass generation
and flavor mixing. Therefore we proceed to make a classification
of the textures of $M_\nu$ allowed by current neutrino oscillation
data. 
\small
\begin{table}[t]
\caption{Leading order textures of the neutrino mass matrix
$M_\nu$, in which $s_\odot \equiv \sin \theta_{\rm sun}$, 
$c_{2\bullet} \equiv \cos 2 \theta_{\rm atm}$, and so on; 
and ``0'' or ``1'' only means ${\cal O}(0)$ or ${\cal O}(1)$ in magnitude.}
\vspace{0.25cm}
\begin{center}
\begin{tabular}{ccccc}\hline\hline 
Masses $\{\lambda_1, \lambda_2, \lambda_3 \}$   
&~& $M_\nu$ (small $\theta_{\rm sun}$)        
&~~& $M_\nu$ (large $\theta_{\rm sun}$) \\ \hline 
(a): $\{0, ~ 0, ~ 1 \} m_3$
&& $ m_3 \left ( \matrix{ 0     & 0     & 0 \cr
0       & s^2_\bullet   & -s_\bullet c_\bullet \cr
0       & -s_\bullet c_\bullet   & c^2_\bullet \cr} \right )$
&& $ m_3 \left ( \matrix{ 0     & 0     & 0 \cr
0       & s^2_\bullet   & -s_\bullet c_\bullet \cr
0       & -s_\bullet c_\bullet   & c^2_\bullet \cr} \right )$ \\ \\ 
(b): $\{1, -1, 0 \} m_1$ 
&& $ m_1 \left ( \matrix{ 1     & 0     & 0 \cr
0       & -c^2_\bullet   & -s_\bullet c_\bullet \cr
0       & -s_\bullet c_\bullet   & -s^2_\bullet \cr} \right )$
&& $ m_1 \left ( \matrix{ c_{2\odot}    & s_{2\odot} c_\bullet  & s_{2\odot} c_\bullet \cr
s_{2\odot} c_\bullet    & -c_{2\odot} c^2_\bullet       & -c_{2\odot} s_\bullet c_\bullet \cr
s_{2\odot} c_\bullet    & -c_{2\odot} s_\bullet c_\bullet        
& -c_{2\odot} s^2_\bullet \cr} \right )$ \\ \\
(b): $\{1, ~1, ~0 \} m_1$ 
&& $ m_1 \left ( \matrix{ 1     & 0     & 0 \cr
0       & c^2_\bullet   & s_\bullet c_\bullet \cr
0       & s_\bullet c_\bullet    & s^2_\bullet \cr} \right )$
&& $ m_1 \left ( \matrix{ 1     & ~ 0     & 0 \cr
0       & ~ c^2_\bullet   & s_\bullet c_\bullet \cr
0       & ~ s_\bullet c_\bullet    & s^2_\bullet \cr} \right )$ \\ \\ 
(c): $\{1, ~ 1, ~ 1 \} m_1$
&& $ m_1 \left ( \matrix{ 1     & ~~~ 0 ~~~     & 0 \cr
0       & ~~~ 1 ~~~     & 0 \cr
0       & ~~~ 0 ~~~     & 1 \cr} \right )$
&& $ m_1 \left ( \matrix{ 1     & ~~~~ 0 ~~~~     & 0 \cr
0       & ~~~~ 1 ~~~~     & 0 \cr
0       & ~~~~ 0 ~~~~     & 1 \cr} \right )$ \\ \\ 
(c): $\{-1, 1, 1 \} m_1$ 
&& $ m_1 \left ( \matrix{ -1    & ~ 0 ~~     & ~ 0 \cr
0       & ~ 1 ~~     & ~ 0 \cr
0       & ~ 0 ~~    & ~ 1 \cr} \right )$
&& $ m_1 \left ( \matrix{ -c_{2\odot}   & -s_{2\odot} c_\bullet & -s_{2\odot} c_\bullet \cr
-s_{2\odot} c_\bullet   & s^2_\bullet   & -s_\bullet c_\bullet \cr
-s_{2\odot} c_\bullet   & -s_\bullet c_\bullet   & c^2_\bullet \cr} \right )$ \\ \\
(c): $\{1, -1, 1 \} m_1$ 
&& $ m_1 \left ( \matrix{ 1     & 0     & 0 \cr
0       & -c_{2\bullet}   & -s_{2\bullet} \cr
0       & -s_{2\bullet}   & c_{2\bullet} \cr} \right )$ 
&& $ m_1 \left ( \matrix{ c_{2\odot}    & s_{2\odot} c_\bullet  & s_{2\odot} c_\bullet \cr
s_{2\odot} c_\bullet    & s^2_\bullet   & -s_\bullet c_\bullet \cr
s_{2\odot} c_\bullet    & -s_\bullet c_\bullet   & c^2_\bullet \cr} \right )$ \\ \\
(c): $\{1, 1, -1 \} m_1$ 
&& $ m_1 \left ( \matrix{ 1     & ~ 0     & 0 \cr
0       & ~ c_{2\bullet}    & s_{2\bullet} \cr
0       & ~ s_{2\bullet}    & -c_{2\bullet} \cr} \right )$ 
&& $ m_1 \left ( \matrix{ 1     & ~~ 0 ~     & 0 \cr
0       & ~~ c_{2\bullet} ~    & s_{2\bullet} \cr
0       & ~~ s_{2\bullet} ~    & -c_{2\bullet} \cr} \right )$ \\ \hline \hline
\end{tabular}
\end{center}
\end{table}
\normalsize

In doing so we do not consider the origin of $M_\nu$, 
no matter how it could originate from the seesaw mechanism 
or from other flavor symmetries. But for simplicity we shall 
assume $CP$ invariance in the lepton sector, thus $M_\nu$
is a real symmetric matrix for either Dirac- or Majorana-type
neutrinos.
In the flavor basis where $M_l$ is diagonal, $M_\nu$ takes
the following form 
\begin{equation}
M_\nu \; = \; V^* \left ( \matrix{
\lambda_1       & 0     & 0 \cr
0       & \lambda_2     & 0 \cr
0       & 0     & \lambda_3 \cr} \right ) V^{\dagger} \; ,
\end{equation}
in which $\lambda_i$ are the neutrino
mass eigenvalues ($|\lambda_i| = m_i$), 
and $V$ is the $3\times 3$ flavor mixing matrix. 
Taking the neutrino mass spectra in (5.1) and the approximate 
flavor mixing pattern in (5.2) into account, we
obtain fourteen different textures of $M_\nu$ for 
the case of small $\theta_{\rm sun}$ (small-angle MSW solution)
and that of large $\theta_{\rm sun}$ (large-angle MSW solution
or vacuum oscillation solution). 
The results are listed in Table 5.1 for illustration.
Note that all texture zeros in $M_\nu$ only imply small quantities,
and all unity elements of $M_\nu$ only mean ${\cal O}(1)$ in magnitude.
The matrix elements of ${\cal O}(0)$, in particular those in
case (c) with $\lambda_1 \approx \lambda_2 \approx \lambda_3$,
play an important role to generate large flavor mixing angles. 
If $\theta_{\rm atm} = \theta_{\rm sun} = 45^{\circ}$ (bi-maximal
mixing pattern \cite{Barger98}) is taken, the elements of $M_\nu$ 
are proportional to algebraic numbers of ${\cal O}(1)$.
The same situation appears for the special choices
$\theta_{\rm atm} = 45^{\circ}$ and 
$\theta_{\rm sun} =0^{\circ}$ \cite{Altarelli}.

In principle one could also classify the possible forms of the charged lepton
mass matrix $M_l$ in the flavor basis where the neutrino mass matrix
$M_\nu$ is diagonal. Without much prejudice a more general but
incomplete classification of
the leading order textures of lepton mass matrices (both $M_\nu$ and
$M_l$) have been presented in Ref. \cite{Hall99}. 

\subsection{Lepton mixing from flavor democracy breaking}

The idea of subnuclear flavor democracy and its explicit breaking
has been applied to the quark
sector (see section 4.5) to interpret the strong mass hierarchy of 
up- and down-type quarks.
Since the mass spectrum of charged leptons
exhibits a similar hierarchy as that of quarks, it would be natural to
consider the same symmetry limit for the charged lepton mass matrix,
i.e., $M^{(0)}_l$ takes the same form as $M^{(0)}_{\rm q}$ in (4.44).
As for the neutrino sector, we have no direct information about the
absolute values or relative magnitudes of neutrino masses. The observed
large (nearly maximal) mixing angles and the tiny mass-squared differences
in atmospheric and solar neutrino oscillations, however, favor the
possibility that the three neutrino masses are approximately degenerate,
although the other possibilities for the neutrino mass spectrum 
cannot be excluded. We therefore start with the hypothesis of the flavor
democracy for charged leptons and the mass degeneracy for neutrinos to
construct a phenomenological model of lepton mass generation and flavor
mixing \cite{FX96}. 
$CP$ violation can be incorporated into this model, when complex
symmetry breaking terms are explicitly introduced.

In the limits of the flavor democracy and the mass degeneracy, the
charged lepton and neutrino mass matrices can respectively be 
written as 
\begin{eqnarray}
M^{(0)}_l & = & \frac{C^{~}_l}{3} \left (\matrix{
1       & 1     & 1 \cr
1       & 1     & 1 \cr
1       & 1     & 1 \cr} \right ) \; , 
\nonumber \\
M^{(0)}_\nu & = & C_\nu \left (\matrix{
1       & 0     & 0 \cr
0       & 1     & 0 \cr
0       & 0     & 1 \cr} \right ) \; ,
\end{eqnarray}
where $C^{~}_l =m_\tau$ and $C_\nu =m_0$ measure the 
corresponding mass scales.
If the three neutrinos are of the Majorana type,
$M^{(0)}_\nu$ could take a more general form
$M^{(0)}_\nu P_\nu$ with $P_\nu = {\rm Diag} \{ 1,
e^{i\phi_1}, e^{i\phi_2} \}$. As the Majorana phase matrix
$P_\nu$ has no effect on the flavor mixing and
$CP$-violating observables
in neutrino oscillations, it will be neglected in the subsequent discussions.
The neutrino mass matrix $M^{(0)}_\nu$ exhibits an
S(3) symmetry, while the charged lepton mass matrix $M^{(0)}_l$ exhibits an
$\rm S(3)_L \times S(3)_R$ symmetry.

One can transform the charged lepton mass matrix from the 
democratic basis $M^{(0)}_l$ into the hierarchical basis
$M_{0l}$ given in (2.2) through a simple orthogonal transformation
$M_{0l} = U_0 M^{(0)}_l U^{\rm T}_0$, where $U_0$ has
been shown in (4.45).
Clearly $m_e = m_\mu =0$, as seen in $M_{0l}$; and
$m_1 = m_2 =m_3 =m_0$ in $M^{(0)}_\nu$.
There is no flavor mixing in this symmetry limit.

Of course there are a number of different ways to break the flavor
democracy of $M^{(0)}_l$ (see Table 4.3)
and the mass degeneracy of $M^{(0)}_\nu$. Whether the resultant
flavor mixing patterns are compatible with the neutrino oscillation
data is, at present, the only phenomenological criterion to select
the proper symmetry breaking scenarios. In the following we
take three instructive examples, which can respectively lead to the nearly
bi-maximal mixing pattern, the small-versus-large mixing pattern,
and the bi-maximal mixing pattern for
atmospheric and solar neutrino oscillations.

\underline{\it Nearly bi-maximal mixing pattern} ~
A real diagonal breaking of the $\rm S(3)_L \times S(3)_R$
symmetry of $M^{(0)}_l$ and the $\rm S(3)$ symmetry of $M^{(0)}_\nu$
can lead to a nearly bi-maximal flavor mixing pattern, as first
discussed in Ref. \cite{FX96}. 
To incorporate $CP$ violation into this neutrino mixing scenario, 
however, complex perturbative terms are required for 
$M^{(0)}_l$ \cite{FX99CP}. 
Let us proceed with two different symmetry-breaking steps.

(a) Small real perturbations to the (3,3) elements of $M^{(0)}_l$
and $M^{(0)}_\nu$ are respectively introduced:
\begin{eqnarray}
\Delta M^{(1)}_l & = & \frac{C^{~}_l}{3} \left ( \matrix{
0       & 0     & 0 \cr
0       & 0     & 0 \cr
0       & 0     & \varepsilon^{~}_l \cr } \right ) \; , 
\nonumber \\
\Delta M^{(1)}_\nu & = & C_\nu \left ( \matrix{
0       & 0     & 0 \cr
0       & 0     & 0 \cr
0       & 0     & \varepsilon_\nu \cr } \right ) \; .
\end{eqnarray}
In this case the charged lepton mass matrix $M^{(1)}_l =
M^{(0)}_l + \Delta M^{(1)}_l$ remains symmetric under an
$\rm S(2)_L \times S(2)_R$ transformation, 
and the neutrino mass matrix
$M^{(1)}_\nu = M^{(0)}_\nu + \Delta M^{(1)}_\nu$ has
an $\rm S(2)$ symmetry. 
The muon becomes massive (i.e., $m_\mu \approx 2|\varepsilon^{~}_l|
m_\tau /9$), and the mass eigenvalue $m_3$ is no more degenerate
with $m_1$ and $m_2$ (i.e., $|m_3 - m_0| = m_0 |\varepsilon_\nu|$). 
After the diagonalization of 
$M^{(1)}_l$ and $M^{(1)}_\nu$, one finds that the 2nd and 3rd
lepton families have a definite flavor mixing angle
$\theta$. We obtain $\tan\theta \approx -\sqrt{2} ~$ if the
small correction of ${\cal O}(m_\mu/m_\tau)$ is neglected.
Then neutrino oscillations at the atmospheric scale may arise
in $\nu_\mu \rightarrow \nu_\tau$ transitions with 
$\Delta m^2_{32} = \Delta m^2_{31}
\approx 2m_0 |\varepsilon_\nu|$. The corresponding
mixing factor $\sin^2 2\theta \approx 8/9$ is in good agreement 
with current data.

The symmetry breaking term $\Delta M^{(1)}_l$ given in (5.5)
for the charged lepton mass matrix serves as a good illustrative
example. One could consider a more general perturbation, analogous to
$\Delta M^{(1)}_{\rm q}$ given in (4.47) for quarks, in order to
break the $\rm S(3)_L \times S(3)_R$ symmetry of $M^{(0)}_l$
to an $\rm S(2)_L \times S(2)_R$ symmetry. In this case 
we keep the perturbation to $M^{(0)}_\nu$ to be the same as
$\Delta M^{(1)}_\nu$ in (5.5). Then it is easy to check that
the leading order result for lepton flavor mixing (i.e.,
$\tan \theta \approx -\sqrt{2} ~$ or $\sin^2 2\theta \approx 8/9$)
remains unchanged.

(b) Small imaginary perturbations, 
which have the identical magnitude but
the opposite signs, 
are introduced to the (2,2) and (1,1) elements of
$M^{(1)}_l$. For $M^{(1)}_\nu$ the corresponding real
perturbations are taken into account \cite{FX99CP}:
\begin{eqnarray}
\Delta M^{(2)}_l & = & \frac{C^{~}_l}{3} \left ( \matrix{
-{\rm i}\delta_l      & 0     & 0 \cr
0       & {\rm i}\delta_l     & 0 \cr
0       & 0     & 0 \cr } \right ) \; , 
\nonumber \\
\Delta M^{(2)}_\nu & = & C_\nu \left ( \matrix{
-\delta_\nu     & ~ 0   &  0 \cr
0       & ~ \delta_\nu  &  0 \cr
0       & ~ 0   &  0 \cr } \right ) \; .
\end{eqnarray}
We obtaine $m_e \approx |\delta_l|^2 m^2_\tau /(27 m_\mu)$ 
and $m_{1,2} = m_0 (1 \mp \delta_\nu)$. The diagonalization of 
$M^{(2)}_l = M^{(1)}_l + \Delta M^{(2)}_l$ and 
$M^{(2)}_\nu = M^{(1)}_\nu + \Delta M^{(2)}_\nu$ 
leads to a full $3\times 3$
flavor mixing matrix, which links neutrino mass eigenstates 
$(\nu_1, \nu_2, \nu_3)$ to neutrino flavor eigenstates
$(\nu_e, \nu_\mu, \nu_\tau)$ in the following manner:
\begin{equation}
V \; = \; \left ( \matrix{
\frac{1}{\sqrt{2}}      & \frac{-1}{\sqrt{2}}   & 0 \cr
\frac{1}{\sqrt{6}}      & \frac{1}{\sqrt{6}}    & \frac{-2}{\sqrt{6}} \cr
\frac{1}{\sqrt{3}}      & \frac{1}{\sqrt{3}}    & \frac{1}{\sqrt{3}} \cr}
\right ) 
~ + ~ {\rm i} ~ \xi^{~}_V ~ \sqrt{\frac{m_e}{m_\mu}} 
\; + \; \zeta^{~}_V ~ \frac{m_\mu}{m_\tau} \; ,
\end{equation}
where
\begin{eqnarray}
\xi^{~}_V & = &  \left ( \matrix{
\frac{1}{\sqrt{6}}      & ~ \frac{1}{\sqrt{6}} ~        & \frac{-2}{\sqrt{6}} \cr
\frac{1}{\sqrt{2}}      & ~ \frac{-1}{\sqrt{2}} ~       & 0 \cr
0       & ~ 0 ~ & 0 \cr} \right ) \; ,
\nonumber \\
\zeta^{~}_V & = & \left ( \matrix{
0       & 0     & 0 \cr
\frac{1}{\sqrt{6}}      & \frac{1}{\sqrt{6}}    & \frac{1}{\sqrt{6}} \cr
\frac{-1}{\sqrt{12}}    & \frac{-1}{\sqrt{12}}  & \frac{1}{\sqrt{3}} \cr}
\right ) \; .
\end{eqnarray}
Note that the leading term of $V$ is just the orthogonal matrix $U_0$ in
(4.45), used to transform a democratic mass matrix into its counterpart
in the hierarchical basis. 

The diagonal non-Hermitian perturbation $\Delta M^{(2)}_l$ in (5.6)
is certainly not the only way to generate
large flavor mixing and $CP$ violation of the pattern (5.7). One can find 
other two off-diagonal non-Hermitian perturbations as well as three 
Hermitian perturbations, as listed in Table 5.2, which are able to give 
rise to the same flavor mixing pattern $V$ in the next-to-leading order
approximation. All these six perturbations 
have a common feature: the (1,1) elements of their 
counterparts in the hierarchical basis, which can be read off from
Table 4.3, are all vanishing \cite{FX99CP}.
This feature assures that the $CP$-violating effects, resulted from
the complex perturbations in Table 5.2, 
are approximately independent 
of other details of the flavor symmetry breaking and have
the identical strength to a high degree of accuracy.
Hence it is in practice difficult to distinguish one scenario from
another. The simplicity of the diagonal
non-Hermitian perturbation $\Delta M^{(2)}_l$
and its parallelism with $\Delta M^{(2)}_\nu$ 
seem quite instructive for building a possible model of
lepton mass generation.
The Hermitian forms of $\Delta M^{(2)}_l$, on the other hand,
have much similarity with the favored perturbations generating $CP$
violation in the quark sector and could provide us some useful hints
towards an underlying symmetry between the quarks and the charged leptons.
\begin{table}[t]
\caption{Six imaginary perturbations $\Delta M^{(2)}_l$ for the
charged lepton mass matrix.}
\vspace{0.2cm}
\begin{center}
\begin{tabular}{cccc}\hline\hline 
Hermitian form of $\Delta M^{(2)}_l$ 
&~~~~~&~~~~~& Non-Hermitian form of $\Delta M^{(2)}_l$ \\ \hline 
$\displaystyle\frac{C^{~}_l}{3} \left ( \matrix{
~ 0 ~   & -{\rm i}\delta_l ~    & ~ 0 ~~ \cr
~ {\rm i}\delta_l ~     & ~ 0 ~ & ~ 0 ~~ \cr
~ 0 ~   & ~ 0 ~ & ~ 0 ~~ \cr } \right )$
&&&
$\displaystyle\frac{C^{~}_l}{3} \left ( \matrix{
-{\rm i}\delta_l ~~     & 0 ~~  & ~ 0 ~ \cr
0 ~~    & {\rm i}\delta_l ~~    & ~ 0 ~ \cr
0 ~~    & 0 ~~  & ~ 0 ~ \cr } \right )$ \\ \\
$\displaystyle\frac{C^{~}_l}{3} \left ( \matrix{
~ 0     & ~ 0 ~ & {\rm i}\delta_l \cr
~ 0     & ~ 0 ~ & -{\rm i}\delta_l \cr
- {\rm i}\delta_l       &  ~ {\rm i}\delta_l ~  & 0 \cr } \right )$
&&&
$\displaystyle\frac{C^{~}_l}{3} \left ( \matrix{
~ 0 ~   & ~ 0   & {\rm i}\delta_l \cr
~ 0 ~   & ~ 0   & -{\rm i}\delta_l \cr
~ {\rm i}\delta_l ~   & -{\rm i}\delta_l    & 0 \cr } \right )$ \\ \\
$\displaystyle\frac{C^{~}_l}{3} \left ( \matrix{
0       & -{\rm i}\delta_l    & {\rm i}\delta_l \cr
{\rm i}\delta & 0     & -{\rm i}\delta_l \cr
-{\rm i}\delta_l      & {\rm i}\delta_l     & 0 \cr } \right )$
&&&
$\displaystyle\frac{C^{~}_l}{3} \left ( \matrix{
-{\rm i}\delta_l      & 0     & {\rm i}\delta_l \cr
0       & {\rm i}\delta       & -{\rm i}\delta_l \cr
{\rm i}\delta_l       & -{\rm i}\delta_l    & 0 \cr } \right )$ \\ \hline \hline
\end{tabular}
\end{center}
\end{table}
 
The flavor mixing matrix $V$ can in general be parametrized
in terms of three Euler angles and one $CP$-violating phase
\footnote{For neutrinos of the Majorana type, two additional
$CP$-violating phases may enter. But they are irrelevant 
to neutrino oscillations and can be ignored for our present
purpose.}. 
A suitable parametrization, analogous to that in
(3.29) for quark mixing, reads as follows:
\begin{eqnarray}
V & = & \left ( \matrix{
c^{~}_l & s^{~}_l       & 0 \cr
-s^{~}_l        & c^{~}_l       & 0 \cr
0       & 0     & 1 \cr } \right )  \left ( \matrix{
e^{-{\rm i}\phi}      & 0     & 0 \cr
0       & c     & s \cr
0       & -s    & c \cr } \right )  \left ( \matrix{
c_{\nu} & -s_{\nu}      & 0 \cr
s_{\nu} & c_{\nu}       & 0 \cr
0       & 0     & 1 \cr } \right )  \nonumber \\ \nonumber \\
& = & \left ( \matrix{
s^{~}_l s_{\nu} c + c^{~}_l c_{\nu} e^{-{\rm i}\phi} & 
s^{~}_l c_{\nu} c - c^{~}_l s_{\nu} e^{-{\rm i}\phi} &
s^{~}_l s \cr
c^{~}_l s_{\nu} c - s^{~}_l c_{\nu} e^{-{\rm i}\phi} &
c^{~}_l c_{\nu} c + s^{~}_l s_{\nu} e^{-{\rm i}\phi}   &
c^{~}_l s \cr - s_{\nu} s       & - c^{~}_\nu s & c \cr } \right ) \; ,
\end{eqnarray}
in which $s^{~}_l \equiv \sin\theta_l$, $s_{\nu} \equiv \sin\theta_{\nu}$, 
$c \equiv \cos\theta$, and so on. The three mixing angles can all be
arranged to lie in the first quadrant, while the $CP$-violating
phase may take values between $0$ and $2\pi$. 
With the inputs $m_e/m_\mu \approx 0.00484$ and 
$m_\mu/m_\tau \approx 0.0594$ \cite{PDG} we obtain
\begin{equation}
\theta_l \approx 4^{\circ} \; , ~~~~~
\theta_\nu \approx 45^{\circ} \; , ~~~~~
\theta \approx 52^{\circ} \; , ~~~~~
\phi \approx 90^{\circ} \; 
\end{equation}
by comparing between (5.7) and (5.9).
The smallness of $\theta_l$ is a natural consequence of the 
mass hierarchy in the charged lepton sector, just as the
smallness of $\theta_{\rm u}$ in quark mixing \cite{FX99}.
On the other hand, both $\theta_\nu$ and $\theta$ are too large
to be comparable with the corresponding quark mixing angles
(i.e., $\theta_{\rm d}$ and $\theta$ as defined in section 3.6),
reflecting the qualitative difference between quark and lepton
flavor mixing phenomena. It is worth emphasizing that the
leptonic $CP$-violating phase $\phi$ takes a special value 
($\approx 90^{\circ}$) in our model. The same possibility
is also favored for the quark mixing phenomenon in a variety of
realistic mass matrices (see Refs. \cite{FX99,FX95} and
section 4.3). 
Therefore maximal leptonic $CP$ violation, in the sense that
the magnitude of 
${\cal J}_l = s^{~}_l c^{~}_l s_\nu c_\nu s^2 c \sin\phi$
is maximal for the fixed values of three flavor mixing angles, 
appears naturally as in the quark sector.

Some consequences of this lepton mixing scenario are as follows:

(1) The mixing pattern in (5.7), after neglecting small
corrections from the charged lepton masses, is quite similar to that
of the pseudoscalar mesons $\pi^0$, $\eta$ and $\eta'$ in QCD in
the limit of the chiral $\rm SU(3)_L \times SU(3)_R$
symmetry:
\begin{eqnarray}
|\pi^0 \rangle & = & \frac{1}{\sqrt{2}} \left (
|\bar{u}u\rangle - |\bar{d}d\rangle \right ) \; ,
\nonumber \\
|\eta \rangle ~ & = & \frac{1}{\sqrt{6}} \left (
|\bar{u}u\rangle + |\bar{d}d\rangle - 2 |\bar{s}s\rangle
\right ) \; ,
\nonumber \\
|\eta' \rangle & = & \frac{1}{\sqrt{3}} \left (
|\bar{u}u\rangle + |\bar{d}d\rangle + |\bar{s}s\rangle 
\right ) \; .
\end{eqnarray}
A theoretical derivation of the flavor mixing matrix 
$V\approx U_0$ has been given in Ref. \cite{Mohapatra},
in the framework of a left-right symmetric extension of the
standard model with $\rm S(3)$ and $\rm Z(4) \times Z(3) \times Z(2)$
symmetries.

(2) The $V_{e3}$ element, of magnitude
\begin{equation}
|V_{e3}| \; =\; \frac{2}{\sqrt{6}} \sqrt{\frac{m_e}{m_\mu}} \;\; ,
\end{equation}
is naturally suppressed in
the symmetry breaking scheme outlined above.
A similar feature appears in the $3\times 3$ quark flavor mixing
matrix, i.e., $|V_{ub}|$ is the smallest among the
nine quark mixing elements. Indeed the smallness of $V_{e3}$
provides a necessary condition for the decoupling of
solar and atmospheric neutrino oscillations, even though neutrino
masses are nearly degenerate. The effect of small but nonvanishing
$V_{e3}$ will manifest itself in long-baseline $\nu_\mu
\rightarrow \nu_e$ and $\nu_e \rightarrow \nu_\tau$ transitions,
as shown in Ref. \cite{FX98}.

(3) The flavor mixing between the 1st and 2nd lepton families
and that between the 2nd and 3rd lepton families are nearly
maximal. This property, together with the natural smallness
of $|V_{e3}|$, allows a satisfactory interpretation of the 
observed large mixing 
in atmospheric and solar neutrino oscillations. One obtains
\footnote{In calculating $\sin^2 2\theta_{\rm sun}$ we have
taken into account the ${\cal O}(m_e/m_\mu)$ correction to the 
expression of the lepton flavor mixing matrix $V$ in (5.7).}
\begin{eqnarray}
\sin^2 2\theta_{\rm sun} & = & 1 ~ - ~ \frac{4}{3} \frac{m_e}{m_\mu} 
\; , \nonumber \\
\sin^2 2\theta_{\rm atm} & = & \frac{8}{9} \left ( 1 +
\frac{m_\mu}{m_\tau} \right ) \; 
\end{eqnarray}
to a high degree of accuracy; i.e., 
$\sin^2 2\theta_{\rm sun} \approx 0.99$ and $\sin^2 2\theta_{\rm atm}
\approx 0.94$, in agreement with current data \cite{SK}. 
It is obvious that the model
is fully consistent with the vacuum oscillation solution to
the solar neutrino problem \cite{Barger99} and might also be
able to incorporate the large-angle MSW 
solution \cite{Liu}.

(4) The rephasing-invariant strength of $CP$ violation in (3.14), defined in
close analogy with that for quarks, is given as
\begin{equation}
{\cal J}_l \; = \; \frac{1}{3\sqrt{3}} \sqrt{\frac{m_e}{m_\mu}}
\left ( 1 + \frac{1}{2} \frac{m_\mu}{m_\tau} \right ) \; .
\end{equation}
Explicitly we have ${\cal J}_l \approx 1.4\%$. 
The large magnitude of ${\cal J}_l$ for lepton mixing
is very non-trivial, as the same quantity for quark mixing
is only of order $10^{-5}$ (see Refs. \cite{FX99,FX95} and section 4.3).  
If the mixing pattern under discussion 
is reconciled with the large-angle MSW solution to the
solar neutrino problem, then the relevant $CP$- or $T$-violating signals 
should be large enough to be measured from the asymmetry
between $P(\nu_\mu \rightarrow \nu_e)$ and 
$P(\bar{\nu}_\mu \rightarrow \bar{\nu}_e)$ or that
between $P(\nu_\mu \rightarrow \nu_e)$ and $P(\nu_e \rightarrow
\nu_\mu)$ in the long-baseline
neutrino experiments (see section 5.6 for detailed discussions). 

(5) Finally it is worth remarking that our lepton mixing pattern 
is not necessarily in conflict
with current constraints on the neutrinoless double beta 
decay \cite{KK99,Faessler}, if neutrinos are of the Majorana type.
In the presence of $CP$ violation, the effective
mass term of the $(\beta\beta)_{0\nu}$ decay can simply be written as
$\langle m_\nu \rangle = \sum (m_i ~ \tilde{V}^2_{ei})$,
where $\tilde{V} = VP_\nu$
and $P_\nu = {\rm Diag}\{1, e^{i\phi_1}, e^{i\phi_2} \}$ is the
Majorana phase matrix. If the unknown phases are taken to be 
$\phi_1 =\phi_2 =90^{\circ}$, for example, one arrives at
\begin{equation}
\left | \langle m_\nu \rangle \right | \; = \; 
\frac{2}{\sqrt{3}} \sqrt{\frac{m_e}{m_\mu}}
~ m_i \; ,
\end{equation}
in which $m_i \sim 1 - 2$ eV (for $i=1,2,3$) as required by the 
near degeneracy of three neutrinos in our model
to accommodate the hot dark matter of the universe. 
Obviously $|\langle m_\nu \rangle | 
\approx 0.08 m_i \leq 0.2$ eV, the
latest bound of the $(\beta\beta)_{0\nu}$ decay given in (2.16).

\underline{\it Small-versus-large mixing pattern} ~
To generate a small mixing angle for the solar neutrino oscillation
and a large one for the atmospheric neutrino oscillation,
we turn to a different symmetry-breaking scenario for 
the charged lepton mass matrix $M^{(0)}_l$ and the neutrino
mass matrix $M^{(0)}_\nu$. For simplicity we follow the same
symmetry-breaking chain $M^{(0)}_l \rightarrow M^{(1)}_l \rightarrow M^{(2)}_l$ 
as discussed above (one of the six complex perturbations $\Delta M^{(2)}_l$
listed in Table 5.2 can be taken), but introduce the off-diagonal perturbations
to $M^{(0)}_\nu$. To ensure the ``maximal calculability'' for the 
neutrino mass matrix, we require a special form of 
the overall perturbative matrix \cite{Xing98}: 
it consists solely of two unknown small 
parameters in addition to the scale parameter $C_\nu$,  
and can be diagonalized by a {\it constant} orthogonal
transformation independent of the neutrino masses. Then we 
are left with only three perturbative patterns 
satisfying these strong requirements:
\begin{eqnarray}
M^{(2)}_\nu & = & C_\nu \left [ \left ( \matrix{
1       & 0     & 0 \cr
0       & 1     & 0 \cr
0       & 0     & 1 \cr} \right ) ~ + ~
\left ( \matrix{
0       & \varepsilon_\nu       & 0 \cr
\varepsilon_\nu & 0     & 0 \cr
0       & 0     & \delta_\nu \cr} \right ) \right ] \; , \\
M^{(2)}_\nu & = & C_\nu \left [ \left ( \matrix{
1       & 0     & 0 \cr
0       & 1     & 0 \cr
0       & 0     & 1 \cr} \right ) ~ + ~
\left ( \matrix{
\delta_\nu      & 0     & 0 \cr
0       & 0     & \varepsilon_\nu \cr
0       & \varepsilon_\nu       & 0 \cr} \right ) \right ] \; , \\
M^{(2)}_\nu & = & C_\nu \left [ \left ( \matrix{
1       & 0     & 0 \cr
0       & 1     & 0 \cr
0       & 0     & 1 \cr} \right ) ~ + ~
\left ( \matrix{
0       & 0     & \varepsilon_\nu \cr
0       & \delta_\nu    & 0 \cr
\varepsilon_\nu & 0     & 0 \cr} \right ) \right ] \; .
\end{eqnarray}
Three forms of $M^{(2)}_\nu$ 
can be diagonalized by three Euler rotation matrices $R_{12}(\theta_{12})$,
$R_{23}(\theta_{23})$ and $R_{31}(\theta_{31})$ (see also 
(3.25) for illustration), respectively, with a universal rotation angle 
$\theta_{ij} = 45^{\circ}$.
In Ref. \cite{Tanimoto98} the perturbative pattern (5.16) was discussed,
but $CP$ violation was not taken into account. 
For each scenario of $M^{(2)}_\nu$, versus the given form of $M^{(2)}_l$, 
the resultant flavor mixing matrix reads as
$V' = V R_{ij}$, where $V$ has been obtained in (5.7).
It is easy to check that only the pattern of $M^{(2)}_\nu$ in (5.16)
can give rise to a sufficiently large mixing angle for the atmospheric
neutrino oscillation. In this case the flavor mixing matrix
$V' = V R_{12}$ with $\theta_{12} = 45^{\circ}$ takes the following form:
\begin{equation}
V' \; =\; \left ( \matrix{
1       & 0     & 0 \cr
0       & \frac{1}{\sqrt{3}}    & \frac{-2}{\sqrt{6}} \cr
0       & \frac{2}{\sqrt{6}}    & \frac{1}{\sqrt{3}} \cr} \right )
~ + ~ {\rm i} ~ \xi^{~}_{V'} ~ \sqrt{\frac{m_e}{m_\mu}} 
~ + ~ \zeta^{~}_{V'} ~ \frac{m_\mu}{m_\tau} \; ,
\end{equation}
where
\begin{eqnarray}
\xi^{~}_{V'} & = & \left ( \matrix{
0       & \frac{1}{\sqrt{3}}    & \frac{-2}{\sqrt{6}} \cr
1       & 0     & 0 \cr
0       & 0     & 0 \cr} \right ) \; , \nonumber \\
\zeta^{~}_{V'} & = & \left ( \matrix{
0       & 0     & 0 \cr
0       & \frac{1}{\sqrt{3}}    & \frac{1}{\sqrt{6}} \cr
0       & \frac{-1}{\sqrt{6}}   & \frac{1}{\sqrt{3}} \cr} \right ) \; .
\end{eqnarray}
Parametrizing $V'$ in terms of three Euler angles and one $CP$-violating
phase, as already done in (5.9), we arrive numerically at
\begin{equation}
\theta_l \approx 4^{\circ} \; , ~~~~~
\theta_\nu \approx 0^{\circ} \; , ~~~~~
\theta \approx 52^{\circ} \; , ~~~~~
\phi \approx 0^{\circ} \; .
\end{equation}
Note that $\phi \approx 0^{\circ}$ is consistent with $\theta_\nu \approx 0$,
i.e., the $CP$-violating phase $\phi$ can always be rotated away in
the case that one flavor mixing angle vanishes. Therefore $CP$ violation 
is absent, up to the given accuracy of $V'$. The smallness of $|V'_{e3}|$
ensures that the atmospheric and solar neutrino oscillations are
approximately decoupled.

Let us calculate the mixing factors of solar and atmospheric neutrino
oscillations with the help of (5.19). We obtain
\begin{eqnarray}
\sin^2 2\theta_{\rm sun} & = & \frac{4}{3} \frac{m_e}{m_\mu} \; ,
\nonumber \\
\sin^2 2\theta_{\rm atm} & = & \frac{8}{9} 
\left (1 + \frac{m_\mu}{m_\tau} \right ) \; 
\end{eqnarray}
to a good degree of accuracy. 
Numerically $\sin^2 2\theta_{\rm sun} \approx 0.0064$ and
$\sin^2 2\theta_{\rm atm} \approx 0.94$. Thus this
flavor mixing scenario favors the small-angle MSW
solution to the solar neutrino problem. Its consequence
on the atmospheric neutrino oscillations is the same
as the nearly bi-maximal mixing scenario discussed above.

If neutrinos are of the Majorana type, then
the smallness of $|\theta_l|$ and $|\theta_\nu|$ in this
small-versus-large mixing scenario implies that the effective mass factor 
of the neutrinoless double beta decay (i.e., 
$\langle m_\nu \rangle$) is dominated by $m_1$. A strong constraint 
turns out to be $m_1 \leq 0.2$ eV, in view of current data on the 
$(\beta\beta)_{0\nu}$ decay. The sum of three neutrino masses has an upper
bound of 0.6 eV, too small to account for the hot dark matter of the universe. 

\underline{\it Bi-maximal mixing pattern} ~
Now let us take a look at the exactly ``bi-maximal'' mixing scenario 
of three neutrinos \cite{Barger98}.
The relevant flavor mixing matrix, similar to 
the leading term of $V$ in (5.7), reads as follows:
\begin{equation}
V'' \; =\; \left (\matrix{
\frac{1}{\sqrt{2}}      & -\frac{1}{\sqrt{2}}   & 0 \cr
\frac{1}{2}     & \frac{1}{2}   & -\frac{1}{\sqrt{2}} \cr
\frac{1}{2}     & \frac{1}{2}   & \frac{1}{\sqrt{2}} \cr }
\right ) \; . 
\end{equation}
This flavor mixing pattern is independent of charged lepton
and neutrino masses, and it leads exactly to
$\sin^2 2\theta_{\rm atm} = \sin^2 2 \theta_{\rm sun} =1$
for atmospheric and solar neutrino oscillations. Therefore it favors the
``Just-So'' solution (perhaps also the large-angle MSW solution) to
the solar neutrino problem. We find 
that $V''$ can be derived from the following charged lepton
and neutrino mass matrices \cite{FX98}:
\begin{eqnarray}
M_l & = & \frac{C^{~}_l}{2} \left [ 
\left ( \matrix{
0       & 0     & 0 \cr
0       & 1     & 1 \cr
0       & 1     & 1 \cr } \right ) + \left ( \matrix{
\delta_l        & ~ 0   & 0 \cr
0       & ~ 0   & \varepsilon^{~}_l \cr
0       & ~ \varepsilon^{~}_l       & 0 \cr } \right ) \right ] \; , 
\nonumber \\
M_\nu & = & C_\nu \left [ 
\left ( \matrix{
1       & 0     & 0 \cr
0       & 1     & 0 \cr
0       & 0     & 1 \cr } \right ) + \left ( \matrix{
0       & \varepsilon_\nu       & 0 \cr
\varepsilon_\nu & 0     & 0 \cr
0       & 0     & \delta_\nu \cr } \right ) \right ] \; , 
\end{eqnarray}
where $|\delta_{l,\nu}| \ll 1$ and $|\varepsilon_{l,\nu}|
\ll 1$. In comparison with the democratic mass matrix
$M^{(0)}_l$ given in (5.4), which is invariant
under an $\rm S(3)_L \times S(3)_R$ transformation, the
matrix $M_l$ in the limit $\delta_l = \varepsilon^{~}_l =0$
only has an $\rm S(2)_L \times S(2)_R$ symmetry. However $M_\nu$ 
in the limit $\delta_\nu = \varepsilon_\nu
=0$ takes the same form as $M^{(0)}_\nu$ in (5.4), 
which displays an $\rm S(3)$ symmetry.
The off-diagonal perturbation of
$M_l$ allows the masses of three charged leptons to be
hierarchical:
\begin{equation}
\left \{ m_e \; , \; m_\mu \; , \; m_\tau \right \}
\; =\; \frac{C^{~}_l}{2} \left \{ |\delta_l| \; , \;
|\varepsilon^{~}_l| \; , \; 2 + \varepsilon^{~}_l \right \} \; . 
\end{equation}
One finds $C^{~}_l = m_\mu + m_\tau \approx 1.88$ GeV,
$|\varepsilon^{~}_l| = 2m_\mu/(m_\mu + m_\tau) \approx 
0.11$ and $|\delta_l| = 2m_e/(m_\mu + m_\tau) 
\approx 5.4 \times 10^{-4}$. The off-diagonal perturbation
of $M_\nu$ makes the three neutrino masses non-degenerate:
\begin{equation}
\left \{ m_1, m_2, m_3 \right \} 
\; =\; C_\nu \left \{ 1+\varepsilon_\nu, 
1 - \varepsilon_\nu, 1 + \delta_\nu \right \} \; .
\end{equation}
For solar and atmospheric neutrino oscillations, one can take the
mass-squared differences as in (2.21).
We then arrive at $|\varepsilon_\nu|/|\delta_\nu| \approx 
\Delta m^2_{\rm sun}/(2 \Delta m^2_{\rm atm})$, 
of ${\cal O}(10^{-7})$ for the vacuum-oscillation solution
and of ${\cal O}(10^{-3})$ to ${\cal O}(10^{-2})$ for
the MSW solution to the solar neutrino problem.
The diagonalization of $M_l$ and $M_\nu$
leads to the flavor mixing matrix $V''$.
In Ref. \cite{Barger98} a different neutrino
mass matrix has {\it reversely} been derived from the
given $V''$ in the flavor
basis where the charged lepton mass matrix is diagonal.
The emergence of the bi-maximal mixing pattern from
$M_l$ and $M_\nu$ in (5.24) is obviously similar to 
that of the nearly bi-maximal mixing pattern obtained
in (5.7), although the former is irrelevant to lepton masses
and $CP$ violation.

Note that three mixing angles of $V''$ are given as
$\theta_l =0$, $\theta_\nu = 45^{\circ}$ and $\theta =45^{\circ}$.
The vanishing $\theta_l$ leads to vanishing 
probabilities for $\nu_e \rightarrow \nu_\mu$ and
$\nu_e \rightarrow \nu_\tau$ transitions in the long-baseline
neutrino experiments at the atmospheric scale. This feature 
distinguishes the bi-maximal mixing ansatz 
from the nearly bi-maximal mixing ansatz.

If three neutrinos are of the Majorana type, then the
near degeneracy of their masses implies that 
$\langle m_\nu \rangle \approx m_i$ for the
neutrinoless double beta decay in the bi-maximal 
mixing scenario. Therefore
$m_i < 0.2$ eV. The sum of three neutrino masses is
insufficient to account for the hot dark matter of the universe.

Finally it is worth mentioning the so-called tri-maximal 
neutrino mixing scenario, in which the lepton flavor mixing matrix
takes the form \cite{Cabibbo78}
\begin{equation}
V''' \; = \; \frac{1}{\sqrt{3}} \left ( \matrix{
1       & 1     & 1 \cr
1       & \omega        & \omega^2 \cr
1       & \omega^2      & \omega \cr} \right ) \; ,
\end{equation}
with $\omega$ a complex cube-root of unity. This mixing 
matrix could be derived from the permutation symmetry of
the neutrino mass matrix in the flavor basis where the charged lepton
mass matrix is diagonal. A special feature of $V'''$ is that
the rephasing-invariant measure of $CP$ violation takes 
its maximal value, i.e., ${\cal J}^{~}_l = 1/(6\sqrt{3})$.
In such a threefold maximal mixing scenario all survival
and appearance probabilities of neutrino oscillations are
universal, independent of the specific neutrino flavors.
It is obvious that the plain tri-maximal neutrino mixing pattern
is in conflict with current experimental data \cite{SK,CHOOZ}. If
possible terrestrial matter effects are taken into account \cite{Scott}, 
there is a small chance that it might be reconciled with
the experimental data.

In summary the idea of lepton flavor democracy and its
explicit breaking can lead to phenomenologically favored
flavor mixing patterns, yielding satisfactory interpretations
of current experimental data on solar and atmospheric
neutrino oscillations. Some theoretical attempts have recently been
made towards deeper understanding of this interesting
flavor symmetry (see, e.g., Refs. \cite{Mohapatra,TH}). 

\subsection{Seesaw-invariant texture of lepton mass matrices}

If neutrinos are massive and lepton flavors are mixed, one
may wonder whether there exist some similarities or relations between 
the mass and mixing textures of leptons and quarks. A basic 
question in understanding the fermion mass spectra 
is why the masses of three active neutrinos are so small
compared with those of charged leptons or quarks. For the time
being this question remains open, although a number of theoretical
speculations towards a definite answer have been made. 
The smallness of chiral (left-handed) neutrino masses is 
perhaps attributed to the fact that they are electrically neutral fermions,
or more exactly, to the Majorana feature of neutrino fields.
In this picture the left-handed Majorana neutrinos can
naturally acquire their masses through an effective 
seesaw mechanism of the form \cite{Seesaw}
\footnote{For simplicity the notations of Majorana neutrino
mass matrices used here are different from those used in (2.25). 
Obviously $M_\nu = M^{\rm M}_\nu$ and $M_{\rm R} = \tilde{M}^{\rm M}_\nu$.}:
\begin{equation}
M_\nu \; =\; (M^{\rm D}_\nu )^{\rm T} (M_{\rm R} )^{-1} (M^{\rm D}_\nu ) \; ,
\end{equation}
where $M^{\rm D}_\nu$ and $M_{\rm R}$ denote Dirac and
right-handed Majorana neutrino
mass matrices, respectively. 
In most grand unified theories one takes 
$[M^{\rm D}_\nu, M_{\rm u}] = 0$, 
where $M_{\rm u}$ stands for the mass matrix
of the up-type quark sector. In some left-right symmetric
models one takes $[M^{\rm D}_\nu, M_l]=0$, where $M_l$
is the charged lepton mass matrix. The mass matrix of
the heavy right-handed neutrinos ($M_{\rm R}$) is practically
unknown in almost all reasonable extensions of the standard model.
For this reason a specific texture of $M_{\rm R}$, in addition to
that of $M^{\rm D}_\nu$, has to be assumed in order to 
calculate masses of the light left-handed Majorana neutrinos.

A phenomenologically favored texture of quark mass matrices
has been presented in (4.23). The texture zeros
in the (1,1), (1,3) and (3,1) positions of $M_{\rm u}$ and
$M_{\rm d}$ could be
the consequence of an underlying flavor symmetry. 
In the spirit of
lepton-quark similarity we prescribe the same texture for the 
charged lepton and Dirac neutrino mass matrices:
\begin{eqnarray}
M_{l~} & = & \left (\matrix{
{\bf 0} & ~ D_l & ~ {\bf 0} \cr
D^*_l   & ~ C_l & ~ B_l \cr
{\bf 0} & ~ B^*_l               & ~ A_l \cr} \right ) \;\; , \nonumber \\
\nonumber \\
M^{\rm D}_\nu & = & \left (\matrix{
{\bf 0} & D^{\rm D}_\nu & {\bf 0} \cr
D^{\rm D}_\nu   & C^{\rm D}_\nu & B^{\rm D}_\nu \cr
{\bf 0} & B^{\rm D}_\nu & A^{\rm D}_\nu \cr} \right ) \;\; , 
\end{eqnarray}
in which $|D_l| \ll |C_l| \sim |B_l| \ll |A_l|$ holds. 
Without loss of generality, we have
taken $M^{\rm D}_\nu$ to be real.
The phase differences between $M_l$ and $M^{\rm D}_\nu$ 
are therefore denoted as $\varphi^{~}_1 \equiv \arg (D_l)$
and $\varphi^{~}_2 \equiv \arg (B_l)$. They
are the source of leptonic $CP$ violation in neutrino oscillations.
Note that $M_l$ can be decomposed into 
$M_l = P^{\dagger}_l \overline{M}_l P_l$, where
$\overline{M}_l$ is a real symmetric matrix with 
the same texture zeros as $M_l$, and
$P_l = {\rm Diag} \{ 1, e^{{\rm i}\varphi_1}, 
e^{{\rm i}(\varphi_1 + \varphi_2)} \}$
is a Dirac-type phase matrix.

We conjecture that $M_{\rm R}$ could have the same texture as
that of $M^{\rm D}_\nu$ and $M_l$, i.e.,
\begin{equation}
M_{\rm R} \; = \; \left (\matrix{
{\bf 0} & D_{\rm R}     & {\bf 0} \cr
D_{\rm R}       & C_{\rm R}     & B_{\rm R} \cr
{\bf 0} & B_{\rm R}             & A_{\rm R} \cr} \right ) \;\; , 
\end{equation}
which might follow from a universal 
flavor symmetry hidden in the more fundamental theory of
fermion mass generation. The matrix elements of
$M_{\rm R}$ are in general complex. It is worth mentioning that 
the universal texture of $M_{\rm R}$, $M^{\rm D}_\nu$ (or $M_{\rm u}$) 
and $M_l$ can theoretically be obtained in the context of SO(10) grand 
unified theories \cite{Fr75,Matsuda}. Instead of exploring
the theoretical details of these mass matrices, we proceed 
to calculate the left-handed neutrino mass matrix
$M_\nu$ via the seesaw mechanism.

Given the textures of
$M_{\rm R}$ and $M^{\rm D}_\nu$, it is straightforward to
show that $M_\nu$ has the same texture:
\begin{equation}
M_\nu \; = \; \left (\matrix{
{\bf 0} & D_\nu & {\bf 0} \cr
D_\nu   & C_\nu & B_\nu \cr
{\bf 0} & B_\nu         & A_\nu \cr} \right ) \;\; , 
\end{equation}
in which the four matrix elements are given by 
\begin{eqnarray}
A_\nu & = & \frac{(A^{\rm D}_\nu)^2}{A_{\rm R}} \;\; ,
\nonumber \\
\nonumber \\
B_\nu & = & \frac{A^{\rm D}_\nu B^{\rm D}_\nu}{A_{\rm R}}
~ + ~ \frac{B^{\rm D}_\nu D^{\rm D}_\nu}{D_{\rm R}}
~ - ~ \frac{A^{\rm D}_\nu D^{\rm D}_\nu B_{\rm R}}{A_{\rm R} D_{\rm R}} \;\; ,
\nonumber \\
C_\nu & = & \frac{(B^{\rm D}_\nu)^2}{A_{\rm R}}
~ + ~ \frac{2 C^{\rm D}_\nu D^{\rm D}_\nu}{D_{\rm R}}
~ - ~ \frac{(D^{\rm D}_\nu)^2 C_{\rm R}}{D^2_{\rm R}}
~ - ~ \frac{2B^{\rm D}_\nu D^{\rm D}_\nu B_{\rm R}}{A_{\rm R} D_{\rm R}}
~ + ~ \frac{(D^{\rm D}_\nu)^2 B^2_{\rm R}}{A_{\rm R} D^2_{\rm R}} \;\; ,
\nonumber \\
\nonumber \\
D_\nu & = & \frac{(D^{\rm D}_\nu)^2}{D_{\rm R}} \;\; .
\end{eqnarray}
Note that the texture zeros of 
$M^{\rm D}_\nu$ and $M_{\rm R}$ manifest themselves again in
$M_\nu$, as a consequence of the seesaw mechanism. We refer this 
special texture
of Dirac and Majorana neutrino mass matrices to the
{\it seesaw-invariant} texture. As the matrix elements of
$M_{\rm R}$ represent some kinds of superhigh energy scales \cite{Seesaw}
and those of $M^{\rm D}_\nu$ amount to the mass scales of
up-type quark masses ($[M^{\rm D}_\nu, M_{\rm u}]=0$)
or charged lepton masses ($[M^{\rm D}_\nu, M_l]=0$), 
a significant suppression
of the matrix elements of $M_\nu$ is transparent in (5.32).
This provides a natural interpretation of the smallness of
the left-handed neutrino masses. 

In practice, however, useful predictions for light neutrino masses
and lepton flavor mixing angles are prevented due to the unknown parameters 
of $M_{\rm R}$. A phenomenologically acceptable approach is to
calculate the flavor mixing matrix starting directly from $M_l$ and
$M_\nu$, regardless of the details of $M_{\rm R}$. Subsequently
we follow this strategy to examine how a neutrino mixing
pattern with two large mixing angles
can naturally emerge in the present model. For simplicity and
instruction we assume that the symmetric neutrino mass matrix
$M_\nu$ is real, therefore the only $CP$-violating phase 
existing in the lepton flavor mixing matrix is of the Dirac type
and comes from the complex phases of $M_l$.

The real symmetric mass matrices $\overline{M}_l$ and
$M_\nu$ can be diagonalized by two orthogonal transformations:
\begin{eqnarray}
O^{\rm T}_{l} \overline{M}_{l} O_{l} ~ & = & 
{\rm Diag} \{ -m_e, ~ m_\mu, ~ m_\tau \} \; , 
\nonumber \\
O^{\rm T}_\nu M_\nu O_\nu & = & 
{\rm Diag} \{ -m_1, ~ m_2, ~ m_3 \} \; .
\end{eqnarray}
In general the four parameters of $\overline{M}_l$ (or $M_\nu$)
cannot be uniquely determined by inputting
the measured mass eigenvalues. To obtain an analytically
simple solution for $O_l$ or $O_\nu$, we typically
specify $C_l = m_\mu$ and $C_\nu = m_2$ (the similar 
choices $C_{\rm u} = m_c$ and $C_{\rm d} = m_s$
are favored for the quark mass matrices $M_{\rm u}$ and
$M_{\rm d}$ in (4.23) to reproduce the experimental values
of quark flavor mixing angles \cite{FX99,4zero}).
Then the other parameters of $\overline{M}_l$ and $M_\mu$ 
can be determined in terms of the charged lepton and 
neutrino masses, respectively:
\begin{eqnarray}
A_l ~ & = & m_\tau - m_e \;\; , 
\nonumber \\
|B_l| & = & \left [ \frac{m_e m_\tau (m_\tau - m_e - m_\mu)}
{m_\tau - m_e} \right ]^{1/2} \;\; ,
\nonumber \\
|D_l| & = & \left ( \frac{m_e m_\mu m_\tau}{m_\tau - m_e} \right )^{1/2} \;\; ;
\end{eqnarray}
and
\begin{eqnarray}
A_\nu & = & m_3 - m_1 \;\; , 
\nonumber \\
B_\nu & = & \left [ \frac{m_1 m_3 (m_3 - m_1 - m_2)}
{m_3 - m_1} \right ]^{1/2} \;\; ,
\nonumber \\
D_\nu & = & \left ( \frac{m_1 m_2 m_3}{m_3 - m_1} \right )^{1/2} \;\; .
\end{eqnarray}
In this special but interesting case, we obtain the matrix elements of 
$O_l$ and $O_\nu$ in terms of 
the mass ratios $x^{~}_l \equiv m_e/m_\mu$, $y^{~}_l \equiv m_\mu/m_\tau$ and
$x_\nu \equiv m_1/m_2$ and $y_\nu \equiv m_2/m_3$
as follows (the subscripts ``$l$'' and ``$\nu$'' are neglected for
simplicity):
\begin{eqnarray}
O_{11} & = & + \left [ \frac{1}{(1+x)(1-x^2y^2)} \right ]^{1/2} \;\; , 
\nonumber \\
O_{12} & = & + \left [ \frac{x(1-y-xy)}{(1+x)(1-y)(1-xy)} \right ]^{1/2} \;\; ,
\nonumber \\
O_{13} & = & + \left [ \frac{x^2y^3}{(1-y)(1-x^2y^2)} \right ]^{1/2} \;\; , 
\nonumber \\
O_{21} & = & - \left [ \frac{x}{(1+x)(1+xy)} \right ]^{1/2} \;\; ,
\nonumber \\
O_{22} & = & + \left [ \frac{1-y-xy}{(1+x)(1-y)} \right ]^{1/2} \;\; , 
\nonumber \\
O_{23} & = & + \left [ \frac{xy}{(1-y)(1+xy)} \right ]^{1/2} \;\; ,
\nonumber \\
O_{31} & = & + \left [ \frac{x^2y(1-y-xy)}{(1+x)(1-x^2y^2)} \right ]^{1/2} \;\; , 
\nonumber \\
O_{32} & = & - \left [ \frac{xy}{(1+x)(1-y)(1-xy)} \right ]^{1/2} \;\; , 
\nonumber \\
O_{33} & = & + \left [ \frac{1-y-xy}{(1-y)(1-x^2y^2)} \right ]^{1/2} \; \; .
\end{eqnarray}
It is easy to check that $O_l$ is very close to the unity matrix,
because of the smallness of $x^{~}_l$ ($=0.00484$) and
$y^{~}_l$ ($=0.0594$) \cite{PDG}. Instead $O_\nu$ can significantly deviate
from the unity matrix, if three neutrino masses do not have a
strong hierarchy.

The leptonic flavor mixing matrix $V\equiv O^{\rm T}_l P_l O_\nu$
links the neutrino mass eigenstates $(\nu_1, \nu_2, \nu_3)$ to
the neutrino flavor eigenstates $(\nu_e, \nu_\mu, \nu_\tau)$.
Assuming the phase differences between $M_l$ and $M^{\rm D}_\nu$
to have the special values $\varphi^{~}_1 =\pi/2$ and
$\varphi^{~}_2 =\pi$ (a case favored for the quark mass
matrices in (4.23) to reproduce the measured flavor mixing
angles and $CP$ violation \cite{FX99,FX95}), we obtain
the magnitudes of $|V_{i\alpha}|^2$ as follows:
\begin{equation}
\left |V_{\alpha i} \right |^2 \; =\; \left (O^l_{1\alpha} O^\nu_{1i} \right )^2
+ \left ( O^l_{2\alpha} O^\nu_{2i} - O^l_{3\alpha} O^\nu_{3i} 
\right )^2 \;\; ,
\end{equation}
where $\alpha$ runs over $(e, \mu, \tau)$ and
$i$ over $(1, 2, 3)$.
The rephasing-invariant measure of $CP$ violation, ${\cal J}_l$, can
then be calculated by use of the formula given in (3.12).
As one can see from (5.36), $|V_{i\alpha}|$ and ${\cal J}_l$ are
functions of the neutrino mass ratios $x_\nu$ and $y_\nu$. 

Without loss of much generality we take $x_\nu \leq 1$ and
$y_\nu \leq 1$. Taking (2.21) into account, we obtain
the ratio of $\Delta m^2_{\rm sun}$ to
$\Delta m^2_{\rm atm}$ in terms of $x_\nu$ and $y_\nu$:
\begin{equation}
R \;\; \equiv \;\; \frac{\Delta m^2_{\rm sun}}{\Delta m^2_{\rm atm}}
\;\; =\;\; y^2_\nu ~ \frac{1-x^2_\nu}{1-y^2_\nu} \;\; \ll \;\; 1 \; \; .
\end{equation}
Note that $R\ll 1$ is imposed by current atmospheric
and solar neutrino oscillation data. This condition is satisfied if
both $x_\nu \ll 1$ and $y_\nu \ll 1$ hold, or if $x_\nu \approx 1$
and $y_\nu < x_\nu$ hold for proper values of $y_\nu$. For 
vacuum oscillation and large-angle MSW solutions 
to the solar neutrino problem, one obtains
$R_{\rm VO} \sim 10^{-7}$ and $R_{\rm MSW} \sim 10^{-2}$
respectively
\footnote{Here we do not take the small-angle MSW solution
into account. Indeed a numerical study shows that this
solution is not compatible with 
the model under discussion.}.
The mixing factors of solar ($\nu_e \rightarrow \nu_e$ disappearance)
and atmospheric ($\nu_\mu \rightarrow \nu_\mu$ disappearance)
neutrino oscillations
can be given, under the approximate decoupling condition
$|V_{e3}|^2 \ll 1$, as follows:
\begin{eqnarray}
\sin^2 2\theta_{\rm sun} & = & 4 ~ |V_{e1}|^2 ~ |V_{e2}|^2 \; ,
\nonumber \\
\sin^2 2\theta_{\rm atm} & = & 4 ~ |V_{\mu 3}|^2 \left (1- |V_{\mu 3}|^2 \right ) \; .
\end{eqnarray}
Clearly $R$, $|V_{e3}|^2$, $\sin^2 2\theta_{\rm sun}$ and
$\sin^2 2\theta_{\rm atm}$ are all functions of two independent
neutrino mass ratios. We are therefore able to
find out the allowed parameter space for $(x_\nu, y_\nu)$ by
fitting current experimental data on $R$, $|V_{e3}|^2$, and so on.
\begin{figure}[t]
\setlength{\unitlength}{0.240900pt}
\ifx\plotpoint\undefined\newsavebox{\plotpoint}\fi
\sbox{\plotpoint}{\rule[-0.200pt]{0.400pt}{0.400pt}}%
\begin{picture}(1200,1080)(-230,0)
\font\gnuplot=cmr10 at 10pt
\gnuplot
\sbox{\plotpoint}{\rule[-0.200pt]{0.400pt}{0.400pt}}%
\put(202.0,164.0){\rule[-0.200pt]{4.818pt}{0.400pt}}
\put(182,164){\makebox(0,0)[r]{0}}
\put(1159.0,164.0){\rule[-0.200pt]{4.818pt}{0.400pt}}
\put(202.0,339.0){\rule[-0.200pt]{4.818pt}{0.400pt}}
\put(182,339){\makebox(0,0)[r]{0.2}}
\put(1159.0,339.0){\rule[-0.200pt]{4.818pt}{0.400pt}}
\put(202.0,514.0){\rule[-0.200pt]{4.818pt}{0.400pt}}
\put(182,514){\makebox(0,0)[r]{0.4}}
\put(1159.0,514.0){\rule[-0.200pt]{4.818pt}{0.400pt}}
\put(202.0,690.0){\rule[-0.200pt]{4.818pt}{0.400pt}}
\put(182,690){\makebox(0,0)[r]{0.6}}
\put(1159.0,690.0){\rule[-0.200pt]{4.818pt}{0.400pt}}
\put(202.0,865.0){\rule[-0.200pt]{4.818pt}{0.400pt}}
\put(182,865){\makebox(0,0)[r]{0.8}}
\put(1159.0,865.0){\rule[-0.200pt]{4.818pt}{0.400pt}}
\put(202.0,1040.0){\rule[-0.200pt]{4.818pt}{0.400pt}}
\put(182,1040){\makebox(0,0)[r]{1}}
\put(1159.0,1040.0){\rule[-0.200pt]{4.818pt}{0.400pt}}
\put(202.0,164.0){\rule[-0.200pt]{0.400pt}{4.818pt}}
\put(202,123){\makebox(0,0){0}}
\put(202.0,1020.0){\rule[-0.200pt]{0.400pt}{4.818pt}}
\put(397.0,164.0){\rule[-0.200pt]{0.400pt}{4.818pt}}
\put(397,123){\makebox(0,0){0.2}}
\put(397.0,1020.0){\rule[-0.200pt]{0.400pt}{4.818pt}}
\put(593.0,164.0){\rule[-0.200pt]{0.400pt}{4.818pt}}
\put(593,123){\makebox(0,0){0.4}}
\put(593.0,1020.0){\rule[-0.200pt]{0.400pt}{4.818pt}}
\put(788.0,164.0){\rule[-0.200pt]{0.400pt}{4.818pt}}
\put(788,123){\makebox(0,0){0.6}}
\put(788.0,1020.0){\rule[-0.200pt]{0.400pt}{4.818pt}}
\put(984.0,164.0){\rule[-0.200pt]{0.400pt}{4.818pt}}
\put(984,123){\makebox(0,0){0.8}}
\put(984.0,1020.0){\rule[-0.200pt]{0.400pt}{4.818pt}}
\put(1179.0,164.0){\rule[-0.200pt]{0.400pt}{4.818pt}}
\put(1179,123){\makebox(0,0){1}}
\put(1179.0,1020.0){\rule[-0.200pt]{0.400pt}{4.818pt}}
\put(202.0,164.0){\rule[-0.200pt]{235.359pt}{0.400pt}}
\put(1179.0,164.0){\rule[-0.200pt]{0.400pt}{211.028pt}}
\put(202.0,1040.0){\rule[-0.200pt]{235.359pt}{0.400pt}}
\put(55,602){\makebox(0,0){$y_\nu$}}
\put(690,50){\makebox(0,0){$x_\nu$}}

\put(1080,330){\vector(3,2){80}}
\put(1040,280){VO}

\put(1040,526){\vector(0,-1){50}}
\put(1040,565){\makebox(0,0){LA MSW}}

\put(260,600){\framebox(570,395)}

\put(300,935){\footnotesize $0.82 < \sin^2 2\theta_{\rm atm} \leq 1$}
\put(300,860){\footnotesize $0.65 \leq \sin^2 2\theta_{\rm sun} \leq 1$}
\put(300,785){\footnotesize $0 \leq |V_{e3}|^2 \leq 0.05$}
\put(300,710){\footnotesize $5\cdot 10^{-3} \leq R_{\rm\scriptsize MSW} \leq 5\cdot 10^{-2}$}
\put(300,635){\footnotesize $5\cdot 10^{-8} \leq R_{\rm\scriptsize VO} \leq 5\cdot 10^{-7}$}

\put(202.0,164.0){\rule[-0.200pt]{0.400pt}{211.028pt}}
\put(202,164){\rule{1pt}{1pt}}
\put(1179,1040){\rule{1pt}{1pt}}
\put(935,444){\rule{1pt}{1pt}}
\put(945,436){\rule{1pt}{1pt}}
\put(945,444){\rule{1pt}{1pt}}
\put(954,436){\rule{1pt}{1pt}}
\put(954,444){\rule{1pt}{1pt}}
\put(954,453){\rule{1pt}{1pt}}
\put(964,436){\rule{1pt}{1pt}}
\put(964,444){\rule{1pt}{1pt}}
\put(964,453){\rule{1pt}{1pt}}
\put(974,427){\rule{1pt}{1pt}}
\put(974,436){\rule{1pt}{1pt}}
\put(974,444){\rule{1pt}{1pt}}
\put(974,453){\rule{1pt}{1pt}}
\put(974,462){\rule{1pt}{1pt}}
\put(984,427){\rule{1pt}{1pt}}
\put(984,436){\rule{1pt}{1pt}}
\put(984,444){\rule{1pt}{1pt}}
\put(984,453){\rule{1pt}{1pt}}
\put(984,462){\rule{1pt}{1pt}}
\put(993,427){\rule{1pt}{1pt}}
\put(993,436){\rule{1pt}{1pt}}
\put(993,444){\rule{1pt}{1pt}}
\put(993,453){\rule{1pt}{1pt}}
\put(993,462){\rule{1pt}{1pt}}
\put(993,471){\rule{1pt}{1pt}}
\put(1003,427){\rule{1pt}{1pt}}
\put(1003,436){\rule{1pt}{1pt}}
\put(1003,444){\rule{1pt}{1pt}}
\put(1003,453){\rule{1pt}{1pt}}
\put(1003,462){\rule{1pt}{1pt}}
\put(1003,471){\rule{1pt}{1pt}}
\put(1013,427){\rule{1pt}{1pt}}
\put(1013,436){\rule{1pt}{1pt}}
\put(1013,444){\rule{1pt}{1pt}}
\put(1013,453){\rule{1pt}{1pt}}
\put(1013,462){\rule{1pt}{1pt}}
\put(1023,418){\rule{1pt}{1pt}}
\put(1023,427){\rule{1pt}{1pt}}
\put(1023,436){\rule{1pt}{1pt}}
\put(1023,444){\rule{1pt}{1pt}}
\put(1023,453){\rule{1pt}{1pt}}
\put(1023,462){\rule{1pt}{1pt}}
\put(1032,418){\rule{1pt}{1pt}}
\put(1032,427){\rule{1pt}{1pt}}
\put(1032,436){\rule{1pt}{1pt}}
\put(1032,444){\rule{1pt}{1pt}}
\put(1032,453){\rule{1pt}{1pt}}
\put(1032,462){\rule{1pt}{1pt}}
\put(1042,418){\rule{1pt}{1pt}}
\put(1042,427){\rule{1pt}{1pt}}
\put(1042,436){\rule{1pt}{1pt}}
\put(1042,444){\rule{1pt}{1pt}}
\put(1042,453){\rule{1pt}{1pt}}
\put(1042,462){\rule{1pt}{1pt}}
\put(1052,418){\rule{1pt}{1pt}}
\put(1052,427){\rule{1pt}{1pt}}
\put(1052,436){\rule{1pt}{1pt}}
\put(1052,444){\rule{1pt}{1pt}}
\put(1052,453){\rule{1pt}{1pt}}
\put(1062,409){\rule{1pt}{1pt}}
\put(1062,418){\rule{1pt}{1pt}}
\put(1062,427){\rule{1pt}{1pt}}
\put(1062,436){\rule{1pt}{1pt}}
\put(1062,444){\rule{1pt}{1pt}}
\put(1062,453){\rule{1pt}{1pt}}
\put(1072,409){\rule{1pt}{1pt}}
\put(1072,418){\rule{1pt}{1pt}}
\put(1072,427){\rule{1pt}{1pt}}
\put(1072,436){\rule{1pt}{1pt}}
\put(1072,444){\rule{1pt}{1pt}}
\put(1072,453){\rule{1pt}{1pt}}
\put(1081,409){\rule{1pt}{1pt}}
\put(1081,418){\rule{1pt}{1pt}}
\put(1081,427){\rule{1pt}{1pt}}
\put(1081,436){\rule{1pt}{1pt}}
\put(1081,444){\rule{1pt}{1pt}}
\put(1081,453){\rule{1pt}{1pt}}
\put(1091,409){\rule{1pt}{1pt}}
\put(1091,418){\rule{1pt}{1pt}}
\put(1091,427){\rule{1pt}{1pt}}
\put(1091,436){\rule{1pt}{1pt}}
\put(1091,444){\rule{1pt}{1pt}}
\put(1101,409){\rule{1pt}{1pt}}
\put(1101,418){\rule{1pt}{1pt}}
\put(1101,427){\rule{1pt}{1pt}}
\put(1101,436){\rule{1pt}{1pt}}
\put(1101,444){\rule{1pt}{1pt}}
\put(1111,401){\rule{1pt}{1pt}}
\put(1111,409){\rule{1pt}{1pt}}
\put(1111,418){\rule{1pt}{1pt}}
\put(1111,427){\rule{1pt}{1pt}}
\put(1111,436){\rule{1pt}{1pt}}
\put(1111,444){\rule{1pt}{1pt}}
\put(1120,401){\rule{1pt}{1pt}}
\put(1120,409){\rule{1pt}{1pt}}
\put(1120,418){\rule{1pt}{1pt}}
\put(1120,427){\rule{1pt}{1pt}}
\put(1120,436){\rule{1pt}{1pt}}
\put(1120,444){\rule{1pt}{1pt}}
\put(1130,401){\rule{1pt}{1pt}}
\put(1130,409){\rule{1pt}{1pt}}
\put(1130,418){\rule{1pt}{1pt}}
\put(1130,427){\rule{1pt}{1pt}}
\put(1130,436){\rule{1pt}{1pt}}
\put(1130,444){\rule{1pt}{1pt}}
\put(1140,401){\rule{1pt}{1pt}}
\put(1140,409){\rule{1pt}{1pt}}
\put(1140,418){\rule{1pt}{1pt}}
\put(1140,427){\rule{1pt}{1pt}}
\put(1140,436){\rule{1pt}{1pt}}
\put(1150,409){\rule{1pt}{1pt}}
\put(1150,418){\rule{1pt}{1pt}}
\put(1150,427){\rule{1pt}{1pt}}
\put(1150,436){\rule{1pt}{1pt}}
\put(1179,392){\raisebox{-.8pt}{\makebox(0,0){$\Diamond$}}}
\put(1179,401){\raisebox{-.8pt}{\makebox(0,0){$\Diamond$}}}
\put(1179,409){\raisebox{-.8pt}{\makebox(0,0){$\Diamond$}}}
\put(1179,418){\raisebox{-.8pt}{\makebox(0,0){$\Diamond$}}}
\put(1179,427){\raisebox{-.8pt}{\makebox(0,0){$\Diamond$}}}
\put(1179,436){\raisebox{-.8pt}{\makebox(0,0){$\Diamond$}}}
\end{picture}
\vspace{-0.32cm}
\caption{Allowed ranges of the neutrino mass ratios $x_\nu \equiv m_1/m_2$
and $y_\nu \equiv m_2/m_3$ constrained by current data,
where $R \equiv \Delta m^2_{\rm sun} / \Delta m^2_{\rm atm}$
taking different values for the large-angle MSW and 
vacuum oscillation solutions to the solar neutrino problem.}
\end{figure}
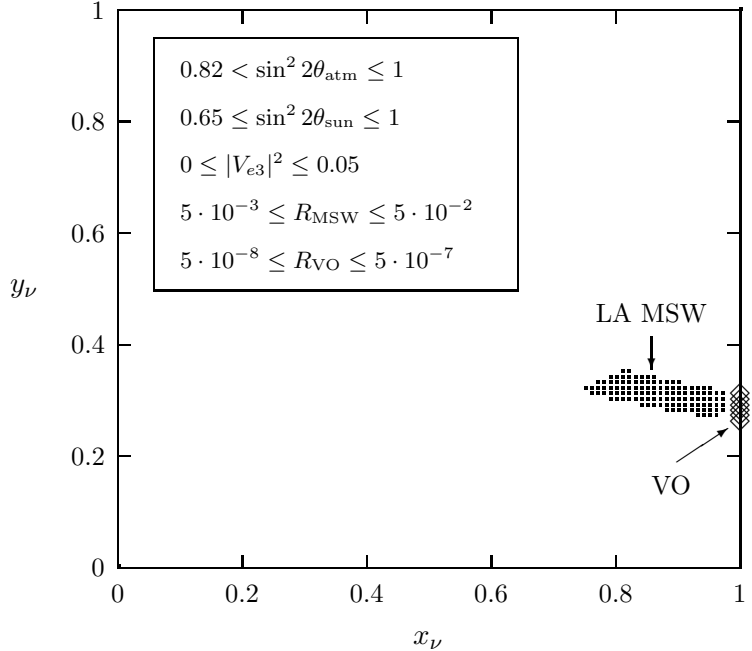
Explicitly we take $0.82 < \sin^2 2\theta_{\rm atm} \leq 1$,
$0.65 \leq \sin^2 2\theta_{\rm sun} \leq 1$, and
$|V_{e3}|^2 \leq 0.05$ \cite{Bahcall98}. In addition, 
$5\times 10^{-3} \leq R_{\rm MSW} \leq 5\times 10^{-2}$
and $5\times 10^{-8} \leq R_{\rm VO} \leq 5\times 10^{-7}$
are adopted. With these inputs (and with $x^{~}_l =0.00484$
and $y^{~}_l =0.0594$ \cite{PDG}), we carry out a 
numerical calculation based on the formulae given in 
(5.36) -- (5.39). The allowed ranges of $x_\nu$ and $y_\nu$
are then illustrated in Fig. 5.1, without any fine tuning.
We observe that $y_\nu \sim 0.3$ for either the large-angle
MSW solution or the vacuum oscillation solution. The former requires
$x_\nu \geq 0.8$, while the latter requires $x_\nu \approx 1$.
The magnitude of ${\cal J}_l$
is in the range $(7 - 8) \times 10^{-3}$, implying that 
a moderate strength of $CP$ violation is allowed by the model.

We see that this lepton mixing scenario can well 
accommodate the large-angle MSW oscillations of solar neutrinos.
It is also compatible with the vacuum oscillation solution
to the solar neutrino problem, although the allowed parameter
space is much smaller. In both cases the atmospheric neutrino
oscillation with a large mixing factor can be incorporated.
Note that the results $0.8 \leq x_\nu < 1$ and $y_\nu \sim 0.3$
indicate a nearly degenerate neutrino mass spectrum, in
contrast with the strong mass hierarchy of charged leptons.
This is not a surprise, because it is technically difficult to obtain
a lepton mixing pattern with two large mixing angles, 
when three neutrino masses are hierarchical. 
The near degeneracy of three neutrino masses in our
model implies that they may be the proper candidate for
the hot dark matter of the universe (in this case
$m_1 + m_2 + m_3 \sim 5$ to 10 eV is required). 
Of course such a neutrino mass spectrum may have no
conflict with the stringent upper limit on the effective mass factor 
$\langle m_\nu \rangle$ of the neutrinoless double beta decay \cite{KK99}. 
The reason is simply that cancellations can take place in the expression of
$\langle m_\nu \rangle$, in particular when $CP$-violating phases of the
Majorana type are introduced.

Once the neutrino masses are determined in the 
future neutrino experiments, one may specify the form of $M_\nu$
in a reverse way. The right-handed neutrino mass matrix
$M_{\rm R}$ can then be obtained via the seesaw mechanism, if 
$[M^{\rm D}_\nu, M_{\rm u}]=0$ or $[M^{\rm D}_\nu, M_l]=0$ is taken. 
In an explicit 
theoretical framework (e.g., the SO(10) grand unified model)
the energy scale, at which the universal texture of 
$M_{\rm u}$, $M_l$, $M_{\rm R}$ and $M_\nu$ holds, should be
specified; and the running effects of these mass matrices
between the superhigh and low energy scales should be taken
into account. To make an order-of-magnitude estimate of $M_{\rm R}$,
we take $M^{\rm D}_\nu = M_{\rm u}$ and
$m_1 \approx m_2 \approx m_3/3 \approx 1.5$ eV, and neglect all
possible scale-dependent running effects 
(see section 5.5 for a detailed discussion).
With the help of (5.32), (5.34) and (5.35), we obtain
\begin{eqnarray}
M_{\rm u} & \sim & \left ( \matrix{
{\bf 0} & ~ {\cal O} (10^{-2})    & {\bf 0} \cr
{\cal O} (10^{-2})      & ~ {\cal O} (1)    & {\cal O} (1) \cr
{\bf 0} & ~ {\cal O} (1)    & {\cal O} (10^2) \cr} \right ) \;\; ,
\nonumber \\ \nonumber \\
M_{\rm R} & \sim & \left ( \matrix{
~ {\bf 0}       & {\cal O} (10^6)    & ~ {\bf 0} \cr
~ {\cal O} (10^6)       & {\cal O} (10^8)    & ~ {\cal O} (10^{10}) \cr
~ {\bf 0}       & {\cal O} (10^{10})    & ~ {\cal O} (10^{13}) \cr} \right ) \;\;
\end{eqnarray}
in units of GeV; and 
\begin{equation}
M_\nu \; \sim \; \left ( \matrix{
0       & {\cal O} (1)  & 0 \cr
{\cal O} (1)    & {\cal O} (1)  & {\cal O} (1) \cr
0       & {\cal O} (1)  & {\cal O} (1) \cr} \right ) \;\; 
\end{equation}
in units of eV. This simple example illustrates that the tiny and
nearly degenerate masses of left-handed neutrinos can naturally
come forth from the hierarchical textures of $M^{\rm D}_\nu$ and
$M_{\rm R}$ through the seesaw mechanism.

Without the assumptions $C_l = m_\mu$ and
$C_\nu = m_2$ as well as $\varphi_1 =\pi/2$ and $\varphi_2 =\pi$, 
one can find out other possible
parameter spaces of $M_l$ and $M_\nu$ to accommodate the 
atmospheric neutrino oscillation data and allow the
small-angle MSW
solution, the large-angle MSW solution or the vacuum oscillation
solution to the solar neutrino problem.
A strong correlation between the mixing factors of
solar and atmospheric neutrino oscillations in the model
is generally expected, unlike those
(nearly) bi-maximal or tri-maximal neutrino mixing
scenarios in which the lepton mixing angles are
essentially algebraic numbers independent of the
neutrino mass ratios.

\subsection{An illustrative scheme of four-neutrino mixing}

In the context of three neutrino species, we have discussed
several instructive schemes of lepton mass matrices which
can interpret current experimental data on atmospheric and
solar neutrino oscillations. The LSND evidence
for $\nu_\mu \rightarrow \nu_e$ and 
$\bar{\nu}_\mu \rightarrow \bar{\nu}_e$ oscillations have
so far been put aside. To make a simultaneous interpretation
of solar, atmospheric and LSND neutrino oscillation data,
which have three different levels of neutrino mass-squared 
differences (i.e., $\Delta m^2_{\rm sun} \sim 10^{-10} -
10^{-5} ~ {\rm eV}^2$, $\Delta m^2_{\rm atm} \sim 
10^{-3} ~ {\rm eV}^2$ and $\Delta m^2_{\rm LSND} \sim 1
~ {\rm eV}^2$), one has to go beyond the conventional
three-neutrino framework by incorporating one sterile
light neutrino ($\nu_{\rm s}$ with mass $m_0$). As already
pointed out in section 2.2, a particularly favored 
four-neutrino mixing scenario is that the
$\nu_e \rightarrow \nu_{\rm s}$, $\nu_\mu \rightarrow \nu_\tau$
and $\nu_\mu \rightarrow \nu_e$ transitions are respectively
responsible for the solar, atmospheric and LSND neutrino
oscillations. Let us denote the mass eigenstates of 
$\nu_{\rm s}$, $\nu_e$, $\nu_\mu$ and $\nu_\tau$ as
$\nu_0$, $\nu_1$, $\nu_2$ and $\nu_3$, respectively.
Then the $4\times 4$ flavor mixing matrix $V$ is explicitly given as
\begin{equation}
\left ( \matrix{
\nu_{\rm s} \cr
\nu_e \cr
\nu_\mu \cr
\nu_\tau \cr} \right ) \; =\; \left ( \matrix{
V_{\rm s 0}     & V_{\rm s 1}   & V_{\rm s 2}   & V_{\rm s 3} \cr
V_{e 0} & V_{e 1}       & V_{e 2}       & V_{e 3} \cr
V_{\mu 0}       & V_{\mu 1}     & V_{\mu 2}     & V_{\mu 3} \cr
V_{\tau 0}      & V_{\tau 1}    & V_{\tau 2}    & V_{\tau 3} \cr}
\right ) \left ( \matrix{
\nu_0 \cr
\nu_1 \cr
\nu_2 \cr
\nu_3 \cr} \right ) \;\; .
\end{equation}
Note that the
four-neutrino mass spectrum is characterized by
\begin{eqnarray}
\Delta m^2_{\rm sun} & = & \Delta m^2_{10} \; \equiv \;
\left | m^2_1 - m^2_0 \right | \; , \nonumber \\
\Delta m^2_{\rm atm} & = & \Delta m^2_{32} \; \equiv \;
\left | m^2_3 - m^2_2 \right | \; , \nonumber \\
\Delta m^2_{\rm LSND} & = & \Delta m^2_{21} \; \equiv \;
\left | m^2_2 - m^2_1 \right | \; .
\end{eqnarray}
In contrast, the mass spectrum of charged leptons remains
unchanged. It should be noted that the four-neutrino mixing
phenomenon is not solely the property of a $4\times 4$ 
neutrino mass matrix. Indeed the realistic lepton flavor
mixing arises from the mismatch between diagonalizing the
charged lepton and neutrino mass matrices in a given flavor basis.
Enlarging the number of charged lepton families to four might
pose a problem, because the fourth charged lepton must be
sufficiently heavy (with mass $> 80$ GeV \cite{PDG}) and the
existence of such a heavy fermion requires credible
theoretical motivation and experimental evidence.
To avoid this ``fourth charged lepton'' problem and the 
related complexity \cite{3By4}, 
most authors have chosen to construct
the four-neutrino mixing scenarios in the flavor basis in
which the charged lepton mass matrix is diagonal. Such a
treatment is of course not perfect, but it may
serve as an acceptable approach to illustrate some features
of the four-neutrino mixing phenomenon. We therefore
follow the same strategy in the following.

For simplicity we assume $CP$ invariance. In the absence of
$CP$ violation a generic $4\times 4$ (Dirac or Majorana)
neutrino mass matrix $M_\nu$ consists of 10 arbitrary parameters.
Some of these entries receive strong restrictions from current
experimental data and may be negligibly small in magnitude.
In the flavor basis $(\nu_{\rm s}, \nu_e, \nu_\mu, \nu_\tau)$
one can diagonalize $M_\nu$ through an orthogonal transformation:
\begin{equation}
V^{\rm T} M_\nu V \; = \; {\rm Diag} \{ m_0, m_1, m_2, m_3 \} \; ,
\end{equation}
where the real lepton flavor mixing matrix $V$ 
links the neutrino mass eigenstates ($\nu_0$, $\nu_1$, $\nu_2$ and $\nu_3$)
to the neutrino flavor eigenstates ($\nu_{\rm s}$,
$\nu_e$, $\nu_\mu$ and $\nu_\tau$). Note that the
strength of active-sterile neutrino mixing is strictly 
constrained by astrophysics and cosmology. A careful analysis of
recent astrophysical data yields an upper limit 
$N^{\rm BBN}_\nu \leq 3.2$ (at the $95\%$ confidence level \cite{BBN})
for the effective number of neutrinos in Big-Bang nucleosynthesis,
although the matter remains quite controversial \cite{Giunti}. This upper bound
implies that atmospheric ($\nu_\mu \rightarrow \nu_\tau$)
and solar ($\nu_e \rightarrow \nu_{\rm s}$) neutrino
oscillations are essentially decoupled in the framework of
four-neutrino mixing. As a consequence, we arrive at the
following leading order result \cite{4N,Giunti99}
\begin{equation}
V \; \approx \; \left ( \matrix{
c_\odot    & -s_\odot    & 0      & 0 \cr
s_\odot    & c_\odot    & 0      & 0 \cr
0          & 0          & c_\bullet    & s_\bullet \cr
0          & 0          & -s_\bullet   & c_\bullet \cr}
\right ) \;\; ,
\end{equation}
where $s_\odot \equiv
\sin \theta_{\rm sun}$, $c_\bullet \equiv \cos\theta_{\rm atm}$,
and so on. To accommodate the LSND neutrino oscillation data 
(i.e., $\nu_\mu \rightarrow \nu_e$ and 
$\bar{\nu}_\mu \rightarrow \bar{\nu}_e$ transitions with
a small mixing factor $\sin^2 2\theta_{\rm LSND} \sim 10^{-3} - 10^{-2}$), 
a small admixture between $(\nu_e, \nu_\mu)$ and $(\nu_1, \nu_2)$
states needs to be introduced. We therefore rotate the
$\nu_1$-$\nu_2$ sector of $V$ by a very small angle 
$\epsilon \sim \theta^{~}_{\rm LSND}$. 
Such a rotation can be done from either the left-handed side of $V$
or the right-handed side of $V$.
The resultant 
flavor mixing matrix $V'$ deviates only slightly from $V$:
\begin{equation}
V' \; \approx \; \left ( \matrix{
~ c_\odot    & ~ -s_\odot    & -\epsilon s_\odot      & 0 \cr
~ s_\odot    & ~ c_\odot    & \epsilon c_\odot      & 0 \cr
~ 0          & ~ -\epsilon c_\bullet          & c_\bullet    & s_\bullet \cr
~ 0          & ~ \epsilon s_\bullet          & -s_\bullet   & c_\bullet \cr}
\right ) \;\; ,
\end{equation}
or 
\begin{equation}
V' \; \approx \; \left ( \matrix{
c_\odot    & -s_\odot    & 0      & 0 \cr
s_\odot   & c_\odot    & \epsilon c_\bullet      & \epsilon s_\bullet \cr
-\epsilon s_\odot  & -\epsilon c_\odot         & c_\bullet    & s_\bullet \cr
0          & 0          & -s_\bullet   & c_\bullet \cr}
\right ) \;\; .
\end{equation}
At this stage the texture of the neutrino mass matrix $M_\nu$
can straightforwardly be recast from its mass eigenvalues and $V'$,
as shown in (5.3). 

The four-neutrino mixing pattern $V'$ with 
$\theta_{\rm atm} \approx 45^{\circ}$ may generally arise from the following
neutrino mass matrix \cite{4N,Mohapatra98}:
\begin{equation}
M_\nu \; = \; m \left ( \matrix{
\delta_1   & \delta_2   & 0    & 0 \cr
\delta_2   & 0         & 0   & \delta_3 \cr
0    & 0     & \delta_4    & 1 \cr 
0    & \delta_3    & 1    & \pm \delta_4 \cr} \right ) \;\; ,
\end{equation}
where $m$ measures the mass scale of the LSND neutrino oscillations
(i.e., $\Delta m^2_{\rm LSND} \approx m^2$), and the other
dimensionless parameters satisfy $|\delta_i| \ll 1$ 
(for $i=1,\cdot\cdot\cdot, 4$). We obtain
$\tan 2\theta_{\rm sun} \approx 2|\delta_2/\delta_1|$ 
and $\epsilon \approx |\delta_3|$, in addition to
$\theta_{\rm atm} \approx 45^{\circ}$.
The relative magnitudes of
$\delta_i$ can properly be chosen in order to describe the small-angle
MSW solution ($|\delta_2| \ll |\delta_1|$; $|\delta_4| \ll |\delta_3|\ll 1$), 
the large-angle MSW solution ($|\delta_1|, |\delta_2|, |\delta_4|
\ll |\delta_3|\ll 1$), or the vacuum oscillation solution 
($|\delta_1|\ll |\delta_2|\ll |\delta_4| \ll |\delta_3| \ll 1$)
to the solar neutrino problem \cite{4N}. In any case the desired mass
hierarchy $m_0, m_1 \ll m_2, m_3$ as illustrated in Fig. 2.1(a)
appears; i.e., the neutrino mass spectrum is characterized by
two nearly degenerate mass pairs separated by a mass gap of
$m \sim 1$ eV. 

The consequences of this simple four-neutrino mixing scenario on the
long-baseline neutrino experiments have been explored in
Ref. \cite{4N} in some detail. Also a theoretical attempt
towards understanding such a specific texture of $M_\nu$ has
been made in Ref. \cite{Mohapatra98}. One can certainly find out some
other interesting textures of the $4\times 4$ neutrino mass matrix  
to accommodate the present solar, atmospheric
and LSND neutrino oscillation data \cite{4Nothers}. 
Whether there exists
such a light sterile neutrino in nature remains an open question
and could be resolved by the future neutrino experiments
(see, e.g., Ref. \cite{Kayser99} for a comprehensive discussion). On
the theoretical side, the sterile neutrino can be introduced
as an extra fermion to the standard model and its mass can
be suppressed by use of the radiative mechanism. It is also
possible to double the particle spectrum and the gauge forces of the
standard model by including a mirror sector to it and inducing
the Majorana neutrino masses via non-renormalizable operators.
In this framework the lightest neutrino of the mirror sector
will be identified as the sterile neutrino. For a review of
the relevant theoretical models of sterile neutrinos, we refer
the reader to Ref. \cite{Mohapatra99}.

\subsection{Scale dependence of the neutrino mass matrix}

So far we have discussed some instructive scenarios of lepton
mass matrices at low energy scales, at which their consequences
on neutrino oscillations can directly be confronted with the
experimental data. From the theoretical point of view, however,
a phenomenologically favored texture of lepton mass matrices
might only serve as the low-scale approximation of a more 
fundamental model responsible for the fermion mass generation
at superhigh energy scales. It is therefore desirable to
investigate the scale dependence of lepton mass matrices,
as that of quark mass matrices, by use of the renormalization-group
equations. The running behaviors of the neutrino mass matrix
is, however, strongly model-dependent. To be specific we only consider
the possibility that the light Majorana neutrino masses are
generated by the conventional seesaw mechanism with a 
singlet-neutrino mass scale $M_0 \sim 10^{13}$ GeV \cite{Ellis}
\footnote{Certainly the heavy Majorana neutrino masses 
are unnecessary to be degenerate, and the structure of 
$M_{\rm R}$ may have non-negligible effects on 
the renormalization-group equations of $M_\nu$. For
the purpose of illustration we simply assume that 
$M_{\rm R}$ is characterized 
by a single mass scale $M_0$.}.
Below this mass scale the Yukawa couplings of Dirac neutrinos
become decoupled. Thus the running of the left-handed
neutrino mass matrix $M_\nu$ with the energy scale $\mu$ 
can be described, in the framework
of the minimal supersymmetric standard model, by \cite{RGE93}
\begin{equation}
16 \pi^2 \frac{{\rm d} M_\nu}{{\rm d} \chi}
\; =\; \left [ - \left (\frac{6}{5} g^2_1 + 6 g^2_2 \right )
+ 6 {\rm Tr} \left (Y_{\rm u} Y^{\dagger}_{\rm u} \right ) \right ] M_\nu 
+ \left (Y_l Y^{\dagger}_l \right ) M_\nu
+ M_\nu \left (Y_l Y^{\dagger}_l \right )^{\rm T} \; ,
\end{equation}
where $\chi = \ln (\mu/M_0)$; $g^{~}_i$ (for $i=1$ and 2) denote
the gauge couplings; $Y_{\rm u}$ and $Y_l$ are the Yukawa 
coupling matrices of the up-type quarks and the charged leptons,
respectively.

In the flavor basis where the charged lepton mass matrix is
diagonal, one can simplify the renormalization-group equation (5.49)
and obtain the running behaviors of individual elements of $M_\nu$
at the weak scale $\mu = M_Z$.
Neglecting the tiny contributions from up and charm quarks and
defining the evolution functions
\begin{eqnarray}
I_g & = & \exp \left [ +\frac{1}{16\pi^2}
\int^{\ln (M_0/M_Z)}_0 \left ( \frac{6}{5}g^2_1(\chi) +
6 g^2_2(\chi) \right ) {\rm d}\chi \right ] \;\; ,
\nonumber \\
I_t & = & \exp \left [ -\frac{1}{16\pi^2}
\int^{\ln (M_0/M_Z)}_0 f^2_t(\chi) {\rm d}\chi \right ] \;\; ,
\nonumber \\
I_\alpha & = & \exp \left [ -\frac{1}{16\pi^2}
\int^{\ln (M_0/M_Z)}_0 f^2_\alpha (\chi) {\rm d}\chi \right ] \;\; ,
\end{eqnarray}
where $f_t$ and $f_\alpha$ (for $\alpha = e, \mu, \tau$)
are the Yukawa coupling eigenvalues of the top quark and
charged leptons, one can solve (5.49) and
arrive at
\begin{equation}
M_\nu (M_Z) \; =\; \left (I_g I^6_t \right ) ~ T_l \cdot M_\nu (M_0) \cdot T_l \;
\end{equation}
with $T_l = {\rm Diag} \{ I_e, ~ I_\mu, ~ I_\tau \}$.
The overall factor $(I_g I^6_t)$ in (5.51) does not affect the 
relative magnitudes of the matrix elements of $M_\nu$. 
Only the matrix $T_l$, which amounts to the unity matrix
at the mass scale $M_0$, can modify the texture of
the light neutrino mass matrix from $M_0$ to $M_Z$.
The mass hierarchy of three charged leptons 
(i.e., $f_e < f_\mu < f_\tau$) implies
$I_e > I_\mu > I_\tau$ at any energy scale below $M_0$.
If three neutrinos were exactly degenerate (i.e.,
$m_1 = m_2 = m_3$) at the scale
$M_0$, they would become non-degenerate and have the
spectrum $m_1 > m_2 > m_3$ at a lower scale like $M_Z$.
Indeed the magnitude of $I_\tau$ may substantially deviate 
from unity, if $\tan\beta_{\rm susy}$ (the ratio of
Higgs vacuum expectation values) takes large values.
In contrast, $I_e \approx I_\mu \approx 1$ is expected to be
a good approximation.

To be more explicit, we take $m_t (M_Z) = 175$ GeV,
$m_b (M_Z) = 2.9$ GeV, $m_\tau (M_Z) = 1.777$ GeV,
$g^2_1(M_Z) =0.127$ and $g^2_2(M_Z) =0.42$ to calculate 
the evolution factors $I_g$, $I_t$ and $I_\tau$ in
the framework of the minimal supersymmetric standard model.
We obtain $I_g \approx 1.6$. The magnitudes of $I_t$ and
$I_\tau$ depend on the input value of $\tan\beta_{\rm susy}$.
For example, $I_t \approx 0.89$ and $I_\tau \approx 1.0$ for
$\tan\beta_{\rm susy} =10$; while $I_t \approx 0.86$ and
$I_\tau\approx 0.9$ for $\tan\beta_{\rm susy} =60$. 
We see that the overall factor of $M_\nu (M_Z)$ takes the value 
$(I_g I^6_t) \approx 0.80$ and 0.65,
respectively, for $\tan\beta_{\rm susy} =10$ and 60. 
The specific texture of $M_\nu (M_Z)$ is different from that
of $M_\nu (M_0)$, only because of the deviation of $I_\tau$ from unity. 

For illustration we take the general symmetric
texture of $M_\nu (M_0)$ in the basis where $M_l (M_0)$
is diagonal:
\begin{equation}
M_\nu (M_0) \; =\; \left ( \matrix{
E_\nu (M_0)   & ~ D_\nu (M_0) ~    & F_\nu (M_0) \cr
D_\nu (M_0)   & ~ C_\nu (M_0) ~   & B_\nu (M_0) \cr
F_\nu (M_0)   & ~ B_\nu (M_0) ~  & A_\nu (M_0) \cr} \right ) \; .
\end{equation}
With the help of (5.51), we obtain the renormalized
neutrino mass matrix at the scale $M_Z$ as follows:
\begin{equation} 
M_\nu (M_Z) \; =\; \left ( \matrix{
E_\nu (M_Z)   & D_\nu (M_Z)    & F_\nu (M_Z) \cr
D_\nu (M_Z)   & C_\nu (M_Z)   & B_\nu (M_Z) \cr
F_\nu (M_Z)   & B_\nu (M_Z)   & A_\nu (M_Z) \cr} \right ) \; ,
\end{equation}
where
\begin{eqnarray}
A_\nu (M_Z) & = & \left (I_g I^6_t \right ) ~ I^2_\tau ~ A_\nu (M_0) \;\; ,
\nonumber \\
B_\nu (M_Z) & = & \left (I_g I^6_t \right ) ~ I_\mu I_\tau ~ B_\nu (M_0) \;\; ,
\nonumber \\
C_\nu (M_Z) & = & \left (I_g I^6_t \right ) ~ I^2_\mu ~ C_\nu (M_0) \;\; ,
\nonumber \\
D_\nu (M_Z) & = & \left (I_g I^6_t \right ) ~ I_e I_\mu ~ D_\nu (M_0) \;\; ,
\nonumber \\
E_\nu (M_Z) & = & \left (I_g I^6_t \right ) ~ I^2_e ~ E_\nu (M_0) \;\; ,
\nonumber \\
F_\nu (M_Z) & = & \left (I_g I^6_t \right ) ~ I_e I_\tau ~ F_\nu (M_0) \;\; .
\end{eqnarray}
An interesting feature of $M_\nu (M_Z)$ would be that its
possible texture zeros remain the same as those of $M_\nu (M_0)$,
up to the degree of accuracy taken in deriving (5.51). 
This result could in some sense support our conjecture 
made in section 5.3; i.e., the seesaw-invariant
texture of lepton mass matrices, which might arise from a
dynamical mechanism of fermion mass generation, holds essentially at
both superhigh and low energy scales.

Of course the evolution of $M_\nu$ from $M_0$ to
$M_Z$ will in general affect the form of the lepton flavor mixing
matrix $V (M_0)$. This is naturally expected, 
as the flavor mixing angles are sensitive to the off-diagonal
elements of the neutrino mass matrix. Therefore 
the mixing pattern $V (M_Z)$, in particular its
$(\nu_\mu, \nu_\tau)$ and $(\nu_e, \nu_\tau)$
sectors, might depart significantly
from $V (M_0)$ if $I_\tau$ deviates substantially from unity.
The explicit running behaviors of the flavor mixing matrix
elements are, however, correlated in a complicated way with those of
the neutrino mass eigenvalues
(see, e.g., Ref. \cite{Casas} for detailed
calculations of $V(M_Z)$ from $V(M_0)$ with different spectra of
neutrino masses).

It is possible to invoke another model in which the
light Majorana neutrino masses are simply generated by
a new non-renormalizable interaction with dimension-5 mass
operators at the scale $\Lambda \sim 10^5$ GeV. 
In this case there is no Dirac neutrino coupling to be 
renormalized \cite{Ellis},
thus the running effect of $M_\nu$ below the scale $\Lambda$
can be described by the same formulae given above. 
As $\ln (\Lambda/M_Z) \sim 7$, the deviation of 
$I_\tau$ (as well as $I_t$) from unity is insignificant 
and even negligible. Then we are left with 
$M_\nu (M_Z) \approx M_\nu (\Lambda)$ and
$V (M_Z) \approx V (\Lambda)$. 

\subsection{$CP$ violation in long-baseline neutrino experiments}

It is known that within the framework of three lepton families
the strength of $CP$ violation is universal
in $\nu_e \rightarrow \nu_\mu$,
$\nu_\tau \rightarrow \nu_e$ and $\nu_\tau \rightarrow \nu_e$
transitions (see Ref. \cite{CP} as well as (3.22) and (3.23) for details).
The asymmetry between the probabilities of two $CP$-conjugate
appearance processes is uniquely given as
\begin{eqnarray}
\Delta_{CP} & = & P(\nu_\alpha \rightarrow \nu_\beta) - P(\bar{\nu}_\alpha
\rightarrow \bar{\nu}_\beta) \; \nonumber \\
& = & -16 {\cal J}_l \sin F_{\rm sun} \sin^2 F_{\rm atm} \; ,
\end{eqnarray}
where $(\alpha, \beta) = (e,\mu)$, $(\mu, \tau)$ or $(\tau, e)$;
${\cal J}_l$ is the universal $CP$-violating parameter;
$F_{\rm sun}$ and $F_{\rm atm}$ measure the frequencies of solar
and atmospheric neutrino oscillations, respectively.
The $T$-violating asymmetry can be obtained in a
similar way:
\begin{eqnarray}
\Delta_T & = & P(\nu_\alpha \rightarrow \nu_\beta)
- P(\nu_\beta \rightarrow \nu_\alpha) \; 
\nonumber \\
& = & -16 {\cal J}_l \sin F_{\rm sun} \sin^2 F_{\rm atm} \; ,
\nonumber \\ \nonumber \\
\bar{\Delta}_T & = & P(\bar{\nu}_\alpha \rightarrow \bar{\nu}_\beta)
- P(\bar{\nu}_\beta \rightarrow \bar{\nu}_\alpha) \; 
\nonumber \\
& = & +16 {\cal J}_l \sin F_{\rm sun} \sin^2 F_{\rm atm} \; .
\end{eqnarray}
These formulas show clearly that $CP$ or $T$ violation is a property
of all three lepton families. The relationship $\Delta_T 
= \Delta_{CP}$ is a straightforward consequence
of $CPT$ invariance, and $\bar{\Delta}_T = -\Delta_T$ indicates
that the $T$-violating measurable is an odd function of time. 

The $CP$- and $T$-violating signals can in principle be
measured in the long-baseline neutrino experiments \cite{LBL}.
As $\Delta_{CP}$ and $\Delta_T$ depend linearly on
the oscillation term $\sin F_{\rm sun}$, the length
of the baseline suitable for detecting $CP$ and $T$ asymmetries
should satisfy the condition $L\sim E/\Delta m^2_{\rm sun}$.
This requirement singles out 
the large-angle MSW solution, which has $\Delta m^2_{\rm sun} \sim 10^{-5}$
to $10^{-4} ~ {\rm eV}^2$ and $\sin^2 2\theta_{\rm sun} \sim 0.65$ 
to $1$ \cite{Bahcall98}, among
three possible solutions to the solar neutrino problem.
The small-angle MSW solution poses a problem, because
it does not give rise to 
a relatively large magnitude of ${\cal J}_l$, which determines 
the significance of practical $CP$- or $T$-violating signals.
The long wave-length vacuum oscillation solution requires $\Delta m^2_{\rm sun}
\sim 10^{-10}$ eV$^2$, too small to meet the realistic 
long-baseline prerequisite.

The observation of $\Delta_T$ might be free from the
matter effects of the earth, which is possible to fake the genuine
$CP$ asymmetry $\Delta_{CP}$ in any long-baseline neutrino
experiment. The joint measurement of $\nu_\alpha \rightarrow
\nu_\beta$ and $\nu_\beta \rightarrow \nu_\alpha$ transitions to
determine $\Delta_T$ is, however, a challenging task in practice.
Probably it could only be realized in a neutrino factory,
where high-quality neutrino beams can be produced with high-intensity 
muon storage rings \cite{Muon}.

To illustrate the earth-induced matter effects, we write out the
effective Hamiltonians for neutrinos and antineutrinos \cite{CPmatter}:
\begin{eqnarray}
{\cal H}_\nu & = & \frac{1}{2E} \left [ ~ V \left (\matrix{
m^2_1     & 0      & 0 \cr
0         & m^2_2  & 0 \cr
0         & 0      & m^2_3 \cr} \right ) V^{\dagger}
~ + ~ \left ( \matrix{
a     & 0     & 0 \cr
0     & 0     & 0 \cr
0     & 0     & 0 \cr} \right ) \right ] \; \; ,
\nonumber \\
{\cal H}_{\bar{\nu}} & = & \frac{1}{2E} \left [ V^* \left (\matrix{
m^2_1     & 0      & 0 \cr
0         & m^2_2  & 0 \cr
0         & 0      & m^2_3 \cr} \right ) V^{\rm T} 
~ - \; \left ( \matrix{
a     & 0     & 0 \cr
0     & 0     & 0 \cr
0     & 0     & 0 \cr} \right ) \right ] \;\; ,
\end{eqnarray}
where $V$ and $m_i$ (for $i=1, 2, 3$) denote the flavor
mixing matrix and neutrino mass eigenvalues, respectively,
in vacuum; 
and $a = 2\sqrt{2} G_{\rm F} N_e E \sim {\cal O}(10^{-4}) ~{\rm eV}^2
\cdot E/[{\rm GeV}]$ describes the
charged current interaction with electrons ($N_e$ and $E$
stand for the background density of electrons and the
neutrino energy, respectively). Once $a \sim \Delta m^2_{21}$
or $\Delta m^2_{32}$ for given values of $E$, significant
matter effects will enter neutrino oscillations. In this
case the diagonalization of ${\cal H}_\nu$ and
${\cal H}_{\bar \nu}$ by two different unitary matrices
leads to the {\it effective} neutrino mass eigenvalues
which deviate somehow from the genuine ones. Moreover 
the opposite signs of $a$ in ${\cal H}_\nu$ and
${\cal H}_{\bar \nu}$ signify that the background matter
is not $CP$ invariant. This provokes a fake $CP$ 
asymmetry between neutrino and antineutrino transitions,
whose magnitude could in some cases be comparable with or dominant over
the genuine $CP$ asymmetry measured by ${\cal J}_l$.
Roughly speaking, the longer the baseline, which in turn requires the
higher neutrino beam energy, the larger the matter effect on
leptonic $CP$ violation \cite{LBL}.

In practical experiments one might prefer to measure the $CP$ 
asymmetry $\Delta_{CP}$ normalized by the sum of two
$CP$-conjugate transition probabilities. Such a signal is
particularly significant for $\nu_\mu \rightarrow \nu_e$ and
$\bar{\nu}_\mu \rightarrow \bar{\nu}_e$ oscillations,
because their $CP$-conserving and $CP$-violating parts
are comparable in magnitude (see (3.23) for illustration).
To be specific we adopt the nearly bi-maximal neutrino
mixing scenario introduced in section 5.2 and calculate the 
$CP$-violating observable in vacuum \cite{FX99CP}:
\begin{eqnarray}
{\cal A} & = & \frac{P(\nu_\mu \rightarrow \nu_e) ~ - ~ P(\bar{\nu}_\mu
\rightarrow \bar{\nu}_e)}{P(\nu_\mu \rightarrow \nu_e) ~ + ~
P(\bar{\nu}_\mu \rightarrow \bar{\nu}_e)} 
\nonumber \\ \nonumber\\
& = & \frac{P(\nu_\mu \rightarrow \nu_e) ~ - ~ P(\nu_e \rightarrow \nu_\mu)}
{P(\nu_\mu \rightarrow \nu_e) ~ + ~ P(\nu_e \rightarrow \nu_\mu)}
\nonumber \\ \nonumber \\
& \approx & \frac{\displaystyle \frac{8}{\sqrt{3}} \sqrt{\frac{m_e}{m_\mu}}}
{\displaystyle \frac{16}{3} \frac{m_e}{m_\mu}
+ \left (\frac{\sin F_{\rm sun}}{\sin F_{\rm atm}} \right )^2} 
~ \sin F_{\rm sun} \; .
\end{eqnarray}
Let us typically take the baseline length to be $L = 732$ km or $L = 7332$ km
for a neutrino source at Fermilab pointing toward the Soudan mine 
in Minnesota or the Gran Sasso underground laboratory in 
Italy \cite{Muon}.
The mass-squared differences are chosen as
(a) $\Delta m^2_{\rm sun} = 5 \times 10^{-5} ~ {\rm eV}^2$ and (b)
$\Delta m^2_{\rm sun} = 10^{-4} ~ {\rm eV}^2$ versus the fixed
$\Delta m^2_{\rm atm} = 10^{-3} ~ {\rm eV}^2$. The behavior of
${\cal A}$ changing with the beam energy $E$ in the range
$3 ~{\rm GeV} \leq E \leq 20 ~{\rm GeV}$ 
is shown in Fig. 5.2. We see that the asymmetry ${\cal A}$ 
can be of ${\cal O}(0.1)$ to ${\cal O}(1)$.
The matter effect on ${\cal A}$ must be 
taken into account, in order to extract the genuine $CP$-odd
parameters. For the model under consideration,
the smallness of $|V_{e3}|$ ($\approx 0.056$) together with the maximal
$CP$ violating phase ($\phi \approx 90^{\circ}$) is expected to make 
the possible matter effect insignificant, and unable to completely
fake the genuine $CP$-violating signals (see, e.g.,
Ref. \cite{FX99CP} and references therein).
\begin{figure}[t]
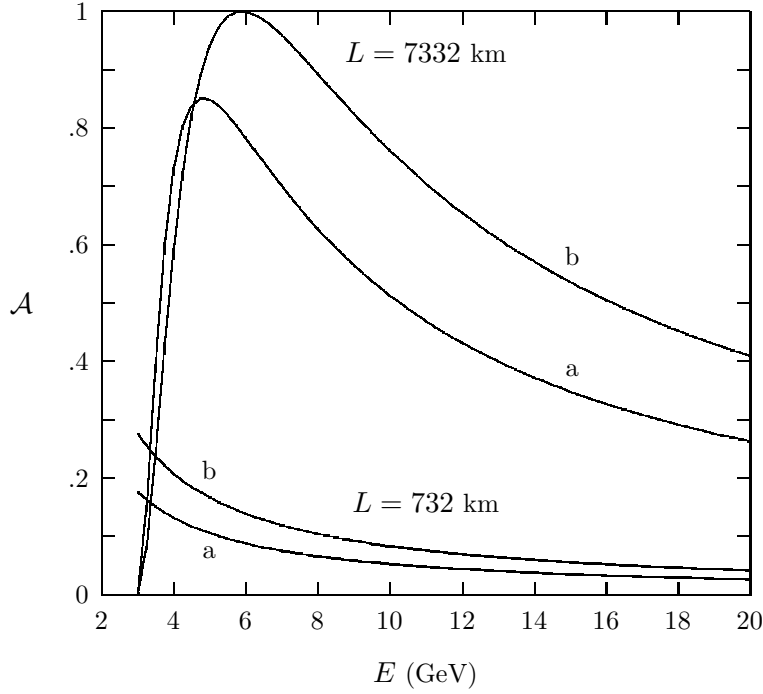

\setlength{\unitlength}{0.240900pt}
\ifx\plotpoint\undefined\newsavebox{\plotpoint}\fi
\sbox{\plotpoint}{\rule[-0.200pt]{0.400pt}{0.400pt}}%

\vspace{0.25cm}
\caption{Illustrative plot for the $CP$-violating asymmetry ${\cal A}$
between $\nu_\mu \rightarrow \nu_e$ and $\bar{\nu}_\mu \rightarrow
\bar{\nu}_e$ transitions in vacuum, 
changing with the neutrino beam energy $E$. Here
(a) $\Delta m^2_{\rm sun} = 5\times 10^{-5} ~ {\rm eV}^2$ 
and (b) $\Delta m^2_{\rm sun} = 10^{-4} ~ {\rm eV}^2$ versus the fixed 
$\Delta m^2_{\rm atm} = 10^{-3} ~ {\rm eV}^2$
have typically been taken in the case of the baseline length
$L=732$ km or $L=7332$ km.}
\end{figure} 

Note again that in the framework of three neutrino families
the leptonic $CP$ violation could be observable, if and only
if the large-angle MSW oscillation is the true solution to
the solar neutrino problem. The situation can dramatically
change, when the four-neutrino mixing scheme is adopted to
interpret solar, atmospheric and LSND neutrino oscillation
data. The most general mixing matrix of four Majorana 
(3 active and 1 sterile) neutrinos consists of 6 rotation angles and
6 $CP$-violating phases, as already counted in section 3.1. However, the
mixing angles of the sterile neutrino $\nu_{\rm s}$ with 
the active neutrinos $\nu_\mu$ and $\nu_\tau$ are expected to be very small
(see the arguments given in section 5.4). In this case one may 
approximate the $4\times 4$ lepton flavor mixing matrix $V$,
which links the neutrino mass eigenstates $(\nu_0, \nu_1, \nu_2, \nu_3)$
to the neutrino flavor eigenstates 
$(\nu_{\rm s}, \nu_e, \nu_\mu, \nu_\tau)$,
by taking $\cos \theta_{02} \approx \cos\theta_{03} \approx 1$.
Explicitly we have the following approximate form of $V$:
\begin{equation}
V \; \approx \; \left ( \matrix{
c_{01}     & s^*_{01}     & s^*_{02}     & s^*_{03} \cr
-s_{01}    & c_{01}       & s^*_{12}     & s^*_{13} \cr
V_{\mu 0}     & V_{\mu 1}       & c_{23}       & s^*_{23} \cr
V_{\tau 0}     & V_{\tau 1}       & -s_{23}      & c_{23}   \cr} \right ) \;\; ,
\end{equation}
where $s_{ij} \equiv \sin \theta_{ij} e^{{\rm i}\phi_{ij}}$ and
$c_{ij} \equiv \cos\theta_{ij}$ (for $i, j =0, 1, 2, 3$ and $i < j$),
and the expressions of $V_{\mu 0}$, $V_{\mu 1}$, $V_{\tau 0}$ and
$V_{\tau 1}$ can be obtained by using the leading order
unitarity conditions. Based on this flavor mixing pattern 
as well as the mass spectrum given in (5.43), one may carry
out an analysis of $CP$ violation in 
$\nu_e \rightarrow \nu_\mu$, $\nu_\mu \rightarrow \nu_\tau$
and $\nu_\tau \rightarrow \nu_e$ oscillations. 
For example, the $CP$ asymmetry between 
$\nu_\mu \rightarrow \nu_e$ and $\bar{\nu}_\mu \rightarrow \bar{\nu}_e$
transitions reads
\begin{eqnarray}
\Delta_{\mu e} & = & P(\nu_\mu \rightarrow \nu_e)  - 
P(\bar{\nu}_\mu \rightarrow \bar{\nu}_e) \; \nonumber \\
& \approx & 16 {\cal J}_l \sin F_{\rm atm} \sin^2 F^{~}_{\rm LSND} \; ,
\end{eqnarray}
where ${\cal J}_l$ is the conventional rephasing-invariant measure of 
$CP$ violation in the three-neutrino mixing framework;
$F^{~}_{\rm LSND} \propto \Delta m^2_{\rm LSND}$ and
$F_{\rm atm} \propto \Delta m^2_{\rm atm}$ 
describe the oscillation frequencies of LSND and atmospheric 
neutrinos, respectively. Note that
the $CP$-violating effect arising from the interference between
sterile and active neutrinos is negligible in $\Delta_{\mu e}$, as a 
straightforward consequence of the approximation made in (5.59).
Maximizing
the contributions of $\phi_{ij}$ to the relevant $CP$-violating
observables in an appropriate way, one can find that the
signal of $CP$ violation is much more significant and the matter 
effect is much smaller than those in the conventional three-neutrino 
mixing framework \cite{LBL}.
The $CP$-violating asymmetry between $\nu_\mu \rightarrow \nu_e$ and
$\bar{\nu}_\mu \rightarrow \bar{\nu}_e$ transitions, i.e.,
the asymmetry $\Delta_{\mu e}$ obtained in (5.60),
remains promising in practical experiments.

Let us end this section with one remark. 
The lepton flavor mixing angles depend in general not only on
the ratios of charged lepton masses but also on those of 
neutrino masses, unless
their contributions to flavor mixing are forbidden by
some special flavor symmetries. Therefore the mixing factors
of various neutrino oscillations are expected to have 
correlations with the corresponding neutrino mass-squared
differences. In other words, $\sin^2 2 \theta$ and $\Delta m^2$
may not be two completely independent parameters in the
true theory of lepton masses and flavor mixing. All of the present
model-independent analyses of $CP$ violation in neutrino
oscillations have tried to maximize the observable effects
by adjusting the unknown parameters arbitrarily, i.e., 
regardless of the possible parameter correlation. We hope that
progress in both theory and experiments would finally allow us
to gain more insights into the problems under discussion
and lead us to a full understanding of
fermion mass generation and $CP$ violation.

\section{Concluding remarks}

We have given an overview of phenomenological
studies of fermion mass and flavor mixing schemes. Particular attention
has been paid to the underlying flavor symmetries which can lead to
realistic textures of quark and lepton mass matrices.

With the current experimental data on quark masses and flavor mixing
angles, we find that only very few specific patterns of quark mass
matrices are realistic. These patterns lead to different predictions
for the sides and angles of the quark unitarity triangles, thus it
is possible to distinguish one from another and to identify the 
``right'' one by improved measurements of the quark mixing angles and 
of $CP$ violation in the near future. 

In comparison, the present experimental results for neutrino
oscillations remain rather preliminary, although they do provide 
strong hints for possible patterns of lepton flavor mixing and
the associated textures of lepton mass matrices. For illustration
we have described several interesting schemes of lepton mass matrices,
whose predictions can be 
examined in a variety of upcoming neutrino oscillation experiments. 
Nevertheless it should not be a surprise, if none of the currently proposed
neutrino mass and mixing scenarios is realized by nature.
But one may hope that at least some of the ideas explored in these
attempts will survive the test of future measurements. 

While the salient features of flavor mixing and $CP$ violation
have well been illustrated by a number of simple and instructive models of
fermion mass matrices, some other interesting models proposed
in the literature cannot be included in this article due to limitation of space.
We therefore refer the reader to Refs. \cite{Langacker98,Raby95,THmore},
in which the theoretical prospects for a deeper understanding of
fermion mass generation and flavor mixing are described.
We also refer the reader to Refs. \cite{Kayser99,BaBar,EXreview},
which elaborate the experimental prospects to determine the
$CP$-violating phases of quark mixing, to pin down the true 
mechanism of neutrino oscillations, and to measure the neutrino
masses and lepton mixing parameters. 

Certainly much more efforts are needed, both experimentally
and theoretically, to gain a deeper
insight into the generation of fermion masses, the pattern of
flavor mixing, and the origin of $CP$ violation in
a unified picture of leptons and quarks. Let us finish
with Max Jammer's remarks \cite{Jammer}:

{\it ``The modern physicist may rightfully be proud of his spectacular
achievements in science and technology. However, he should always be
aware that the foundations of his imposing edifice, the basic notions
of his discipline, such as the concept of {\it mass}, are entangled with
serious uncertainties and perplexing difficulties that have as yet not
been resolved.''}


\begin{thebibliography}{99}
\bibitem{3By4} Of course one may take the lepton flavor mixing
matrix to be of the $3\times 4$ form, in order to incorporate
the extra sterile neutrino with has no partner in the charged
lepton sector.
\end{thebibliography}
\end{document}